# 6G: From Connectivity Infrastructure to Guaranteed Digital Services

**David Soldani (Senior Member, IEEE), Petrit Nahi, Awn Muhammad, Nikhil Dwivedi, Francesco Monaco, and Francis Jebamani**

Rakuten Mobile Inc., Tokyo 158-0094, Japan
Corresponding author: David Soldani (e-mail: david.soldani@rakuten.com).

**ABSTRACT** Sixth-generation (6G) mobile networks offer an opportunity to align network architecture with operator control, verifiable service outcomes, and governed automation. This article develops four complementary contributions. The *Control Compact* allocates ownership and governance responsibilities through an own–federate–consume taxonomy. The *Guarantee Economy* proposes six outcome-priced service tiers aligned with IMT-2030 usage scenarios. The *Network MCP Platform*, the principal contribution, is a conceptual reference architecture that uses the Model Context Protocol (MCP) to connect agentic workflows to operational tools and telecom execution interfaces under operator-defined governance. Two real-network examples from Rakuten Mobile illustrate MCP-enabled agentic workflows: radio access network (RAN) second-line (L2) support combines diagnosis, remediation, and outcome verification, while IP Multimedia Subsystem (IMS) availability investigation combines evidence collection, adaptive investigation, and reporting. These examples support operational feasibility; evaluation of the complete platform, and its performance and security properties, remains future work. A separate TM Forum Level 4 assessment corroborates deployed closed-loop autonomy for a scoped RAN energy-optimization scenario. The fourth contribution presents Rakuten Mobile's public standardization position and a phased deployment roadmap. The analysis distinguishes operational records, external assessment evidence, standards-based requirements, and proposed architectural and commercial extensions. It identifies validation requirements for broader deployment and outcome-based service assurance.

## I. INTRODUCTION

Mobile communication has connected more than five billion unique subscribers [1]. Successive generations have expanded network capabilities, while operators continue to face integration and governance challenges associated with proprietary implementations, restricted data access, and vendor-specific management interfaces. These dependencies can constrain an operator's ability to modify network behavior, evaluate embedded artificial intelligence (AI), and develop differentiated services.

The International Telecommunication Union Radiocommunication Sector (ITU-R) IMT-2030 framework [2]–[4] establishes the direction for sixth-generation (6G) mobile networks. Alongside performance improvements, an important architectural question is how operators can retain effective control of the software, data, and interfaces on which service delivery depends. Procurement and design choices during the 6G study and specification period will influence these capabilities.

This article organizes the discussion around five priorities, illustrated in Fig. 1: *Control First*, operator governance of the software-defined control plane; *Customer First*, verifiable service outcomes; *Business First*, outcome-based commercial models; *Operations First*, software-driven operations with governed agentic AI; and *Technology Last*, requirements-led selection of enabling technologies.

Technology Last denotes requirements-led technology selection, not a reduction in the importance of technical innovation. Candidate technologies include sub-terahertz (sub-THz) spectrum, reconfigurable intelligent surfaces (RIS), and integrated sensing and communication (ISAC). The argument is about *sequencing*: technology selection should follow from the four preceding priorities.

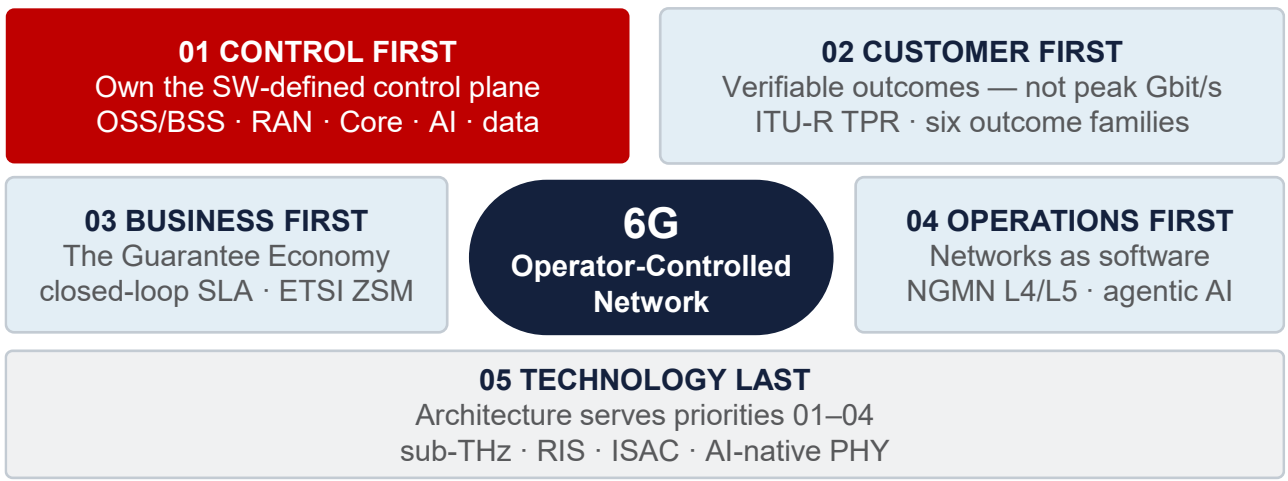

**FIGURE 1. Five-axis priority reordering for the 6G era. Technology Last is a sequencing principle, not a downgrade. Each block represents an operator investment and architectural decision set.**

Rakuten Mobile provides evidence that Control First is achievable at scale: its network runs on a fully disaggregated, software-defined architecture controlling the complete stack from the distributed unit (DU) to the core and orchestration layer. Public FY2025 full-year earnings before interest, taxes, depreciation, and amortization (EBITDA) disclosure [5] provides contextual commercial evidence, but it is not treated here as proof of the article's broader architectural thesis, as presented in the following sections.

The article makes four principal contributions, each graded by its evidentiary status:

- The *Network MCP Platform* is the principal contribution: a conceptual, MCP-enabled reference architecture integrating six operator control functions with telecom execution interfaces. The functions cover tool mediation, workload identity, telemetry and enforcement, lifecycle governance, risk-tiered validation, and intent translation. A proposed tool catalog and migration path connect this architecture to operational deployment. The radio access network (RAN) second-line (L2) support and IP Multimedia Subsystem (IMS) availability-investigation examples from Rakuten Mobile illustrate agentic workflows using MCP tools for operational execution, while platform-wide performance, safety, charging, and lawful-intercept integration remain to be validated. MCP, Secure Production Identity Framework for Everyone (SPIFFE), extended Berkeley Packet Filter (eBPF), digital twins, and agent-based management are existing building blocks; the claimed contribution is their telecom-specific integration and governance.
- *The Guarantee Economy*: a six-tier, outcome-priced commercial catalog – one tier per IMT-2030 usage scenario, with Sovereign Trust as a cross-cutting assurance premium – extending ETSI Zero-touch Network and Service Management (ZSM) and TM Forum IG1252 [6] with specific service-level objectives (SLOs), breach consequences, billing mechanisms, and the three enforcement prerequisites that condition commercial deployment.
- *A standards-grounded strategic framework*: The 6G Control Compact ownership mapping (Own | Federate | Consume), a six-family mapping of the IMT-2030 usage scenarios onto commercial tiers with quantified key performance indicators (KPIs) anchored in 3GPP TR 22.870 [7] and the ITU-R Technical Performance Requirements (TPR) [4], an integrated trust-sovereignty-sustainability compliance argument, and a three-phase deployment roadmap.
- *Rakuten Mobile's public position* on seven principal standards development organization (SDO) architecture decisions.

The evidence model is deliberately tiered. Public operational records document the cloud-native Open RAN substrate, large language model (LLM)-assisted infrastructure optimization, and eBPF observability [5], [8]–[10]. Standards and ecosystem documents ground the service, control, and roadmap requirements. Section III.C separately reports a scoped external TM Forum assessment, while Section VII.D presents two author-supplied operational workflow records. Both workflows use MCP tools, but, at the time of writing, neither is a controlled experiment or an independent audit. The complete Network MCP Platform and Guarantee Economy therefore remain proposals whose performance, security, and commercial properties require validation.

The remainder of this article is organized as follows. Section II positions the contributions against related work. Sections III–VI develop the four positive "reorderings": the Control Compact, the six outcome families, the Guarantee Economy, and the agentic AI operating model. Section VII presents the Network MCP Platform, MCP-enabled RAN and IMS agentic workflow evidence, and the remaining validation requirements. Section VIII introduces Rakuten Mobile's public position on the 6G standardization landscape. Section IX describes the end-to-end 6G architecture (Technology Last). Section X addresses zero-trust security, sustainability, and data sovereignty. Section XI illustrates the phased roadmap and stakeholder implications. Section XII identifies open research questions. Conclusions are drawn in Section XIII.

## II. RELATED WORK

The proposed framework draws on six related research areas. The comparison concerns integration of operator governance, agentic tool use, and outcome-oriented service design. The analysis distinguishes demonstrated capabilities from standards-based requirements and proposed extensions; it does not claim that the constituent technologies are new.

*6G vision, requirements, and AI integration*. ITU-R IMT-2030 [2]–[4], 3GPP TR 22.870 [7], and the AI-RAN Alliance reports [11]–[13] establish the requirements and AI-RAN baseline; NTT DOCOMO [14] supplies a cyber-physical evolution view. Recent surveys and demonstrations cover AI across the physical layer and network management [15], [16]. This article adds an operator-governance synthesis: it allocates ownership, connects AI-RAN capabilities to outcome-oriented services, and defines how agent-facing tools remain bounded by telecom execution interfaces.

*Open RAN and RAN-Core convergence*. Salmi et al. [17] address real-time adaptation and conflict resolution in AI-

native Open RAN: their Conflict Management Engine exemplifies the mechanism any operator-owned RAN intelligent controller (RIC) fleet must implement to honor Guarantee-Economy SLOs across overlapping xApp scopes. The O-RAN next Generation Research Group (nGRG) reports on the near-real-time (Near-RT) RIC [18], distributed applications (dApps) [19], and scalable RAN [20] provide a research and prototype baseline for the three-tier intelligence architecture of Section IX (nGRG reports are contributed research reports, not O-RAN normative specifications). Harkous et al. [21] show that RAN-Core convergence – a distributed user plane function (UPF) co-located with the centralized unit user plane (CU-UP) – addresses latency constraints relevant to the proposed sub-10 ms edge-inference targets; this article frames that choice within the Control Compact.

*Autonomous network management and agentic AI.* ETSI ZSM [22]–[24], TM Forum IG1252 [6], and NGMN [25] define closed-loop and autonomy frameworks; related work covers zero-touch management [26], explainability [27], intent-to-code operation [28], and bounded LLM agents above deterministic control planes [29]. Rakuten Mobile's production study [8] demonstrates human-governed LLM-assisted resource recommendations, not an independently assigned autonomy level. Together, this literature supports policy-bounded agents and heterogeneous model placement; it does not establish the performance of the proposed platform.

*Security for autonomous and AI-native networks.* NIST SP 800-207 [30] establishes zero-trust principles and ETSI GR ZSM 017 [31] applies them to closed-loop automation. Choudhary et al. [32] frame the 5G-to-6G post-quantum transition as a multi-year sequencing problem, combining NIST-finalized module-lattice-based key-encapsulation mechanism (ML-KEM) [33] and module-lattice-based digital signature algorithm (ML-DSA) [34] with classical primitives. Altintaş et al. [35] categorize adversarial threats specific to AI-native 6G – physical-world, input-manipulation, model-level, and cross-domain – specifying the attack surface the ZSM 017 mechanisms must defend. Karahan et al. [36] demonstrate engineering-grade, multi-method explainable AI (XAI) – permutation importance, Shapley additive explanations (SHAP), local interpretable model-agnostic explanations (LIME), and partial dependence – that can inform model-interpretability assessment.

*6G standardization.* 3GPP TR 38.914 [37], the 3GPP 6G workshop contributions from Nokia [38] and Qualcomm [39], and Japan MIC spectrum planning [40] ground Section VIII. Rakuten Mobile's operational publications [5], [8], [9], [10] provide operator-specific deployment context.

The contribution is an architectural integration informed by operator workflows. Existing work addresses AI-native RAN functions, intent handling, tool exposure, closed-loop management, security, and agentic communication [17], [23], [25], [28], [29], [41]. The Network MCP Platform connects agent-facing MCP tools to telecom execution paths and makes operator responsibilities for identity, authorization, validation, audit, and recovery explicit. The RAN and IMS case studies illustrate this integration pattern through agentic workflows that invoke MCP tools for different operational tasks. They do not establish measured superiority over prior architectures or invention of the constituent technologies.

Network-slicing and service-level agreement (SLA) economics, intent-based networking, and broader LLM-agent network management directly overlap the Guarantee Economy and intent translator. They therefore belong within the comparison scope. The novelty claim is limited to the proposed operator-governance synthesis, its MCP-enabled telecom integration, and its connection to outcome-oriented service design.

*Intent-based networking and adjacent literatures.* Intent-based networking supplies the concept on which the intent translator of Section VII depends: intent as a declarative statement of operational goals and outcomes rather than of configuration, together with the lifecycle and assurance obligations that follow from it [42], surveyed systematically in [43]. The distinction drawn here is narrow. Intent handling is one subsystem, and the claimed contribution is the operator-side apparatus around it – workload identity, risk-tiered validation, audit, and named rollback ownership – together with the binding of intents to commercial remedies rather than to configuration alone. Appendix D compares this framework against ETSI ZSM, TM Forum IG1252, the NGMN agentic AI framework, the NTT DOCOMO vision, the O-RAN nGRG research baselines, the AI-RAN Alliance architecture, intent-based networking, and the LLM-agent network-management literature, stating for each what it provides, its evidence status, and what this article adds. The network-slicing and SLA-economics literature is represented there by two peer-reviewed optimization studies, with a broader review declared outstanding; Appendix C, (8)–(10), states the break-even condition that such a review would inform.

## III. CONTROL FIRST – THE 6G CONTROL COMPACT

### A. Operator Control and Integration Dependencies

Open standards do not necessarily imply interchangeable implementations or complete operational visibility. Across mobile generations, proprietary core and operations and business support system (OSS/BSS) implementations have required vendor-specific integration. The 4G transition to all-IP networking and the 5G introduction of service-based architecture (SBA) and network functions virtualization (NFV) expanded programmability, but management integration and access to operational data remain important concerns. The NGMN Framework for Network Simplification [44] identifies fragmentation as an operational-cost issue. Open RAN addresses part of this problem; control of the complete service lifecycle also requires interoperable core, cloud, transport, and management interfaces.

In this article, operator control means the ability to access relevant network data, inspect and modify operational policies, evaluate AI behavior, and migrate implementations within contractual and technical constraints. Licensed spectrum, radio-edge infrastructure, service accountability, and subscriber relationships provide a basis for differentiation, but their value also depends on *software* and data governance. The Control Compact proposes an allocation of these responsibilities rather than assuming *that* ownership alone guarantees operational or commercial benefits.

### *B. Three Phases of Telecom Evolution*

As illustrated in Fig. 2, the industry's trajectory comprises three phases.

- Phase 1 – *Connectivity (1G–4G)* – emphasizes coverage, capacity, and service availability, with differentiation strongly influenced by deployment scale and pricing.
- Phase 2 – *Programmability (5G/5G-Advanced)* – replaced appliances with software on commodity compute, but the transition has been partial: commercial network slicing depends on the maturity of orchestration, assurance, and service integration. Earlier work framing mobile networks as a software operating system [45] anticipated this transition; the Control Compact operationalizes it.
- Phase 3 – *Intelligent Growth (6G)* – casts the operator as a digital service platform owner: a network that senses application requirements, guarantees outcomes, exposes capabilities through application programming interfaces (APIs), and prices on value delivered [46]. Phase 3 emphasizes three capabilities: real-time AI-driven SLA assurance, an operator-owned data layer, and a business model mapping network outcomes to customer outcomes. Rakuten Mobile's evidence that the Phase 3 foundation is in place is architectural: a TM Forum-validated Level 4 assessment for scoped RAN energy optimization [47], a nationwide RIC deployment, and LLM-based infrastructure optimization in production [5], [8]. The FY2025 commercial results – revenue of JPY 374.7 billion (+32.0% year-on-year), first full-year EBITDA profitability, 10+ million subscribers, and a 48.8% higher annual average Rakuten Ichiba gross merchandise sales (GMS) for Rakuten Mobile subscribers than non-subscribers, with net average revenue per user (ARPU) of JPY 2,467 (+JPY 59 year on year) [5] – provide commercial context for that foundation rather than proof of the architectural thesis. The outcome-priced revenue model that fully defines Phase 3, however, remains in early commercialization (Sections V, IX.B, and XII).

The three phases provide an analytical model, not a uniform account of every operator's evolution. The proposed transition emphasizes AI integration, data governance, and verifiable service outcomes. Its feasibility depends on procurement, open-source participation, skills, integration costs, and the ability to demonstrate lifecycle value under the operator's deployment conditions.

### *C. Own, Federate, Consume*

The central design question for 6G operator strategy is not "*how much should we build versus buy?*" but "*what must we own to remain sovereign?*" Fig. 3 distinguishes the proposed Own, Federate, and Consume categories, according to strategic control requirements, opportunities for shared governance, and lifecycle economics; Table 1 maps the taxonomy across the full 6G stack.

The Control Compact proposes that operators own: (1) the *control plane* – the software layer expressing policy, enforcing service guarantees, managing spectrum, and coordinating cross-domain behavior; (2) the *AI substrate* – data pipelines, fine-tuning and adaptation, retrieval corpora, inference infrastructure, and feedback loops – AI the operator cannot audit, modify, or redirect cannot serve operator-specific objectives [25], [46]; frontier foundation-model weights are the deliberate exception: they are federated, licensed under the contractual protections defined below, while everything that grounds them in the network remains owned; (3) the *data layer* – telemetry, analytics, customer behavior, and performance records, the raw material for AI training and SLA verification; (4) the *OSS/BSS stack*; and (5) *spectrum policy and licensing*.

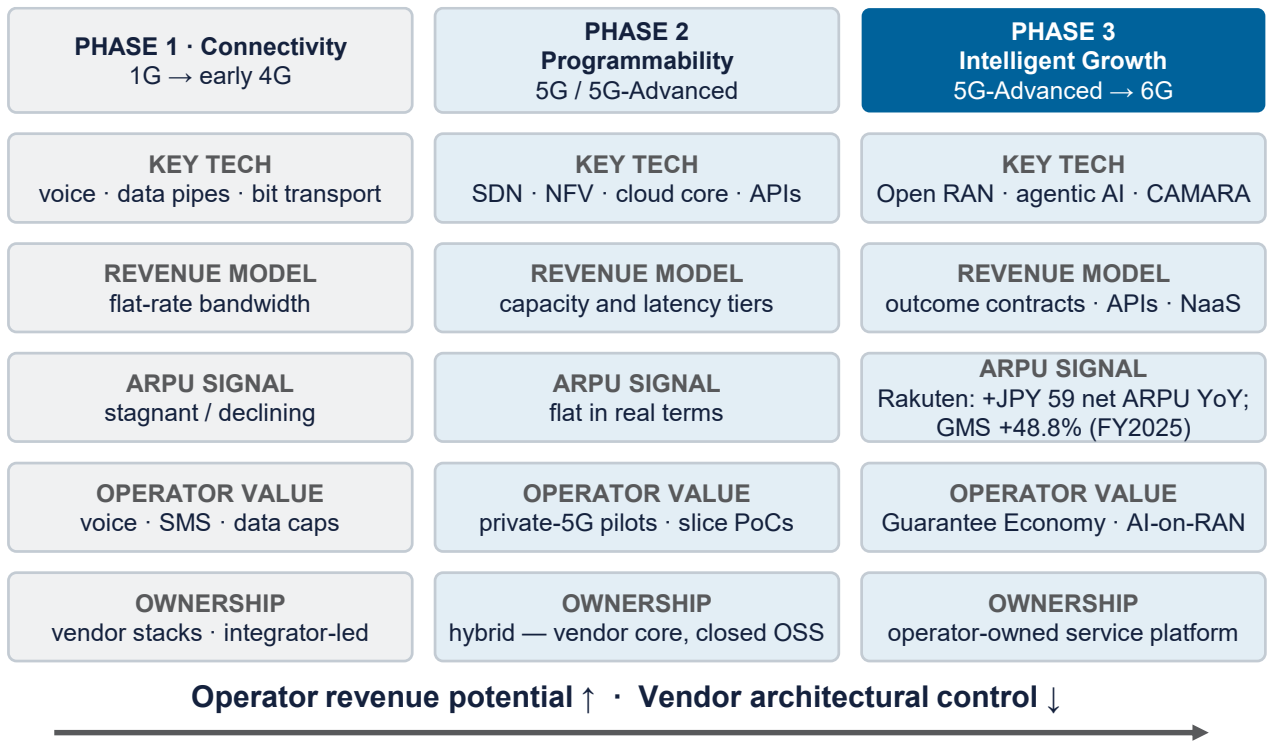

**FIGURE 2. Three Phases of Telecom Evolution: From Pipe to Platform. In the proposed model, operator revenue potential rises as vendor control falls; Phase 3 reverses two decades of disaggregation drift.**

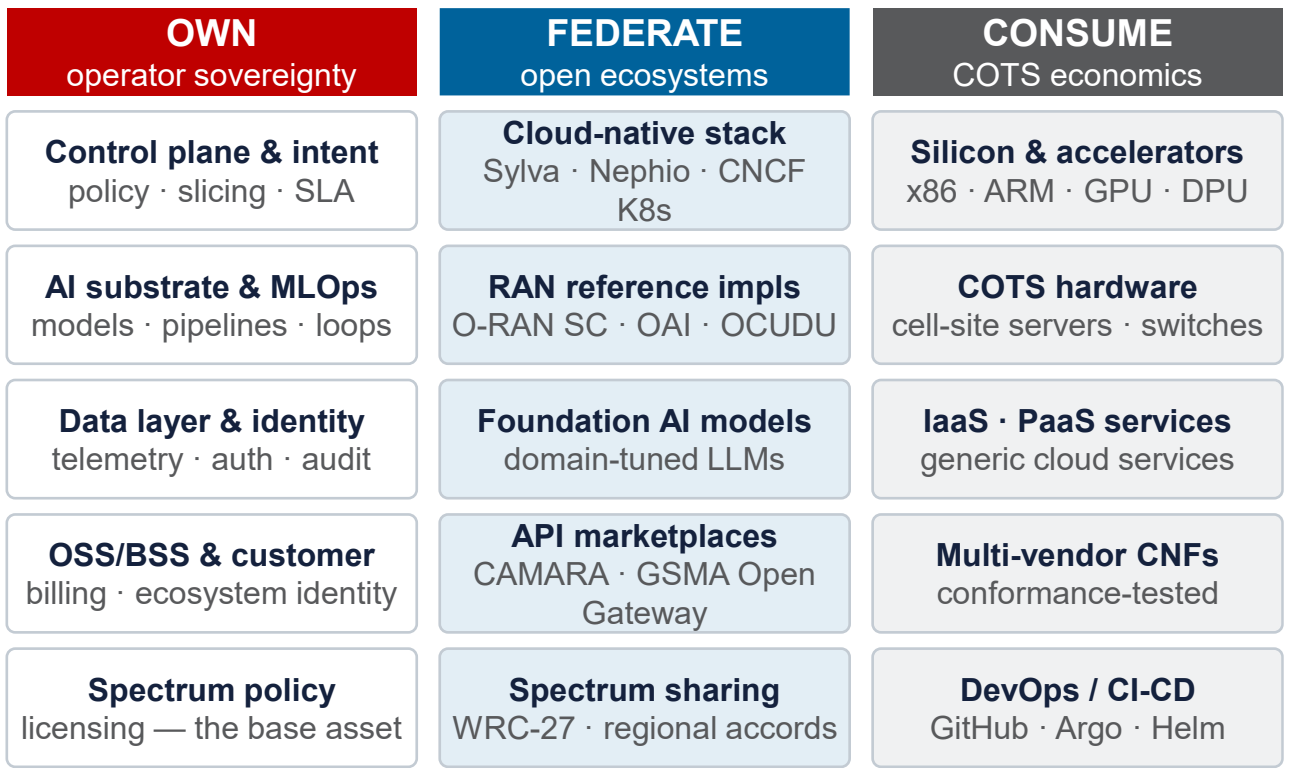

**FIGURE 3. The 6G Control Compact. Operator sovereignty is allocated by layer: the control plane is owned, ecosystems are federated, hardware and IaaS are commoditized.**

Operators can federate roaming and interconnect, API marketplaces (GSMA Open Gateway [48]), open-source platform stacks, and shared rural infrastructure; they should consume cloud infrastructure-as-a-service (IaaS), conformance-tested multi-vendor cloud-native network functions (CNFs), and hardware behind open interfaces at commercial off-the-shelf (COTS) economics.

Rakuten Mobile's deployment provides operational context for the Compact: major core, RAN-software, and management functions are virtualized or containerized and managed by Kubernetes with open APIs at every layer, while radio-frequency (RF), real-time physical-layer (PHY), firmware, and acceleration functions continue to rely on specialized hardware and deterministic execution environments. Commercial context includes more than 10 million subscribers and FY2025 full-year EBITDA profitability [5]. For the live Open RAN energy-optimization deployment, the joint announcement of 19 February 2026 reports estimated RAN energy conservation of approximately 20% [49]. This is an operator-reported estimate, not an independently measured benefit established by this article. The earlier 15–20% RIC power-reduction figure, an operator-reported estimate for which no published locator is available, is not added to this estimate or treated as an independent result; reconciling their baselines, periods, and network boundaries requires the underlying measurement records. Separately, the LLM-enhanced Kubernetes optimization study [8] reports 61% average potential cost reduction under human approval, and [9] documents eBPF adoption.

*External assessment evidence*. TM Forum's report dated 11 February 2026 records a score of 4.0 under GB1059H for intra-RAT intra-frequency and inter-RAT multi-site energy saving; intra-RAT inter-frequency saving is outside the assessed scope [47]. The public register labels the scenario partial and the evaluation tool alpha [50]. The assessment report describes 15-minute performance and configuration management (PM/CM) data collection, AI/machine learning (ML) forecasting, per-cell decisions, automated configuration-management-as-a-service (CMaaS) API execution, KPI-triggered rollback/backoff, and ticket escalation for repeated failures [47]. This is stronger corroboration of a deployed closed loop than an architecture proposal alone. TM Forum's Autonomous Networks Level Assessment Validation (ANLAV) validates the assessment method and supporting evidence, not actual network performance [50]. Neither this assessment nor the energy estimate validates MCP, LLM-agent decision quality, or the complete Network MCP Platform.

*Implementation evidence*. Author-supplied assessment-review material [51] describes a bounded action space: 4G multiple-input multiple-output (MIMO) transmit muting from 4Tx4R to 2Tx4R, and 5G administrative lock/unlock. The 5G policy also uses underlying 4G performance metrics when locked 5G cells provide no PM data. The material shows intent-derived configuration, forecast outputs, CMaaS Get/Set calls and completion notifications, with daily/weekly reporting. Performance-triggered reversion and backoff protect service quality; repeated failures trigger tickets and suspension of further actions on the affected element. These records illustrate governed operational execution, not unrestricted optimization or MCP-specific validation. The later public report [47], rather than the preliminary review slides, establishes the validated assessment outcome. For brownfield incumbents, migration is a *multi-year sequencing program*: new spectrum deployments adopt Control Compact architecture from the outset while existing infrastructure migrates as vendor contracts expire. This article does not evaluate a completed brownfield migration against the full Control Compact; the sequencing below is a proposed path, not a validated precedent. The minimum viable first step is threefold: require Cloud Native Telco initiative (CNTi) CNF conformance [52] and open API documentation as contractual conditions for all new network function (NF) procurement from 2026; deploy Cloud Native Computing Foundation (CNCF)-conformant machine-learning-operations (MLOps) tooling on a parallel data plane; and insert data-access rights as a non-negotiable clause in all vendor renewals. Where the structurally limited supply of combined telecom/cloud-native/AI talent [44] forces tactical reliance on vendor-managed AI platforms, four contractual protections should be evaluated: *full access to training data and model architecture documentation; standard open inference APIs; the right to migrate models without vendor approval; and real-time API access to all telemetry generated by vendor NFs*.

TABLE 1. THE 6G CONTROL COMPACT: OPERATOR-OWNED, FEDERATED, AND COMMODITY LAYERS

| **Operator OWNS (proposed)** | **Operator should FEDERATE** | **Operator may CONSUME as Commodity** |
|---|---|---|
| End-to-end operating model and system architecture | Foundation AI models (LLMs, domain-adapted) | Silicon (SoCs, GPUs, FPGAs) |
| AI strategy: model selection, training data, MLOps governance | Shared spectrum capacity (neutral host, national roaming) | Optical transport components |
| Customer relationship, identity, and trust layer | Network API exposure marketplaces (CAMARA, Open Gateway [48]) | Generic cloud-native services (undifferentiated compute) |
| Data layer: telemetry pipelines, model training data, inference context | Sustainability accounting and carbon reporting platforms | Standard undifferentiated network functions |
| Security governance: zero-trust policies, PQC key management | Partner settlement and onboarding infrastructure | Off-the-shelf COTS hardware (servers, switches) |
| Network observability: eBPF telemetry, real-time analytics | Spectrum-sharing and dynamic allocation frameworks (WRC-27) | Standardized conformance testing tools |
| OSS/BSS stack; spectrum policy and licensing | Multi-operator federation for roaming and API interoperability | Public cloud regions for non-latency-sensitive workloads |
| Agentic AI operations: closed-loop automation, digital twin | Cross-operator research programs (Hexa-X-II, SNS JU, Next G) | Generic DevOps tooling (CI/CD pipelines) |

*The cost of ownership*. Each owned component carries maintenance, security, integration, skills, and obsolescence costs. Federation adds interoperability testing and shared-accountability overhead; consumption adds switching, data-egress, and supplier-concentration risk. The Control Compact is therefore applied per component and contract cycle, with ownership justified only where the retained control and differentiation exceed the full lifecycle cost (Appendix E).

### D. What 6G Must Not Repeat

The most important 6G design inputs are recurring failure modes: vendor-controlled AI and data, proprietary management interfaces, delayed programmability, fragmented assurance, security added after deployment, and business cases built around peak capability rather than verifiable outcomes. These failures motivate operator-controlled data and policy, standards-based execution interfaces, measurable service boundaries, and governance designed into automation from the outset [37].

## IV. CUSTOMER FIRST – THE SIX OUTCOME FAMILIES

### A. The Specification Trap

The specification trap arises when peak rates, isolated laboratory gains, or feature counts substitute for customer outcomes. Draft IMT-2030 requirements remain necessary engineering baselines [4], but commercial value depends on sustained performance at a declared boundary, under representative load, with a remedy when the commitment is missed. The six outcome families below translate that requirement into service-level objectives rather than treating every technical maximum as a marketable guarantee.

### B. The Six Outcome Families

Aligned with the six ITU-R IMT-2030 usage scenarios and the ITU-R WP 5D draft new Report M. [IMT-2030.TECH PERF REQ] (Feb. 2026) [4], this article defines six outcome families – the 6G service design space illustrated in Fig. 4 and mapped in Table 2 – with Sovereign Trust as a cross-cutting assurance foundation spanning all six: post-quantum cryptography (PQC) readiness, zero-trust operation, data residency, trusted AI lifecycle, and auditability. The assurance foundation rests on the full TPR dimension set: peak rate, spectral efficiency, area capacity, bandwidth (≥400 MHz), connection density, mobility, latency, reliability, positioning, sensing, AI, energy efficiency, and resilience.

*Immersive Experience* (ITU-R Immersive Communication) requires low-variance, consistently maintained throughput: TPR [4] establishes a 5th-percentile user-experienced data rate of 300 Mbit/s downlink / 50 Mbit/s uplink (Dense Urban) and ≤4 ms user-plane latency for the interactive extended-reality (XR) and spatial-media service class.

*Mission-Critical Determinism* (ITU-R hyper-reliable and low-latency communication, HRLLC) industrializes deterministic connectivity for autonomous systems, remote machinery, and emergency control: ≤1 ms user-plane latency with $1-10^{-5}$ reliability for a 32-byte layer-2 protocol data unit (PDU) at the coverage edge [4]; safety-critical contracts can specify stricter targets commercially (Section V).

TABLE 2. MAPPING OF SIX OUTCOME FAMILIES TO IMT-2030 USAGE SCENARIOS, ITU-R TPR KPIs, AND GUARANTEE ECONOMY TIERS

| Outcome | Scenario source | Targets / status | Tier | Ph. |
|---|---|---|---|---|
| Immersive | IMT-2030 immersive comms | 5th-percentile DL 300 / UL 50 Mbit/s (Dense Urban); user-plane latency ≤4 ms; peak DL 36 / UL 18 Gbit/s | Consumer | 1→2 |
| Critical | HRLLC / critical comms | User-plane latency ≤1 ms; reliability $1-10^{-5}$ (32-byte L2 PDU, coverage edge); jitter <1 µs (author target) | Enterprise | 2 |
| Massive IoT | Massive comms / ambient IoT | Connection density $10^6$ devices/km²; QoS-bound message delivery; low-overhead control | IoT | 1→2 |
| AI-native | AI and comms / distributed intelligence | Edge inference latency <10 ms (author target); AI-native UL share ~26% [53]; data/training/inference support | AI service | 2 |
| Sensing | ISAC / positioning | Positioning ≤0.75 m indoor / ≤6 m urban macro; detection 95% at 5% false alarm; update <1 s (author/vertical target) | Sensing | 2 |
| Sustainable | Ubiquitous connectivity | Energy-efficient low-load operation (unloaded, ≤30% load) [4]; energy-per-bit improvement an operator target (not a TPR figure); extended reach (HIBS, relays) | Sustainable | 1→2 |

*Massive IoT Fabric* (ITU-R Massive Communication, including Ambient Internet of Things, IoT) provides $10^6$

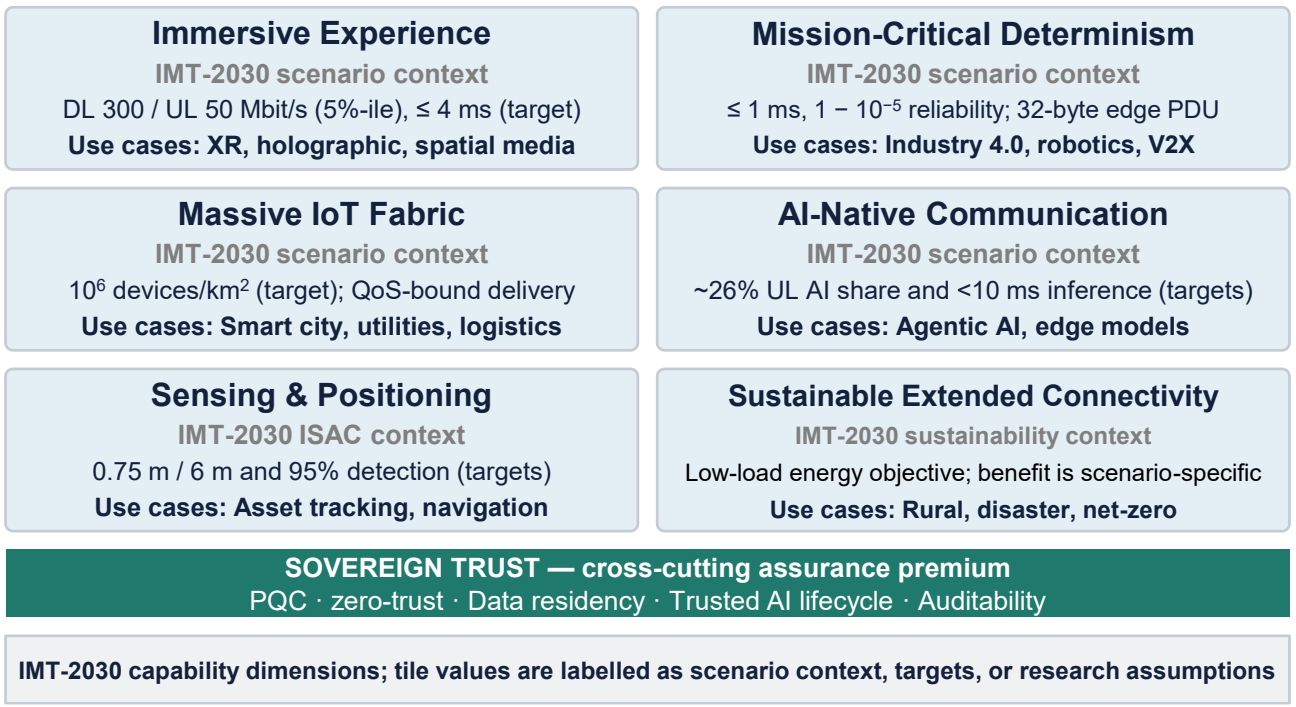

**FIGURE 4. Six Outcome Families – The 6G Service Design Space. Six tiles aligned with the IMT-2030 usage scenarios – Immersive Experience, Mission-Critical Determinism, Massive IoT Fabric, AI-Native Communication, Sensing and Positioning, and Sustainable Extended Connectivity – with Sovereign Trust as the cross-cutting assurance foundation.**

devices/km² connection density with quality-of-service (QoS)-bound message delivery and low-overhead, sensor/actuator-centric access for smart-city, utility, logistics, and industrial telemetry.

*AI-Native Communication* (ITU-R AI and Communication) marks the point at which AI becomes a radio-interface capability – network for AI and AI for network – spanning data collection, distributed learning, compute, and model execution; Ericsson [53] documents that AI-native sessions generate an uplink-to-total-traffic ratio of approximately 26%, versus around 10% for conventional broadband, driven by continuous upload of context, sensor data, and reasoning artifacts.

*Sensing and Positioning* (ITU-R ISAC) elevates the air interface to a shared sensing fabric: 95% object-detection probability at 5% false-alarm rate, and horizontal positioning accuracy of 0.75 m (indoor factory) and 6 m (urban macro) [4].

*Sustainable Extended Connectivity* (ITU-R Ubiquitous Connectivity plus sustainability) combines energy-efficient low-load operation – evaluated for unloaded and $\leq$30% load conditions against a full-load reference [4] – with resilience and extended reach through high-altitude IMT base stations (HIBS), relays, repeaters, and shared infrastructure for rural coverage, disaster recovery, and net-zero networks.

Each family is anchored in quantified targets with explicit evidence status: draft ITU-R TPR evaluation minima where [4] applies, 3GPP TR 22.870 use-case requirements where [7] applies, and author commercial SLOs where the target deliberately exceeds the standards baseline. This enables the Guarantee Economy tiers of Section V to translate outcome requirements into enforceable SLOs without treating every figure as a final deployment guarantee, and positions operators to demonstrate standards alignment and, where requirements are finalized, compliance through live network operations with independently auditable data; the Innovative Optical and Wireless Network Global Forum (IOWN GF) MetricEES review [54] surveys the fragmented energy-efficiency metrics (including Alliance for Telecommunications Industry Solutions (ATIS) telecommunications energy efficiency ratio (TEER), ITU-T network carbon intensity energy (NCIe), and ETSI ES 203 228) and proposes their harmonization toward auditable, comparable energy performance across operators and vendors.

*Derivation of the mapping*. A separate family is defined where an IMT-2030 scenario produces a distinct, measurable combination of objectives at a service boundary. The mapping is not a claim that the scenarios are commercial products; it is an author taxonomy for testing whether performance can be specified, verified, and priced. Overlapping technical enablers are expected, while Sovereign Trust remains a cross-cutting assurance premium rather than a seventh family.

### C. Customer Segments and the Outcome Matrix

The families form a configurable matrix across segments, as illustrated in Fig. 5. Consumers in dense urban markets prioritize Immersive Experience, AI-Native Communication, and trust assurances; enterprises in manufacturing, logistics, energy, and healthcare weight Mission-Critical Determinism, AI-Native Communication, Massive IoT Fabric, and Sensing most heavily; public-sector organizations place the Sovereign Trust foundation, resilience, and Sustainable Extended Connectivity at the top, reflecting accountability to citizens and geopolitical risk exposure. Two scoping dimensions condition the matrix:

- The first is *market economics*: the Guarantee Economy applies most immediately in high-ARPU markets with enterprise density; for operators serving low-ARPU subscriber bases, as in parts of South Asia, Sub-Saharan Africa, and Southeast Asia, the near-term priority is connectivity-first, with three Control Compact elements applying directly – spectrum-policy ownership, national data-layer control, and open-source adoption to reduce capital cost – which the Bharat 6G Vision [55] frames as a national strategic priority.
- The second is *jurisdiction*. In the EU, the EU AI Act [56], the General Data Protection Regulation (GDPR) [57], the European Health Data Space (EHDS) [58], the NIS2 Directive [59], and the Cyber Resilience Act [60] bind service design – EU AI Act Article 14 human oversight must be satisfied at service activation, not at audit. The UK operates a separate regime – UK GDPR, the Data (Use and Access) Act, and a sector-led 'pro-innovation' AI approach rather than the EU AI Act. Japan and the Republic of Korea are governed by the Act on the Protection of Personal Information (APPI) and the Personal Information Protection Act (PIPA), respectively, and telecommunications business law; the United States by the Secure and Trusted Communications Networks Act [61] and CHIPS Act [62] supply-chain provisions, with state privacy laws layered on top. India operates under the Digital Personal Data Protection (DPDP) Act and the trusted-supplier framework [55], Australia under the Security of Critical Infrastructure (SOCI) Act and Privacy Act amendments, and Singapore under Infocomm Media Development Authority (IMDA) codes.

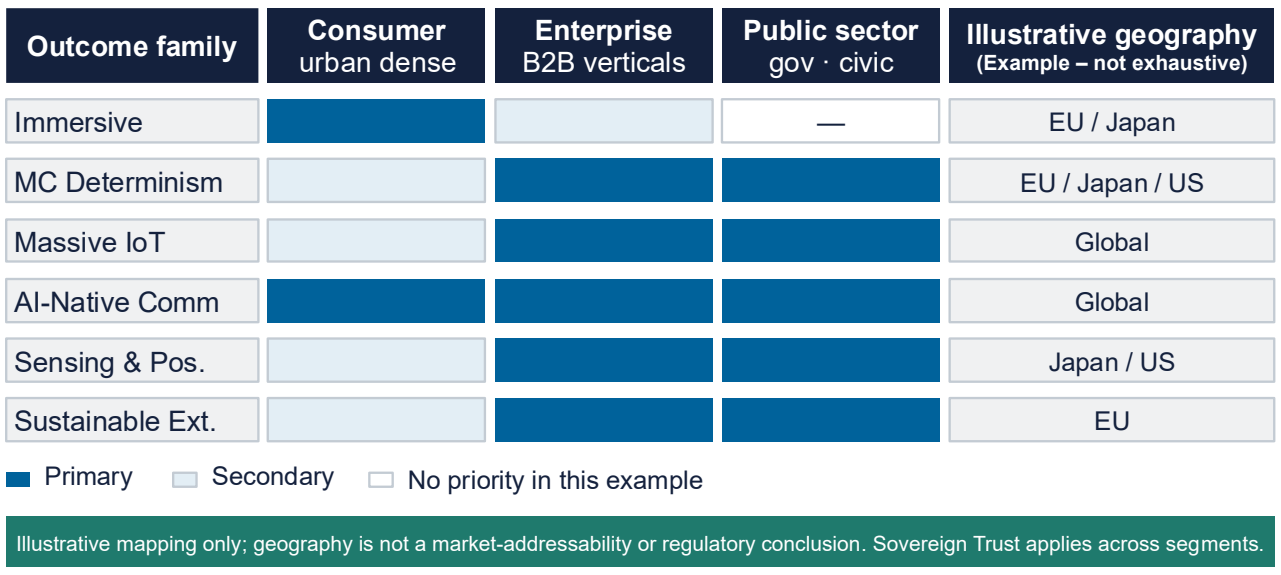

**FIGURE 5. Customer Segment × Outcome Matrix. Three segments × six outcome families with Primary/Secondary/– designations; regional dimensions add a third axis reflecting divergent national 6G priorities.**

These differences shape which outcome families can be offered, at what tier, under what audit obligation: a Sovereign-Trust-assured tier for a Japanese government customer cannot be operated identically to one for an Australian critical-infrastructure customer. The commercially scalable approach is a regionally parameterized template – one product with jurisdiction-specific configuration enforced by the operator-owned policy engine – not dozens of distinct catalogs.

## V. BUSINESS FIRST – THE GUARANTEE ECONOMY

### A. The Connectivity Trap

Operators continue to carry large network-investment obligations while seeking growth beyond undifferentiated connectivity [1]. This motivates service models based on programmable exposure and verifiable outcomes. The argument is not that connectivity ceases to matter, but that any incremental premium must be tied to a measurable service commitment rather than to peak capability alone.

Rakuten Mobile reports that its subscribers generate approximately 48.8% higher annual average gross merchandise sales (GMS) on Rakuten Ichiba than non-subscribers, and FY2025 net ARPU of JPY 2,467, up JPY 59 year on year [5]. These figures provide context for an ecosystem-based commercial model; the association does not establish that connectivity caused the spending difference or validate demand for the proposed service tiers. Operators without a comparable ecosystem would need to assess API monetization, enterprise contracts, or service partnerships against their own customer base and cost structure.

### B. Three Revenue Engines

Three candidate revenue mechanisms are considered: outcome-based service contracts, programmable network APIs, and Network-as-a-Service. Their individual and combined revenue effects require commercial validation.

The Guarantee Economy is proposed as an outcome-based service model: AI-RAN Alliance WG3 research [13] indicates nearly half of global 5G users experience connectivity challenges during peak hours and approximately half indicate willingness to pay for guaranteed reliability. It requires three architectural components – granular SLO definition, real-time closed-loop enforcement, and transparent auditable verification; without all three, a guarantee is a marketing claim. (This willingness-to-pay evidence derives from alliance member research rather than independent demand studies; survey-stated willingness does not imply willingness to sign enforceable SLAs – the private-5G bypass precedent is the caution.)

Programmable Network APIs expose network capabilities as developer-consumable services: CAMARA [63] and GSMA Open Gateway [48] established the 5G vocabulary (Quality on Demand (QoD), Location Verification, Location Retrieval, Geofencing Subscriptions, Number Verification, SIM Swap, and Device Reachability APIs); 6G extends to AI-inference offload, Sensing-as-a-Service, and network digital-twin APIs, governed by ETSI ZSM [22]–[24]. A representative CAMARA/Open Gateway candidate set is cataloged in Table 11 (Appendix A). Network-as-a-Service (NaaS) delivers end-to-end network functionality as an on-demand, intent-driven service – slices provisioned in minutes rather than months – an operator design target for Nephio [64]-based orchestration. First-generation NaaS is a velocity improvement, not yet an outcome-contracted product; the transition requires adding real-time SLA monitoring, automated breach detection, and consequence execution, with ETSI ZSM 019 [22] as the vendor-agnostic abstraction. Cloud-service growth provides commercial context: Amazon Web Services grew to over USD 100 billion in annual revenue in under two decades [65]; mobile infrastructure adds licensed spectrum, physical edge proximity, and carrier-grade accountability that may support differentiated deployment models.

### C. The Guarantee Economy in Practice

Table 3 deliberately defines six commercial tiers – one per IMT-2030 outcome family – with guaranteed SLOs, breach consequences, and billing models; this is the canonical taxonomy used in this article, with Massive IoT Fabric treated as a distinct tier rather than folded into sustainable connectivity. SLO values draw from draft ITU-R TPR [4], 3GPP TR 22.870 [7], and explicitly marked author commercial targets. Sovereign Trust is not a seventh tier but a cross-cutting assurance premium attachable to any tier: PQC readiness, zero-trust operation, data residency, trusted AI lifecycle, and auditability, contracted as an uplift on the underlying SLO package.

Commercial SLOs may exceed draft TPR minima where the customer pays for stricter targets: the proposed Enterprise Determinism tier specifies reliability up to 99.9999% and jitter below 1 μs against the TPR baseline of $1-10^{-5}$ [4].

SLA enforcement spans five dimensions: latency (mean and variance), reliability, throughput (peak, sustained, and fifth percentile), availability, and security posture. At the service layer, TMF641 [66] provides service ordering and TMF657 [67] represents and exposes service-quality information, including Service Level Specifications (SLSs) and associated objectives and thresholds. Assurance, breach detection, and consequence execution are separate operator functions subject to applicable charging, lawful-intercept, and legal controls. Operational use requires a standardized measurement method accepted by both parties, jurisdiction-specific legal enforceability of breach consequences, and unambiguous attribution across operator, device, and application domains. TM Forum IG1252 [6] is an autonomous-networks evaluation methodology, not a master SLA template; bilateral contracts must define the service specification, measurement authority, breach attribution, remedies, and liability.

A Guarantee Economy claim is commercially testable only when a named buyer accepts a disclosed premium for a specified SLO and remedy. Each pilot should pre-register its measurement authority, breach-attribution procedure, liability cap, renewal decision, and exit criterion; until such results are published, tier-level willingness-to-pay claims remain hypotheses.

Commercial deployment still requires common measurement methods, enforceable remedies, and cross-domain breach attribution. Industry studies may help frame hypotheses [13], [68], but they do not establish buyer willingness to pay or capital recovery.

The proposed SLA model initially assumes domestic services under a single operator's control. International roaming introduces additional measurement, attribution, and contractual dependencies; its inclusion is a later-phase design objective rather than a demonstrated extension.

The catalog assumes a facilities-based operator: mobile virtual network operator (MVNO), neutral-host, and spectrum-sharing arrangements introduce multi-party SLA pass-through chains; Section XI.C states what such chains require, and their quantitative evaluation is outside this article.

A worked contract, and what it is not. Table 3 defines service tiers, while Appendix C, (8)–(10), defines the break-even relation among premium, delivery cost, breach probability, and remedy. A contract must also specify the measurement boundary, observation window, credit formula, and dispute path. The model identifies the quantities a pilot must measure; it does not supply an attach rate or demonstrate willingness to pay.

TABLE 3. THE GUARANTEE ECONOMY: SIX-TIER SERVICE CATALOG WITH SLOs, BREACH CONSEQUENCES, AND BILLING MODELS (SOVEREIGN TRUST AS CROSS-CUTTING PREMIUM)

| Tier | Target Segment | Guaranteed SLOs | Breach Consequence | Billing Model |
|---|---|---|---|---|
| Premium Consumer Immersive | Consumer: gaming, XR, live events | DL ≥500 Mbit/s; E2E latency ≤20 ms; availability 99.99% | Bill credit; automatic QoS re-routing | Monthly premium subscription |
| Enterprise Determinism | Manufacturing, logistics, robotics, remote surgery (subject to separate clinical/safety /regulatory validation) | Latency ≤1 ms (device-to-edge); reliability to 99.9999% (author commercial target ≈$1-10^{-6}$, vs TPR baseline $1-10^{-5}$); jitter <1 µs (author target; packet-delay-variation, vertical-sourced) | SLA penalty; incident report; automated remediation | Per-SLO outcome contract; framework + usage |
| Massive IoT Fabric | Smart city, utilities, logistics, industrial telemetry | Density to $10^6$ devices/km²; QoS-bound message delivery; battery-life and deep-coverage SLOs | Message-delivery credits; fallback access (incl. NTN IoT) | Per-device / per-message; volume commitments |
| AI Inference Edge | Enterprise AI, real-time CV/NLP at edge | Inference latency <10 ms (device-to-edge); model availability 99.99%; throughput per agreed tokens/s | Fallback to cloud; bill credit | Per-inference; committed capacity |
| Sensing-as-a-Service | Smart city, autonomous logistics, monitoring | Positioning ≤0.75 m indoor / ≤6 m urban; update <1 s; coverage SLA | API SLA credit; fallback sensing source | Per-API-call or per-area subscription |
| Sustainability-certified NaaS | Hyperscaler, enterprise private networks, net-zero-committed enterprises | Slice provisioned <5 min (operator product target); ≥99.9% slice integrity; guaranteed BW/latency isolation; per-slice energy and carbon-intensity reporting (research objective; not an operational standard) | Provisioning credit; automated re-instantiation; carbon-report credit | Usage + committed capacity; outcome contract |

## VI. OPERATIONS FIRST – AGENTIC AI AND INTELLIGENT GROWTH

### A. The Operations Gap

Network operations combine automated functions with manual configuration, ticket-based workflows, and human incident response. The degree of autonomy varies by operator, domain, and scenario. The NGMN agentic-AI framework [25] provides a basis for discussing adoption, but this article does not establish an industry-wide autonomy ceiling.

Two level frameworks appear in this article and are kept distinct: the NGMN agentic-AI levels [25] describe a target operating model, while TM Forum IG1252 [6] supplies the methodology for scenario-specific autonomous network (AN) level assessments. The scoped result in Section III.C uses the latter framework; it does not establish a one-to-one mapping between the two scales or a rating of the proposed MCP platform.

Outcome-based services and NaaS require operational processes that can sustain their specified SLOs. For a dense national deployment, an illustrative engineering assumption of $10^5$–$10^6$ cells and approximately $10^6$ configuration decisions per day motivates scalable automation; it is not a measured workload from the case studies. The NGMN scale describes a maturity trajectory, not a uniform economic conversion factor [25]. Section III.C separates the scenario-specific autonomy assessment from the operator-reported energy estimate; neither establishes a generic autonomy return.

Automation itself is not new: self-organizing networks (SON), software-defined networking (SDN), and rule-based closed loops already handle repeatable decisions. Agentic AI is relevant where intent decomposition or cross-domain diagnosis cannot be fully enumerated in advance, but it expands the attack surface and failure modes. The operating model must therefore retain deterministic controls for routine actions and admit agents only through bounded authority, validation, audit, and recovery.

### B. Networks as Software

The architectural foundation is a cloud-native operating model in which the virtualized and containerized portions of the network stack – from DU/centralized unit (CU) and core functions to OSS – execute as software on shared, horizontally scalable compute orchestrated by Kubernetes, while RF, hard real-time PHY, firmware, and acceleration paths remain specialized deterministic functions, as illustrated in Fig. 6. Kubernetes provides orchestration for the containerized functions; the networking and security layer is eBPF-based – e.g., Cilium/Tetragon [69] or the Sauron eBPF platform [9], [10] – providing kernel-level policy, observability, and security without proprietary hardware; Nephio [64] provides intent-driven lifecycle management through Kubernetes-native reconciliation.

A production study [8] reports a 61% average potential cost reduction across 68 production microservices in five namespaces (range 21–88% per namespace), using 30-day telemetry and deterministic right-sizing rules, with an LLM assisting document and telemetry interpretation under human approval. The result supports the feasibility of LLM-assisted operations; it is not evidence of autonomous control or of platform-wide performance.

### C. Agentic AI: The Operating Model

In the operating model of Fig. 7, LLM-based agents interpret intent, select and invoke MCP tools, reason over returned evidence, and determine subsequent actions within an assigned scope. This agentic workflow complements deterministic automation. The four stages below have task-dependent timescales; the indicative values are design assumptions, not measurements from the RAN or IMS cases.

- *Intent* – an operator goal in natural language (minutes).
- *Observe* – telemetry, counters, probes, eBPF events, and customer KPIs (100 ms-1 s).
- *Decide* – LLM reasoning, multi-agent planning, and validation (seconds).

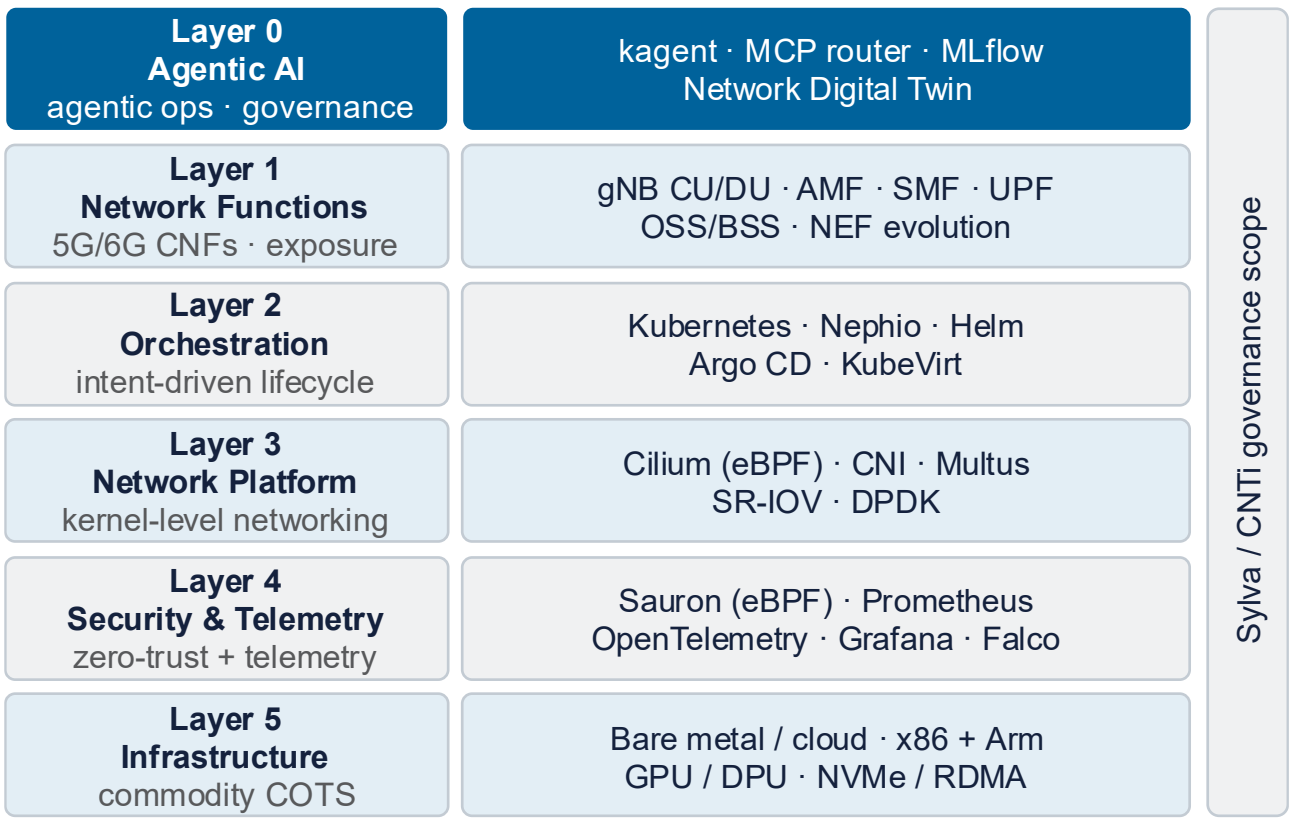

**FIGURE 6. Cloud-Native Telecom Stack – Networks as Software. Six layers (Layer 0 Agentic AI at top, Layer 5 Infrastructure); Sylva and CNTi brackets indicate open-source governance scope.**

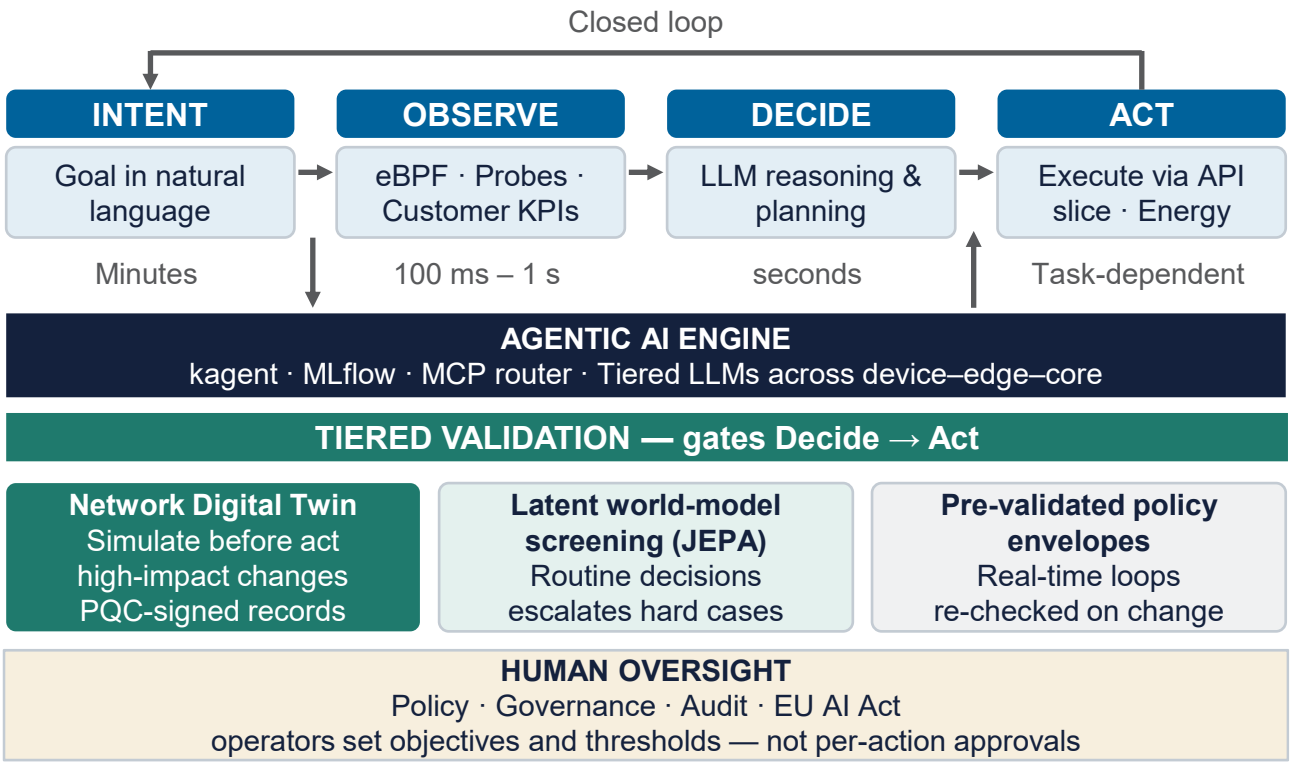

**FIGURE 7. Agentic AI – Closed-Loop Operations Model. Intent → Observe → Decide → Act loop with human oversight (policy, governance, audit, EU AI Act) and a tiered validation layer gating Decide → Act: Network Digital Twin for high-impact changes, latent world-model screening for routine decisions, and pre-validated policy envelopes for real-time loops.**

- *Act* – execution through authorized APIs for operations, steering, energy management, and remediation. Action timescales and validation depth are task-dependent design choices. High-impact changes should pass explicit pre-execution checks; any digital-twin or latent-model gate requires measured prediction quality and defined escalation rules. Latency-critical domain control should remain within separately validated policy envelopes rather than depend on per-action LLM or MCP mediation. The RAN and IMS cases in Section VII.D concern support and management workflows, not validation of radio-control timing.

The operational target is NGMN Level 4 within Level 5 governance [25]: bounded execution under policy, human authority over exceptions and lifecycle decisions, and a tested ability to halt and recover. This structure may contribute evidence relevant to EU AI Act oversight requirements [56] where applicable, but it does not itself establish legal conformity.

A candidate implementation stack comprises four layers, using CNCF [70], Linux Foundation, and O-RAN Software Community (O-RAN SC) [71] components where appropriate. Production readiness depends on integration, authorization, and evaluation:

- An *agent orchestration layer* (kagent [72]) coordinating specialized agents named consistently with Section VII – the Optimizer (RAN and resource optimization), the SLA Enforcer (Guarantee-Economy SLO assurance), the Fault Remediator (detection, diagnosis, remediation), the Energy Saver, plus infrastructure-lifecycle (Nephio) and security (ZSM 017 closed-loop) agents – each within an authorization scope encoded in its trust certificate.
- A *tool-calling layer* exposed through the MCP router whose audit trails can support EU AI Act Article 12 record-keeping evidence, where the use case falls within scope, and ZSM 017-aligned access-control evidence simultaneously.

- A *model-serving layer* (vLLM or equivalent) on operator-owned hardware.
- A *governance layer* – MLflow versioning, Digital Twin validation, and a policy engine encoding regulatory obligations.

### D. AI/ML Infrastructure and Observability

Production agents require full-lifecycle MLOps: unified cross-domain telemetry; MLflow experiment tracking, versioning, and deployment [73]; and Network Digital Twins for pre-execution evaluation where validated for the task. Governance aligned with the EU AI Act [56] requires every model to be version-controlled, auditable, and documented for training data, evaluation, and performance boundaries; alignment risk – technically correct but commercially or regulatorily unacceptable optimization under distribution shift – is addressed at two levels: multi-objective reward design encoding regulatory SLA floors and fairness constraints is a first line of defense; because reward specification is itself susceptible to reward-hacking, the binding control is the runtime pre-execution validation and enforcement of Section VII.B, not the reward function alone. Three capabilities distinguish scale-grade MLOps:

- *Federated learning* where training data cannot leave jurisdictions, coordinated at the RAN layer via the Near-RT RIC [18].
- *Continuous retraining pipelines* with drift-triggered retraining and shadow deployment.
- *Explainability tooling* – grounded in auditable decision provenance: the natural-language intent, retrieved context, tool calls, and pre-execution validation verdict recorded per decision, producing decision records for subsequent technical and regulatory review, complemented by intent-to-code translation validated with human feedback [28]; where individual component models are feature-based, established XAI methods (permutation importance, SHAP, partial dependence) of the maturity Karahan et al. [36] demonstrate supplement this provenance, subject to their known sensitivity to configuration.

Sauron [9], [10] provides eBPF-based kernel telemetry for observability. Its overhead depends on probe placement, event volume, processing, and workload. The workflow cases below use operational diagnostics and logs; they do not establish zero instrumentation overhead or the effectiveness of an LLM anomaly detector.

## VII. NETWORK MCP PLATFORM FOR GOVERNED AGENTIC OPERATIONS

The preceding priorities converge on one engineering question: how can LLM-driven agents use live-network tools without weakening operator control? This section proposes the Network MCP Platform, its tool catalog, migration path, operational workflow evidence, and validation plan. MCP is an operator-internal orchestration and adapter layer: it is not a new 3GPP network function, a replacement for the 3GPP Service-Based Architecture, or a substitute for O-RAN, Network Configuration Protocol (NETCONF)/YANG, or standardized operations, administration, and maintenance (OAM) interfaces. Where a standardized execution interface exists, it remains the authoritative execution path. The six-subsystem platform is a *proposal*, while the RAN and IMS cases illustrate selected MCP-enabled workflow capabilities.

### A. Five Unresolved Control Problems

Agent-to-agent (A2A) and MCP interfaces can support agent integration, but an operator deployment also needs explicit control requirements. Table 4 maps five requirements to proposed subsystems. These are engineering responsibilities to implement and test, not problems resolved merely by adopting the protocol.

First, routing and trust management: direct integration between $N$ agents and $M$ independently governed services can require up to $N \times M$ relationships under that topology. A shared MCP router can reduce duplicated integration work, but the resulting complexity depends on trust domains, adapters, and policy rules. An unconditional $O(N^2)$-to-$O(N)$ improvement (see Appendix C, (1)–(3)) is not established here.

Second, workload identity: authentication of the subscriber and subscription through 5G authentication and key agreement (5G AKA) [74] does not by itself attest the identity or authorization of every agent runtime. Cross-domain agent identity, credential issuance, revocation, and delegated authority require an explicit deployment design.

Third, audit and telecom procedure correlation: agent-facing tool events must be correlated with the underlying network procedure where charging, lawful-intercept, or other regulatory obligations apply [75], [76]. Existing network controls remain in place. MCP invocation logs alone neither satisfy nor invalidate those obligations.

TABLE 4. OPERATOR CONTROL REQUIREMENTS AND PROPOSED SUBSYSTEMS

| Control requirement | Proposed subsystem(s) | Intended role and validation boundary |
|---|---|---|
| Routing and trust management | (A) MCP router and tool registry | Shared mediation; complexity depends on topology, adapters, and policy rules |
| Cross-domain workload identity | (B) SPIFFE-based workload identity | Attestation and revocation are deployment responsibilities; federation and timing unvalidated |
| Audit and telecom-procedure correlation | (A) Router + (C) eBPF telemetry | Metering/audit inputs; compliant charging and LI mediation require separate implementation |
| Bounded enforcement | (E) Validation + (C) telemetry + (D) lifecycle manager | Pre-execution checks and runtime response; no demonstrated sub-second bound |
| Adversarial defense | (C) Telemetry + (E) validation | Layered proposed controls; attack detection, semantic safety, and recovery require testing |

Fourth, bounded enforcement: infrastructure anomalies and semantic action errors require different detection and response mechanisms, with separately measured delays.

Fifth, adversarial defense: prompt injection, tool poisoning, compromised identity, drift, and conflicting intents require layered controls and adversarial testing. The operational cases do not quantify mitigation effectiveness [35].

### B. Six Cooperating Subsystems

Figure 8 shows six proposed subsystems of the Network MCP Platform. MCP supplies an agent-facing tool binding; the surrounding operator services supply authorization, audit, validation, and recovery controls. Their integration is the architectural contribution. Guarantees require implementation and evaluation of the combined system, not just the presence of its components.

(A) *MCP router and tool registry*. MCP servers expose approved tools; governed adapters invoke authorized SBI, O-RAN, NETCONF/YANG, OAM, element management system (EMS), or platform operations. The router validates schemas, enforces authorization and quotas, and records requests, results, and post-action checks. It does not replace procedure-specific charging or lawful-intercept functions, and the available evidence does not establish an MCP transport-latency advantage.

(B) *Cross-domain workload identity*. The proposed design uses operator-governed workload identities, for example SPIFFE and the SPIFFE Runtime Environment (SPIRE) [77], with attestation and short-lived credentials where supported. Protected-resource metadata [78] supports discovery; it does not by itself establish workload attestation or authorization. Federation, delegated authority, attach/detach integration, and revocation behavior require explicit implementation. The trust-scaling benefit depends on topology and policy assumptions, and neither RAN nor IMS case validates cross-operator federation.

*(C) eBPF telemetry and enforcement plane.* Probes on network-function containers and tool servers can supply per-flow, syscall, and resource telemetry [9], [10]. Operator controls may use these signals for throttling, enforcement of policies, quarantine, or escalation. Resource observations are metering inputs, not complete charging records. Overhead and end-to-end enforcement delay depend on probes, event rates, classifiers, policies, and distributed execution. (A 30–50% overhead reduction and sub-second containment are not established by the supplied records. Kernel telemetry cannot by itself establish whether a well-formed network action is semantically correct.)

*(D) Closed-loop AI lifecycle manager.* Consumes eBPF telemetry, maintains per-agent behavioral baselines, and performs drift detection, lifecycle management, and conflict resolution, emitting real-time policy updates to the router and the policy control function (PCF) – operating at NGMN Level

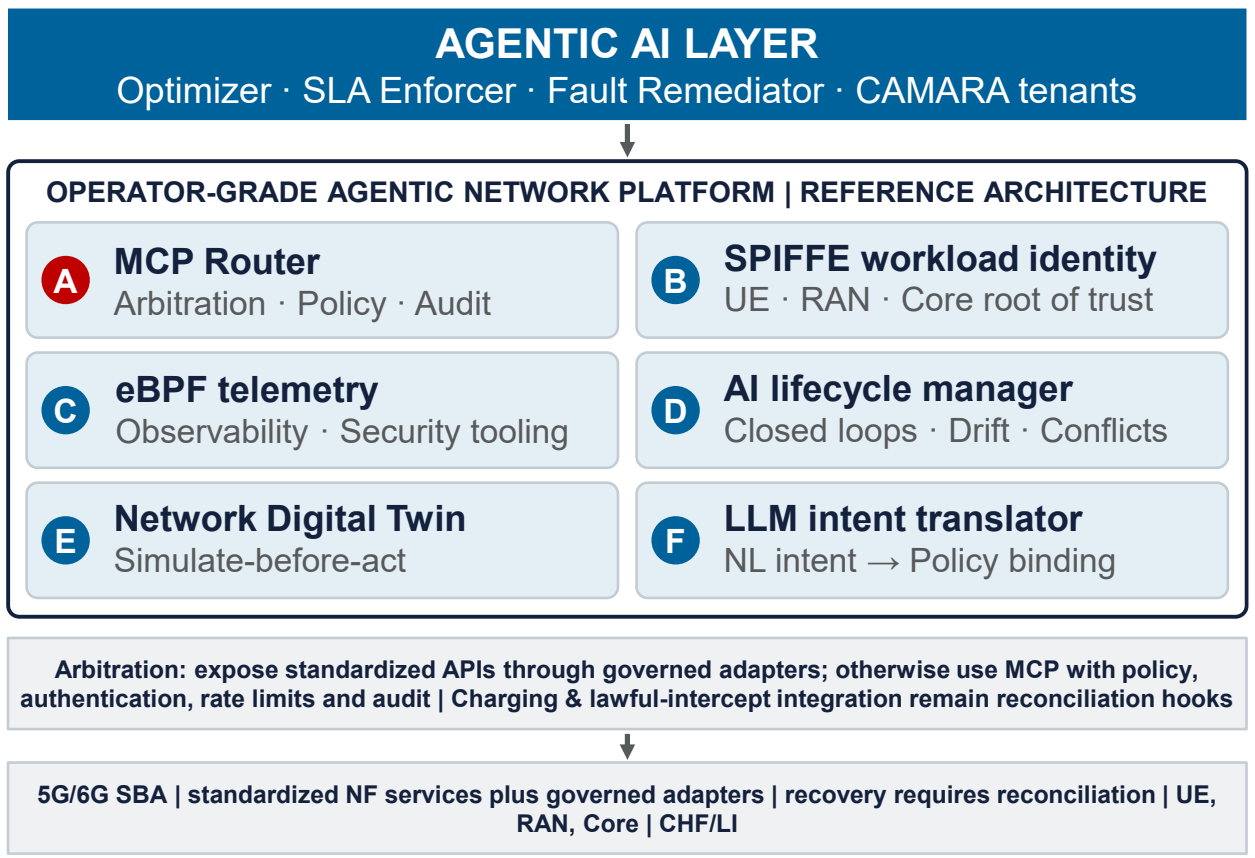

**FIGURE 8. Proposed Network MCP Platform with six operator control subsystems. MCP mediates agent-facing tools above telecom execution interfaces; identity, policy, validation, audit, and recovery require operator implementation. This diagram is a reference architecture, not an experimentally validated stack. MCP is an operator-internal orchestration and adapter layer; standardized 3GPP SBI, O-RAN and OAM interfaces remain the execution paths.**

4 under Level 5 human governance [25], and supporting high-risk-AI oversight evidence [56] at the policy level where the deployed use case falls within the EU AI Act scope. A priority-weight scheduler is a proposed arbitration mechanism; it does not establish freedom from deadlock, starvation, or unsafe policy interaction. Model drift, tool versions, policy changes, and escalation outcomes require auditable lifecycle records and tests.

(E) *Risk-tiered validation with digital-twin and latent screening gates*. Each proposed action is assigned a risk tier and checked against trusted state and application-specific criteria. High-impact actions use a synchronized digital twin where the task can be simulated [79]; routine actions may use a calibrated latent screening model only under declared bounds for state error and age, prediction horizon, KPI interval coverage, support/out-of-distribution (OOD) score, blast radius, and escalation. An exceeded or unknown bound fails closed. Joint-Embedding Predictive Architecture (JEPA), as illustrated in Fig. 7, and network world models [80], [81] and [82] are candidate mechanisms, not evidence of safe telecom control.

Because the screening tier admits routine actions without full simulation, Table 17 in Appendix F specifies it by what it must output, not by how it is built: a calibrated prediction interval for each KPI in the action's blast radius, and a scalar support score showing whether the proposed state lies within the model's training data. The blast radius is the set of managed objects reachable from the action's target in the configuration dependency graph maintained by subsystem (D); the safety envelope is the set of KPI values that meet the Section V.C service-level objectives of every tenant in that radius. JEPA [80], within the latent-variable energy-based framing of [83], is one candidate model, and latent predictive models have been demonstrated on network data: radio-channel dynamics from channel-state information [82] and, closest to this use, the effects of configuration changes on operator data [81]. These models predict in an abstract representation space, not in KPI units, so the calibrated interval must come from an added predictive head; it is the calibration of that head, not the representation, that remains unproven in the network domain (research question (2) in Section XII).

(F) *LLM intent translator – Transactional intent binding*. The translator grounds natural-language intent in schemas and telemetry, then produces an authorized sequence of MCP tool calls and domain primitives. Preconditions, bounded leases, idempotency, and compensating actions limit partial application where target interfaces support them; ambiguous intent aborts, and residual inconsistency is escalated with an auditable decision record. Fault-injection testing is still required.

### C. MCP Tool Exposure and Telecom Interface Mapping

The Network MCP Platform exposes approved operational capabilities as discoverable tools above existing domain interfaces. A RAN diagnostic tool may invoke an EMS query or a command wrapper; an IMS investigation tool may use a managed shell session and Kubernetes APIs. Existing rule-based and AI/ML automation can coexist with these agents. Figure 9 presents a schematic migration from isolated integrations toward a shared tool catalog, not a claim that current automation is universally RAN-only or that all network domains are already integrated. MCP remains an operator-internal orchestration and adapter layer; standardized 3GPP SBI, O-RAN, NETCONF/YANG, and OAM interfaces remain the execution paths. Operator implementation should standardize interfaces and control requirements, retain governance of network data, and evolve adapters where lifecycle value is demonstrated.

In Table 5, the tool catalog proposes representative READ, WRITE, and STREAM capabilities across access, core, observability, and cloud-control domains. Its role is architectural rather than vendor-specific: it identifies candidate governed tools, not an inventory demonstrated by the two cases, regardless of whether a capability is reached through a legacy wrapper, a standards-based API, or a native agent interface.

The migration therefore follows three abstract stages: expose existing operational systems through governed tool wrappers; replace wrappers with native tool surfaces as network functions and platforms mature; and finally allow 6G-ready functions to advertise self-describing capabilities directly into the governed catalog. These stages align with the phased roadmap in Section XI.

### D. Operational Evidence: MCP-Enabled RAN and IMS Agentic Workflows

The following examples are drawn from operational workflows in Rakuten Mobile's live network environment. In both cases, agents use MCP tools to obtain evidence and carry out task-dependent steps. They complement the externally assessed RAN energy-optimization evidence in Section III.C, but do not imply deployment or validation of every component of the proposed Network MCP Platform.

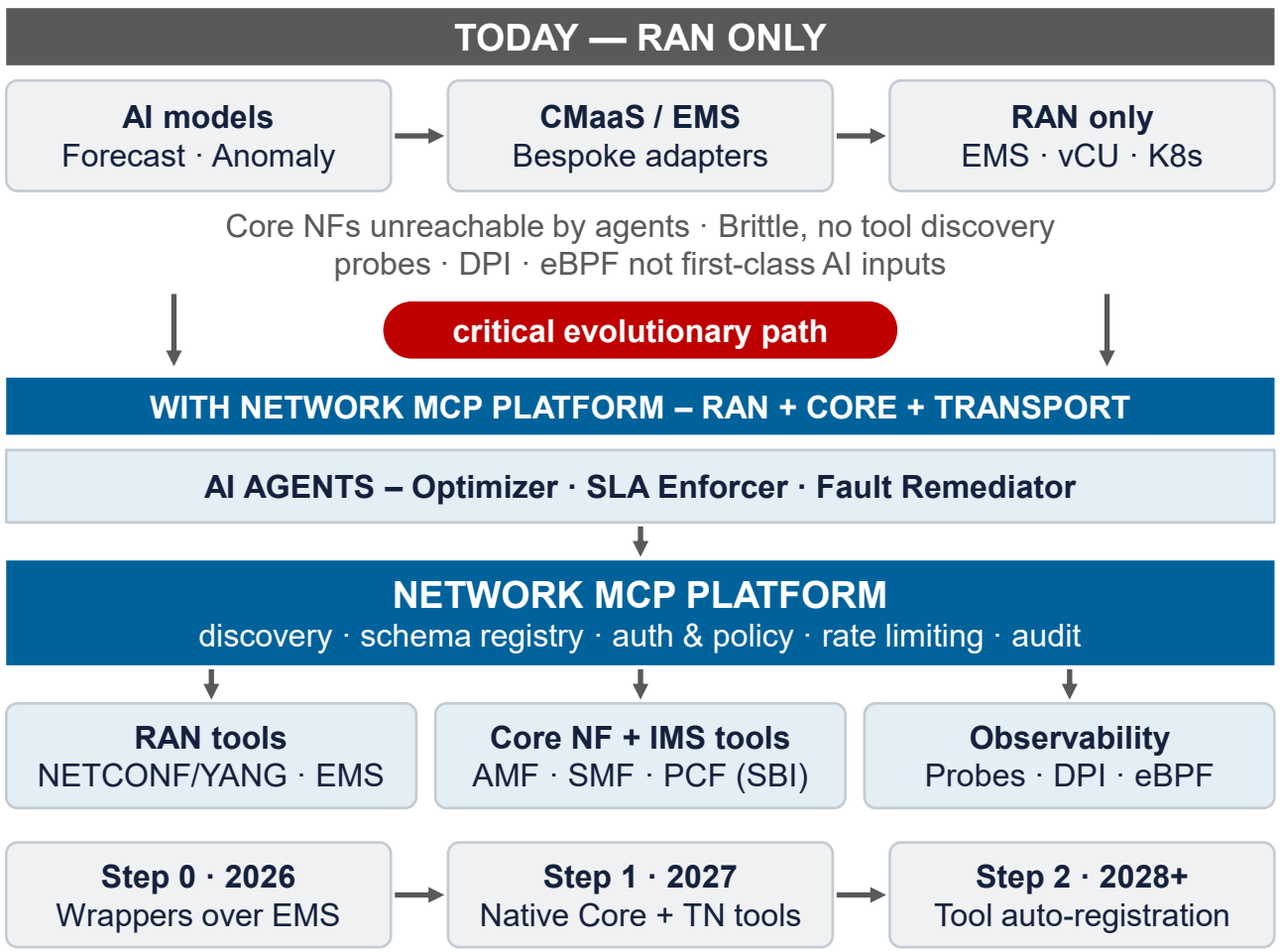

**FIGURE 9. Proposed migration from isolated integrations to a cross-domain MCP tool plane. The RAN-only starting state is schematic, not universal. MCP servers adapt approved domain operations; existing telecom interfaces retain execution authority. The RAN and IMS cases illustrate selected workflows, not full-network deployment.**

*RAN L2 Agentic Workflow*. Here, L2 denotes second-line support in the L1/L2/L3 support model, not protocol Layer 2. Specialized agents use MCP tools to collect RAN and software evidence, determine an appropriate recovery action, and verify the network outcome before completing the support workflow. Human review can confirm or correct the resulting root-cause record. This example illustrates MCP-mediated diagnosis, action, and verification while retaining human oversight.

TABLE 5. NETWORK MCP TOOL CATALOG: PROPOSED READ, WRITE, AND STREAM CAPABILITIES

| Read | Write | Stream / event |
|---|---|---|
| vCU/NETCONF GET: running config, cell parameters, neighbor lists, bearer state (interim CMaaS/EMS API) | vCU config push (NETCONF edit-config): PSM, CIO offset, antenna parameters, PCI | NETCONF notifications: config-change, fault, SW-upgrade lifecycle events |
| Authorized standardized NF service operations and events, including NWDAF analytics, plus operational telemetry exposed through governed OAM adapters; no assumption of agent access to NF-internal state | Authorized PCF/SMF procedures and governed OAM adapters; UPF forwarding and QoS rules are applied by the SMF over N4/PFCP. Any MCP wrapper is an operator-provided, non-standard adapter | NWDAF analytics/events and authorized NF/OAM notifications; availability, event scope, and retention are deployment-specific |
| Network performance data: PM counters, active/passive probes, subscriber experience, DPI flow inspection | DPI traffic shaping: rate-limit profiles, QoS re-marking, flow blocking/throttling | Probe streams: RTT, loss, jitter per link/segment; DPI event flows and QoS-violation triggers |
| FMaaS, service desk, inventory; RAN Commander as lightweight digital twin | Kubernetes resource operations: scale CNF replicas, drain nodes, ConfigMaps, rolling restarts | eBPF kernel events (Sauron): AF_XDP RAN L2 telemetry, flow events, runtime security syscall traces |

*IMS Agentic Workflow*. An agent uses MCP tools to collect and correlate availability evidence across the live IMS environment, adapts the investigation when access is constrained, and produces a report that preserves uncertainty where direct verification is unavailable. The investigation does not alter the network elements under analysis. This example illustrates MCP-mediated investigation and reporting without presenting the workflow as an autonomous remediation system.

*Evidence scope and limitations*. These are real-network examples from one operator, presented at a level intended to avoid disclosing operational identifiers or implementation-sensitive detail. They illustrate MCP tool use within agentic workflows; they report no counts, outcome rates or latencies and are not controlled benchmarks or independent audits of accuracy, security, or platform-wide performance. Comparative evaluation and broader validation remain necessary.

*Transferability*. The common pattern is to expose approved domain operations as MCP tools, constrain agent authority to the task, and verify results against network evidence. Deployment in other environments depends on local interfaces, identity controls, data quality, and operating policy. Relation to human-in-the-loop extraction. The IMS workflow resembles uncertainty-preserving relation discovery [84], but its telecom context requires the same validation and audit obligations for reports that may drive later operational action.

### E. Engineering Constraints and Validation Plan

The Network MCP Platform has no validated platform-wide operating point in the supplied evidence. The following design choices and evaluation requirements distinguish proposed controls from observed workflow behavior.

*Standardization boundary*. The Network MCP Platform is an operator-internal orchestration and adapter layer. An MCP endpoint invokes the applicable standardized interface without bypassing its authentication, authorization, transaction, charging, lawful-intercept, audit, or lifecycle controls. Direct MCP exposure is limited to operator-internal or otherwise non-standardized tooling.

*LLM cost*. Model tiering, caching, and deterministic automation are candidate cost controls, consistent with the device–edge–core feasibility concerns in [29]. Their benefit depends on task complexity, model use, context size, retries, and execution volume. The previously proposed 1–5% LLM-engagement rate is not established by the two cases and is not retained as a supported operating assumption. Report measured model-call rates, tokens, compute, and operator effort per completed task before projecting scale.

*Scalability*. An MCP router may be replicated and partitioned by domain, but throughput depends on session state, shared policy services, downstream limits, and dependency ordering. Not all calls are independent or safely parallel. Adding a domain requires validated adapters, schemas, permissions, and recovery behavior, not only registering an intent type. The two cases establish neither concurrency limits nor independence of validation compute from network workload.

*Conflict resolution*. Resource locks, priority rules, pre-execution checks, and human escalation are proposed mechanisms. Evaluation must include conflicting intents, stale observations, priority inversion, starvation, and failed compensation. A fixed priority hierarchy or twin check does not alone demonstrate safe concurrent operation.

*Evaluation design*. Compare MCP-mediated execution with direct invocation of the same underlying tool, matching network path, payload, authentication, and logging. Report six dimensions: router-added latency; throughput/scalability; availability under component and link faults; enforcement-loop delay separated into detection, classification, decision, execution, and verification; failure-recovery success and duration; and agent decision quality against expert-labelled outcomes. Use representative RAN, IMS/core, and cloud incidents, disclose sample sizes and uncertainty, and predeclare acceptance criteria. Separate model reasoning, transport, and adapter times. Appendix D compares the MCP profiling literature, which reports that planning and schema injections dominate agent cost while tool execution itself is a negligible fraction of it; none of that literature measures MCP overhead inside a 3GPP or O-RAN enforcement loop. Single-digit-millisecond overhead and availability parity with SBA functions remain unverified, not demonstrated design properties.

*Failure and recovery*. MCP router failure should halt new agent actions while established network functions continue under their domain controls; recovery must reconcile in-flight operations before retrying writes. Stale twin states should trigger a declared synchronization-error check and escalation. Partial multi-domain commits require reconciliation and compensating actions where possible. Compromised identity requires credential revocation and review of actions already taken. Unreachable network functions require bounded retries and explicit incomplete outcomes, as illustrated by the IMS access failure. Failed rollback requires an operator recovery procedure because disabling an agent cannot restore distributed network state. The RAN record illustrates one successful recovery, and IMS illustrates an access failure; the remaining scenarios are design requirements awaiting fault-injection tests.

*Threats and the safety-gate contract*. Appendix F sets out the threat and mitigation analysis for prompt injection and tool poisoning, confused-deputy authority escalation, model drift and silent degradation, conflicting concurrent intents, compromised or replayed agent identity, and unsafe action on stale state. Each threat is paired with an attack surface, proposed control, detection signal, recovery step, residual risk, and validation status. The controls are specified, but none has been evaluated adversarially in the evidence available here. The safety-gate contract fixes the parameters an operator must declare and measure before autonomous execution: state synchronization error, state age, prediction horizon, calibrated confidence or interval coverage, support/OOD score, blast radius, and escalation path. A failed or unknown condition is fail-closed for the proposed action, except the pre-authorized containment templates of Table 17. The contract is therefore a precondition for future evaluation, not a claim of validated safety. The required measurements include injection-corpus block rate, drift-detection latency, conflict rates under concurrent intents, router overhead, recovery success, and decision quality against expert-labelled outcomes; research question (2) in Section XII states the corresponding tests for the safety gate.

### F. Regulatory and Standards Alignment

Because the router can emit charging-relevant events and lawful-intercept (LI) design hooks on SBI and MCP paths, and the eBPF plane can source consumption telemetry from the kernel for router-bypassing real-time tiers, the audit and metering visibility that conventional signaling exposes can be extended across both router-mediated and bypass paths. However, compliant charging records, lawful-intercept handover interfaces, retention rules, warrants, and mediation functions remain open standardization and implementation items. Validation records and explanation logs may support record-keeping and oversight evidence, but their legal sufficiency requires a use-case-specific assessment [56].

Containment requires separate evidence for local observation, classification, policy decision, and distributed action. eBPF and AF_XDP can supply infrastructure telemetry [10]; latency-critical RAN functions remain within the domain execution layer. The RAN support and IMS investigation records do not measure sub-millisecond telemetry or sub-second containment. Here, protocol Layer-2 telemetry is distinct from L2 support in the RAN case study.

A conservative failsafe revokes agent identity and returns execution to standard 3GPP procedures. Restoration still requires snapshots, reconciliation, rollback ownership, dependency ordering, and recovery objectives. MCP is the selected agent-to-tool binding; a replacement would require adapter and interoperability testing, and safety-critical chains remain subject to applicable functional-safety regimes.

The proposed platform should implement applicable domain controls through existing standards and ecosystem specifications: 3GPP SBI [85], [86], security [74], charging [75], and lawful intercept [76]; O-RAN RIC and E3 [19], [18]; SPIFFE/SPIRE [77] and RFC 9728 [78]; MCP [87] and A2A [88]; and the OCUDU Ecosystem Foundation [89] for multi-vendor interoperability. The 3GPP SA2 6G study's intent-handling key issue (KI#18) [90] provides study context for the problem space while leaving agent identity, charging, lawful intercept, enforcement latency, and transactional fulfillment open, highlighting a live standardization opportunity.

## VIII. 6G STANDARDIZATION: SDO LANDSCAPE AND RAKUTEN'S PUBLIC POSITION

Standardization of 6G is progressing from framework definition toward detailed studies and specifications: the IMT-2030 framework (Recommendation M.2160) is approved while the Technical Performance Requirements remain a WP 5D draft pending Study Group 5 approval (expected Dec. 2026) [2]–[4]; 3GPP Release 20 6G study items are active and Release 21 will carry the first normative 6G specifications, and first commercial 6G deployments are expected around the 2030 window. These decisions affect interoperability, operator governance, and implementation flexibility. This section presents Rakuten Mobile's public position across seven dimensions, summarized in Table 6.

For 3GPP study-status claims, plenary or working-group records are the primary evidence. RAN#112 records Technical Specification Group (TSG)-level checkpoints [91], the RAN1#125 Chair's notes record the communication-waveform agreement [92], RP-261559 defines the approved study scope [93], and 3GPP TR 38.760-1 records the current RAN1 study baseline [94]. The text distinguishes normative specifications, study agreements, operator positions, and research proposals.

### A. Agentic AI: Standardize Enablers, Not Architectures

The AI-agent standardization landscape, illustrated in Fig. 10, is distributed across SDOs without a unified coordination framework. 3GPP SA2 studies AI-embedded core NFs and user equipment (UE) AI-agent registration and discovery (KI#18 [90]), while SA5 covers 6G operations and maintenance, intents, data management, and autonomous-management enablers.

The O-RAN Alliance specifies service management and orchestration (SMO)/R1/rApps, RIC/xApps, and O-Cloud automation, with agentic-AI evolution progressing through nGRG [18], [41]. ETSI ZSM/Experiential Networked Intelligence (ENI) covers agents in autonomous networks and closed-loop automation [23], [31]; NGMN defines the autonomy levels [25]; TM Forum contributes the autonomous-networks levels framework and its evaluation methodology [6], with agent interfaces (TMF939 Agent Management, TMF785 Copilot) proposed rather than published; CAMARA and the exposure layer sit above.

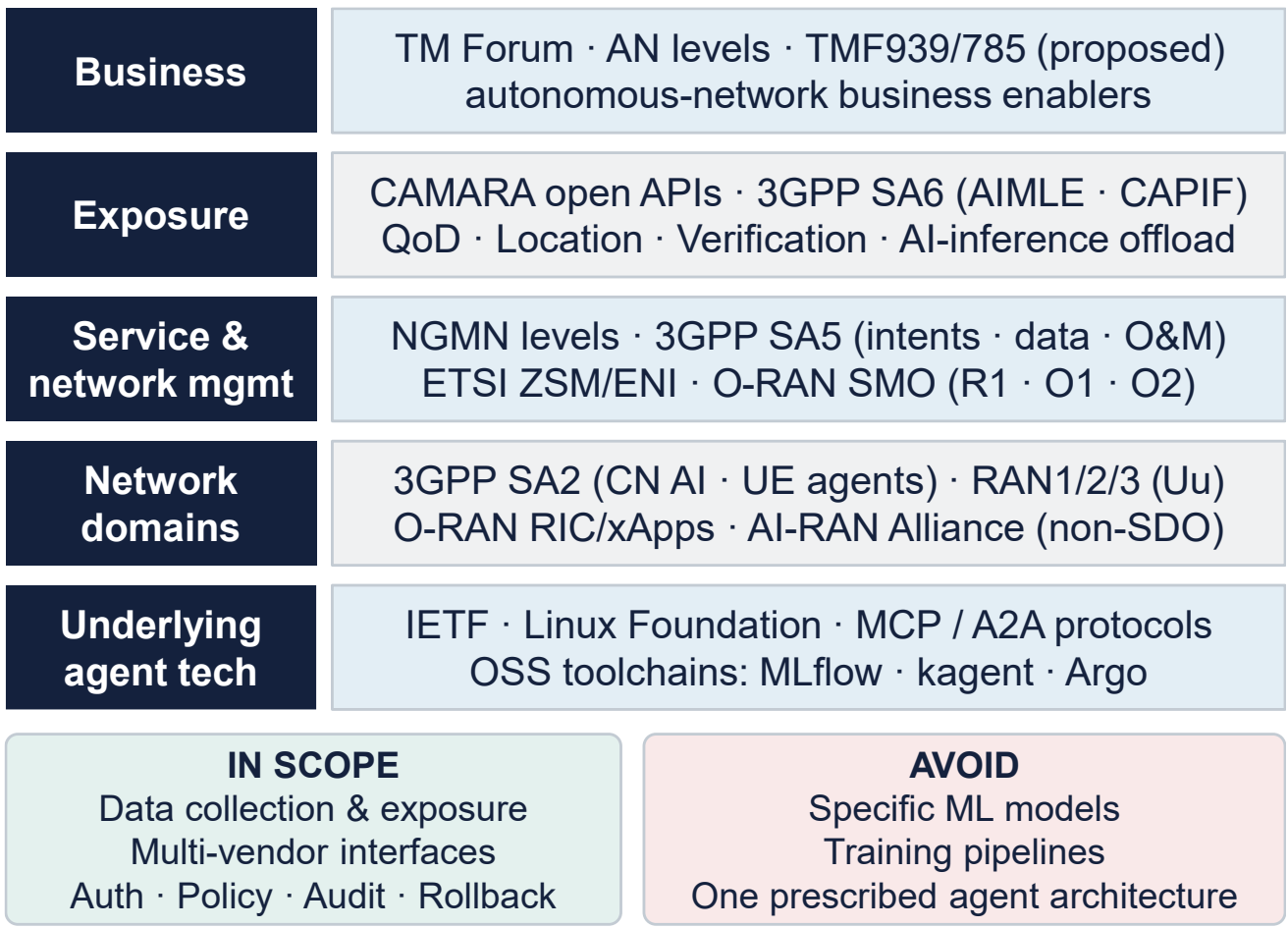

FIGURE 10. Standardization scope for agentic AI in 6G: which SDO defines what – 3GPP SA2/SA5/SA6, O-RAN Alliance, ETSI ZSM/ENI, NGMN, TM Forum, CAMARA, IETF, and open-source toolchains – with the boundary principle that standards expose, secure, and govern capabilities for agents without freezing one AI-agent architecture.

The underlying agent technology – MCP/A2A-like protocols and open-source AI toolchains – evolves in the IETF, the Linux Foundation, and the Agentic AI Foundation.

Rakuten's position: standards should define the enablers and boundaries – the seams rather than the implementation: data collection and exposure, business-critical multi-vendor interfaces, authorization, policy, management and exposure enablers, audit and rollback, and UE/RAN/Core performance control – while avoiding over-standardization of specific ML models, training pipelines, or one frozen agent architecture. Existing 3GPP/O-RAN interfaces remain the execution path; a single frozen agent architecture standardized in 3GPP could constrain implementation diversity and future evolution.

The Network MCP proposal complements NGMN, NTT DOCOMO, 3GPP, and O-RAN frameworks [14], [25]. None of these frameworks requires MCP to replace a standardized southbound telecom interface. Its differentiating element is an operator-internal mediation and governance layer that composes capabilities across standards domains while leaving normative radio, core, management, and exposure interfaces under their existing SDO ownership. The layer targets Level 4 autonomy under Level 5 governance, not Level 5 autonomy: operators set the objectives and thresholds. Trust in it is accumulated through evidence: the operator's own inspectable data and tiered rather than uniform validation, because the bottleneck is not the AI algorithm but the operator's ability to prove how the algorithm behaves.

### B. RAN Architecture, Migration, NTN, PHY, Core, and Spectrum

*RAN-core and the high-level split (HLS).* The centralized-unit/distributed-unit (CU-DU) split is not only a signaling-efficiency question: it determines openness, automation, and troubleshooting capability. The HLS must be standardized to preserve multi-vendor interoperability, with F1 retained and enhanced; a vendor-specific HLS would undermine the tool catalog of Section VII – if the virtualized CU (vCU) NETCONF/YANG surface is proprietary, the READ/WRITE tools of Table 5 cannot deploy multi-vendor, recreating the CMaaS/EMS fragmentation the Network MCP Platform is intended to remove. Rakuten can accept a simple point-to-point RAN-core interface, provided that the standards also supply observability and extension points, exposed through the service-based interface (SBI), for non-connectivity services such as ISAC and data and AI services.

*Migration and interworking.* Rakuten's default is 6G Standalone plus Multi-RAT Spectrum Sharing (MRSS) for spectrum reuse and New Radio (NR)/6G radio (6GR) mobility for coverage continuity, as illustrated in Fig. 11. Of the options under study, a 6G-anchor booster may help day-1 performance (conditionally acceptable); a 5G-anchor risks repeating the non-standalone (NSA) complexity of the 5G transition; dual stack outside the RAN helps continuity but raises UE/core complexity; and dual/split bearer raises unresolved Packet Data Convergence Protocol (PDCP) anchor, latency, policy, and charging questions – none should become mandatory without total-cost-of-ownership (TCO) proof. On core variants: single registration is the baseline; combined access and mobility management function (AMF)/session management function (SMF) may simplify context transfer but risks complexity; separate functions preserve clean architecture but need robust mobility. Migration options should be selected according to demonstrated lifecycle value and manageable operational complexity.

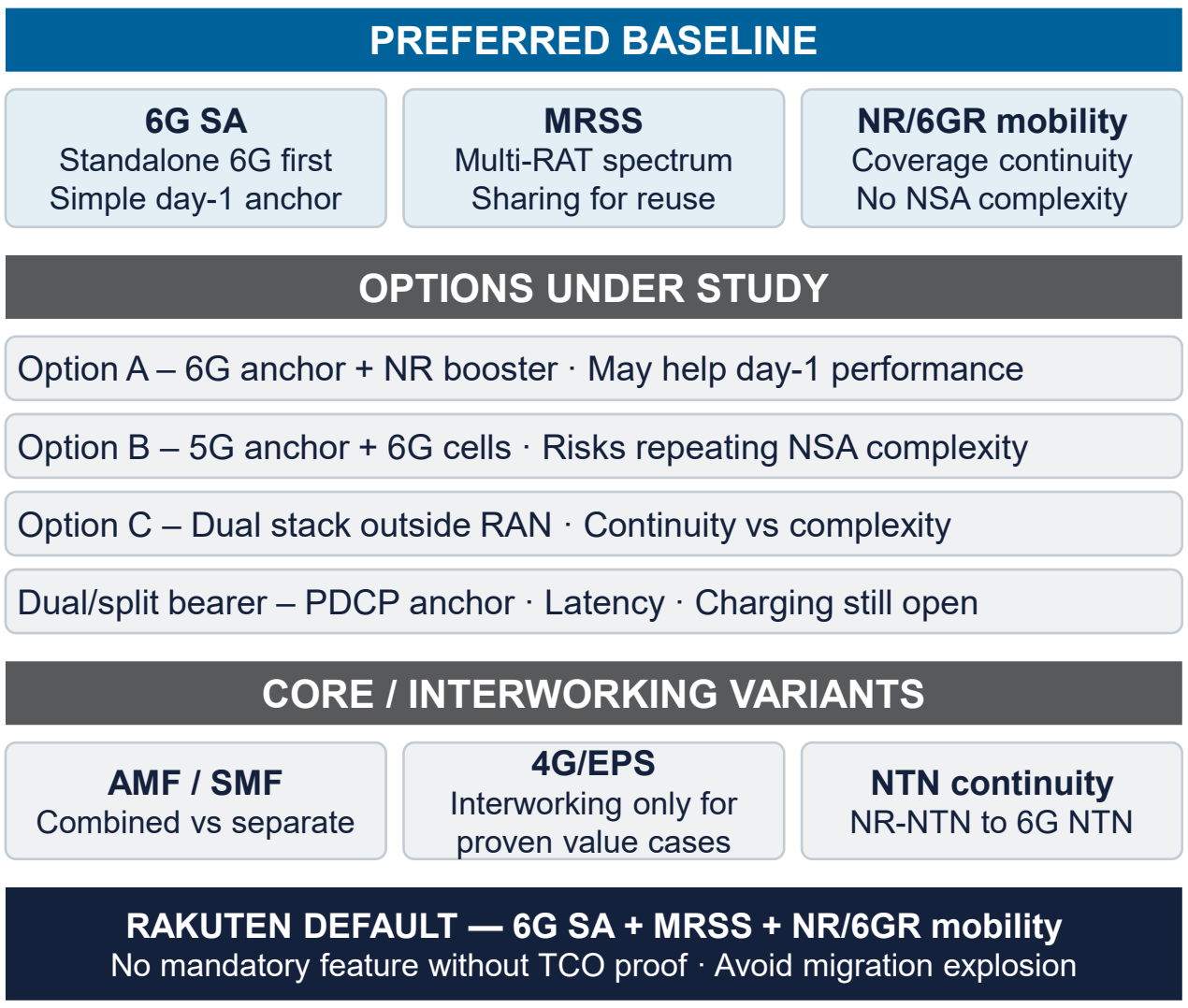

**FIGURE 11. 5G-to-6G migration and interworking: preferred baseline (6G SA + MRSS + NR/6GR mobility), additional options under study (6G-anchor booster, 5G-anchor, dual stack, dual bearer), and core/interworking variants (AMF/SMF combined vs. separate, 4G/EPS interworking, NTN continuity).**

TABLE 6. RAKUTEN MOBILE'S PUBLIC POSITION ON SEVEN PRINCIPAL 6G SDO ARCHITECTURE DECISIONS

| Dimension | Established predecessor baseline or strong study direction | Open items | Rakuten position |
|---|---|---|---|
| Agentic AI | Enablers: data, exposure, authorization, audit, rollback | AI-in-NF vs. subsystem; UE agent registration; cross-SDO data framework | Standardize enablers and governance, never one frozen agent architecture |
| RAN-Core / HLS | O-RAN-aligned disaggregation; open F1 | Non-connectivity service interfaces (ISAC, data/AI) | Standardized HLS; P2P acceptable with observable, documented SBI extensions |
| Migration | 6G SA first | 6G-anchor booster; 5G-anchor; dual stack; dual bearer | 6G SA + MRSS + NR/6GR mobility; extras only with TCO proof |
| NTN | Rel-17→19 NR-NTN evolution to 6G NTN | Feeder/ISL assurance; TN-NTN convergence detail | Day-1 basic support; maximum TN/NTN commonality; handset-first; no overdesign |
| PHY / devices | NR waveform + coding reuse; FR1/FR2 focus | Low-PAPR UL, QAM/shaping, sync/raster, low-tier UE | Coverage first; evidence-gated efficiency; no hidden compute/power/site cost |
| SA/Core | 5G SBA/NAS as starting point; PFCP retained | NAS evolution; UPF SBI; AI subsystem placement; Compute-aaS | Evolve first; open SBA/SBI; governed data layer; no proprietary control islands |
| Spectrum (Japan) | 7125–8400 MHz and 14.8–15.35 GHz under study [40] | WRC-27 outcomes; day-1 bandwidth | Combined legacy + new spectrum; MRSS as operational necessity |

*Non-Terrestrial Network (NTN).* Evolution runs from the Rel-17 NR-NTN and narrowband IoT (NB-IoT)/enhanced machine-type communication (eMTC) NTN baselines through Rel-18/19 enhancements to a 6G NTN target of maximum terrestrial network (TN)/NTN commonality, as illustrated in Fig. 12. The 6G target requires day-1 NTN-ready basic support, handset-first design before very-small-aperture-terminal (VSAT)-heavy optimization, and MRSS with NR NTN alongside global-navigation-satellite-system (GNSS) resilience, positioning-navigation-and-timing (PNT) enhancement, and priority mode. NTN is for resilience and coverage extension – disaster recovery, maritime, IoT deep coverage – not satellite-specific overdesign that inflates device cost; 6G NTN is an evolution, not a start from zero.

*PHY and devices*. RAN#112 cleared the communication-waveform checkpoint [91], and RAN1#125 agreed cyclic-prefix orthogonal frequency-division multiplexing (CP-OFDM) for downlink and CP-OFDM/discrete Fourier transform-spread OFDM (DFT-s-OFDM) for uplink, without precluding uplink enhancements [92]. Current work therefore down-selects low peak-to-average power ratio (PAPR) and frequency-domain spectrum shaping (FDSS) options within that baseline [94]. Vendor projections such as a 6–10 dB uplink gain [38] remain to be independently validated. The current RAN1 study record [94] reports no consensus on constellation shaping or downlink 4096QAM, recommends uplink 1024QAM for fixed wireless access, and records working-group positions on BG3 channel coding and synchronization-signal structure. These are Release-20 study outcomes, not completed normative 6G specifications [95], [96]. The operator decision rule is to preserve the OFDM interoperability baseline unless an enhancement produces deployment-relevant coverage, capacity, or efficiency gains under realistic RF, UE/DU compute, energy, mobility, and multi-vendor constraints. Peak-rate features should not become universal device obligations when their gain is confined to narrow operating conditions.

*Integrated Sensing and Communication (ISAC).* Sensing becomes useful only if measurements, fusion, exposure, and governance are standardized consistently across the *Sense* → *Measure* → *Fuse* → *Expose* → *Assure* chain. Open questions span where fusion is done (RAN node, sensing function, core, SMO, or distributed), which measurement level is standardized (object-level results versus lower-level radio observables for AI fusion), how UEs are involved, and how data is governed – authorization, consent, privacy, and charging (3GPP Rel-20 studies, TR 23.801-01/TR 38.914 discussions). ISAC decides who owns sensing data, fusion logic, and exposure control; fragmented vertical solutions would prevent Sensing-as-a-Service from reaching commercial scale.

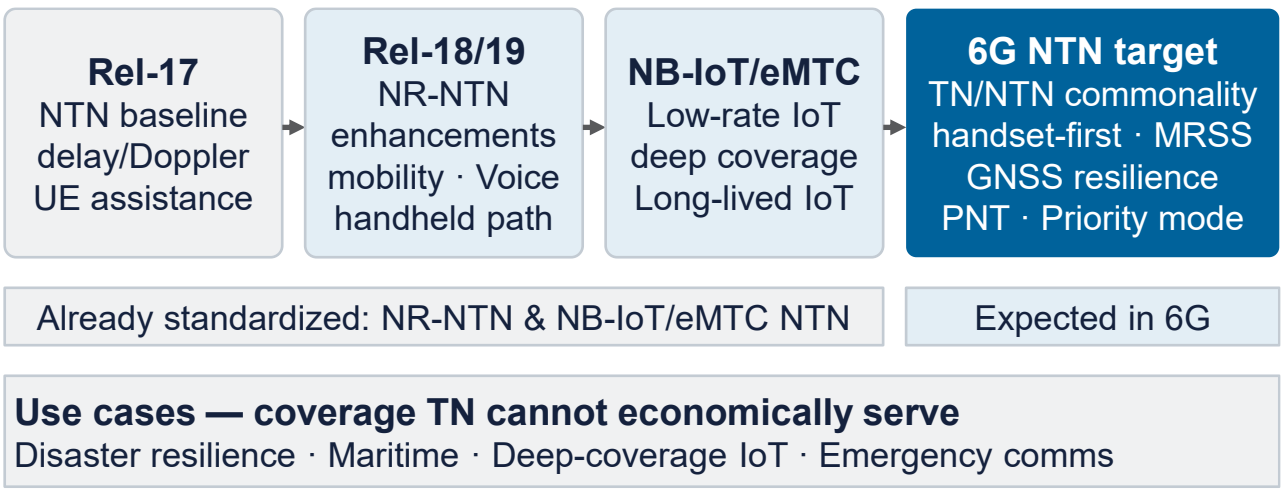

**FIGURE 12. NTN evolution: NR-NTN and NB-IoT/eMTC NTN Rel-17 baselines → Rel-18/19 enhancements → 6G NTN target with maximum TN/NTN common design, handset-first evolution, and MRSS/GNSS-resilience/PNT/priority-mode requirements.**

*Standalone (SA)/Core.* SA2 is deciding who controls 6G intelligence, data, exposure, and continuity. The safe baseline: start from 5G SBA and non-access-stratum (NAS) experience, retain the Packet Forwarding Control Protocol (PFCP) for SMF-UPF stability, and add exposure, policy, and authorization for AI and compute as incremental extensions. Open items include new versus evolved NAS, UPF SBI versus PFCP, AI-embedded core versus a separate AI subsystem, and UE AI-agent registration/discovery. Rakuten's position: evolve first, redesign only where lifecycle value is measurable; standardize enablers, not one AI-agent architecture; keep compute tied to QoS, edge availability, and policy; own a governed data layer across RAN, Core, Cloud, and OAM; prefer open SBA/SBI evolution and avoid proprietary 6G control islands.

*Spectrum and timeline.* Japan is investigating 7125–8400 MHz and 14.8–15.35 GHz for 6G [40]. Since sufficiently wide new spectrum may not be available for day-1 deployment around 2030, combined use of existing and new frequencies will be necessary – and 5G (in some scenarios 4G) will still be operating in both coverage and capacity bands in 2030, making MRSS an operational necessity rather than an option.

The high-level timeline – M.2160 framework (2023), draft TPR (Feb. 2026, pending SG5 approval Dec. 2026), Release 21 6G Stage 1 freeze (Mar. 2027), Stage 3 functional freeze (Dec. 2028), Abstract Syntax Notation One (ASN.1)/OpenAPI freeze (Mar. 2029), commercial deployments around 2030 – aligns with the roadmap of Section XI; the 2027 World Radiocommunication Conference (WRC-27) is the critical external dependency.

## IX. TECHNOLOGY LAST – THE END-TO-END 6G ARCHITECTURE

### A. The End-to-End Architecture

Technology Last argues that every technology choice must be derived from the operator, customer, business, and operational requirements of Sections III–VI, with research roadmaps evaluated against those requirements. The derivation is explicit: *Immersive Experience*'s end-to-end application budget of ≤20 ms [97] – of which the network contribution must respect the TPR ≤4 ms user-plane latency [4] – motivates dense mid-band/mmWave and dApp co-deployment; *Mission-Critical Determinism*'s $1{-}10^{-5}$ TPR reliability floor (with stricter contractual sub-cases) motivates diversity, repetition and predictive maintenance; *AI-Native Communication*'s <10 ms inference latency requires AI-on-RAN edge compute (the strict sub-10 ms bound applies to RAN-local, dApp-tier inference; edge-cloud paths serve the remainder of the latency budget); and *Sensing*'s sub-meter positioning calls for wideband signal processing.

The architecture in Fig. 13 extends cloud-native disaggregation to the full stack – AI substrate on top (kagent, MLflow, MCP, foundation models, multi-agent manager), service exposure (CAMARA, Open Gateway, ISAC-aaS, inference offload, digital twin), UE, Open RAN (O-RU/O-DU/O-CU, distributed UPF, AI-RAN node), three-tier intelligence (Non-RT RIC/SMO, Near-RT RIC, dApps), service-based core (5G core SBA → 6G SBA), edge, central cloud, management and orchestration (ETSI ZSM, Nephio, NGMN L4/L5), and transport.

Three characteristics distinguish it from 5G. First, RAN-AI compute convergence: RAN functions co-hosted with AI inference on shared accelerated hardware [11]; a distributed UPF (dUPF) co-located with the RAN node enables local breakout to AI-on-RAN workloads within the sub-10 ms budget; Harkous et al.'s flat-UP architecture [21] goes further, collapsing the GPRS Tunneling Protocol (GTP) tunnel between CU-UP and UPF with measured throughput, latency, and compute gains. Second, hierarchical multi-timescale intelligence: Non-RT RIC (seconds-minutes), Near-RT RIC [98] (10 ms–1 s), and dApps [19] (<10 ms). Third, unified standards-governed service exposure extending Open Gateway [48] and CAMARA [63] with 6G interfaces. An implementation limitation remains: Open RAN can still trail integrated RAN in throughput in some dense multi-vendor configurations; the argument is not present equivalence but that AI-for-RAN optimization, integration tooling, and interface stability are converging on a trajectory that can make Open RAN a credible long-term choice for operators prioritizing control, with Salmi-style conflict management [17] inside the RIC as a precondition for honoring Guarantee-Economy SLOs across overlapping xApp scopes.

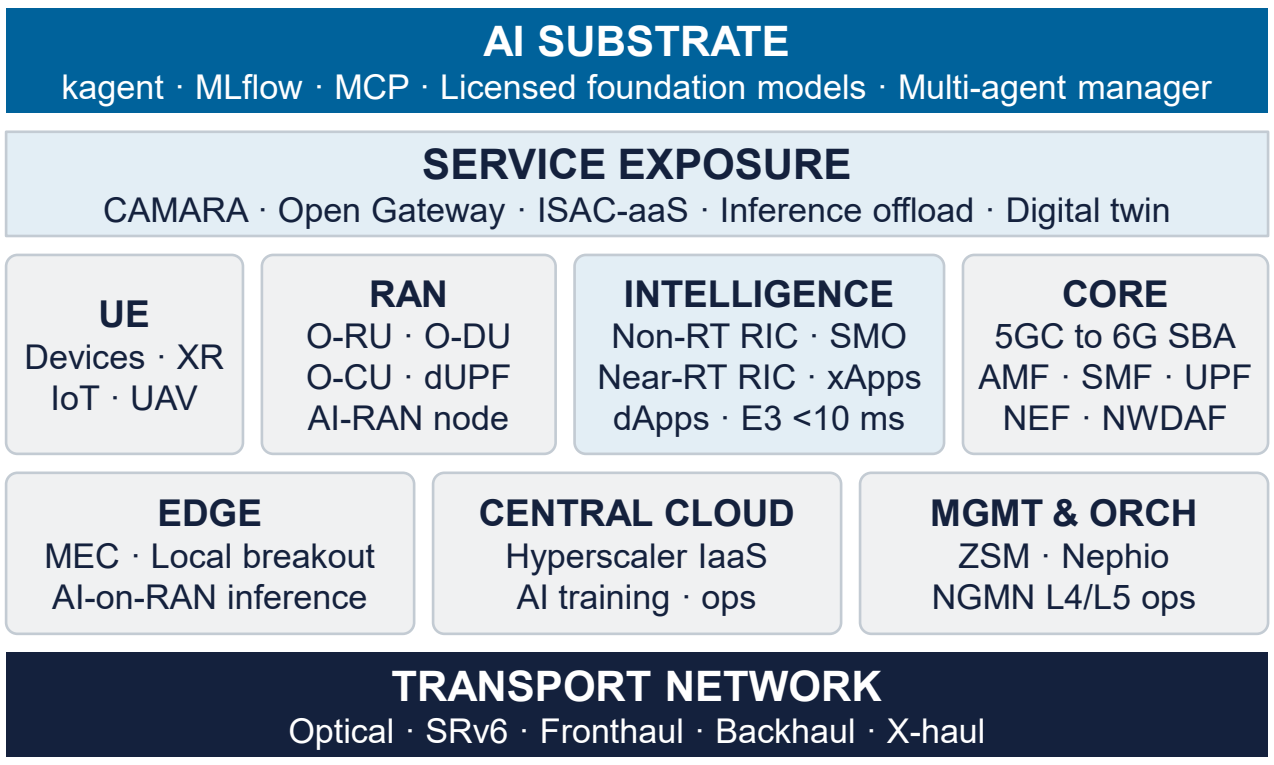

**FIGURE 13. End-to-End 6G Architecture. AI Substrate (top) → Service Exposure → UE → RAN → Intelligence (Non-RT/Near-RT RIC/dApps) → Core → Edge → Central Cloud → Management, with the transport network at the bottom. MCP is an operator-internal orchestration and adapter layer; standardized 3GPP SBI, O-RAN and OAM interfaces remain the execution paths.**

### B. AI-RAN: The Intelligent Radio Platform

As shown in Fig. 14, AI-RAN [11] formalizes two complementary function classes on one accelerated cloud-native platform, creating the possibility that the cell site contributes revenue as well as reduces cost. AI-for-RAN optimizes the radio: neural channel estimators outperform minimum-mean-square-error (MMSE) estimation in complex massive-MIMO channels; learned channel-state-information (CSI) feedback compresses at higher fidelity and lower uplink overhead; transformer beam predictors anticipate mobility; deep-RL schedulers co-optimize throughput, fairness, and energy; and AI cell-sleep and antenna muting target deployment-specific energy-per-bit reductions, consistent with the 6G direction set by TPR [4], but no universal 10× improvement should be inferred; the operational energy-optimization example in Section III.C provides scenario-specific evidence, not validation of all these AI-for-RAN mechanisms. ML co-design is a research option for distributed MIMO (D-MIMO), RIS, and extremely large-scale MIMO (XL-MIMO), not an established prerequisite for their feasibility. AI-on-RAN monetizes the co-located accelerated compute for edge AI workloads [11], [13], [41], [73], [99] across five deployment profiles [13]: interactive GenAI (low-latency uplink voice/multimodal), always-on agentic AI (sustained uplink, context persistence), physical AI (sub-10 ms safety-critical actuation), remote inspection (high-quality video uplink with real-time analysis), and batch broadcast inference (high-throughput downlink). Each maps to a Guarantee Economy tier, suggesting a commercial architecture from AI workload to network product. That commercial architecture remains a projection from demonstrated capability and published demand research rather than a reported large-scale market outcome.

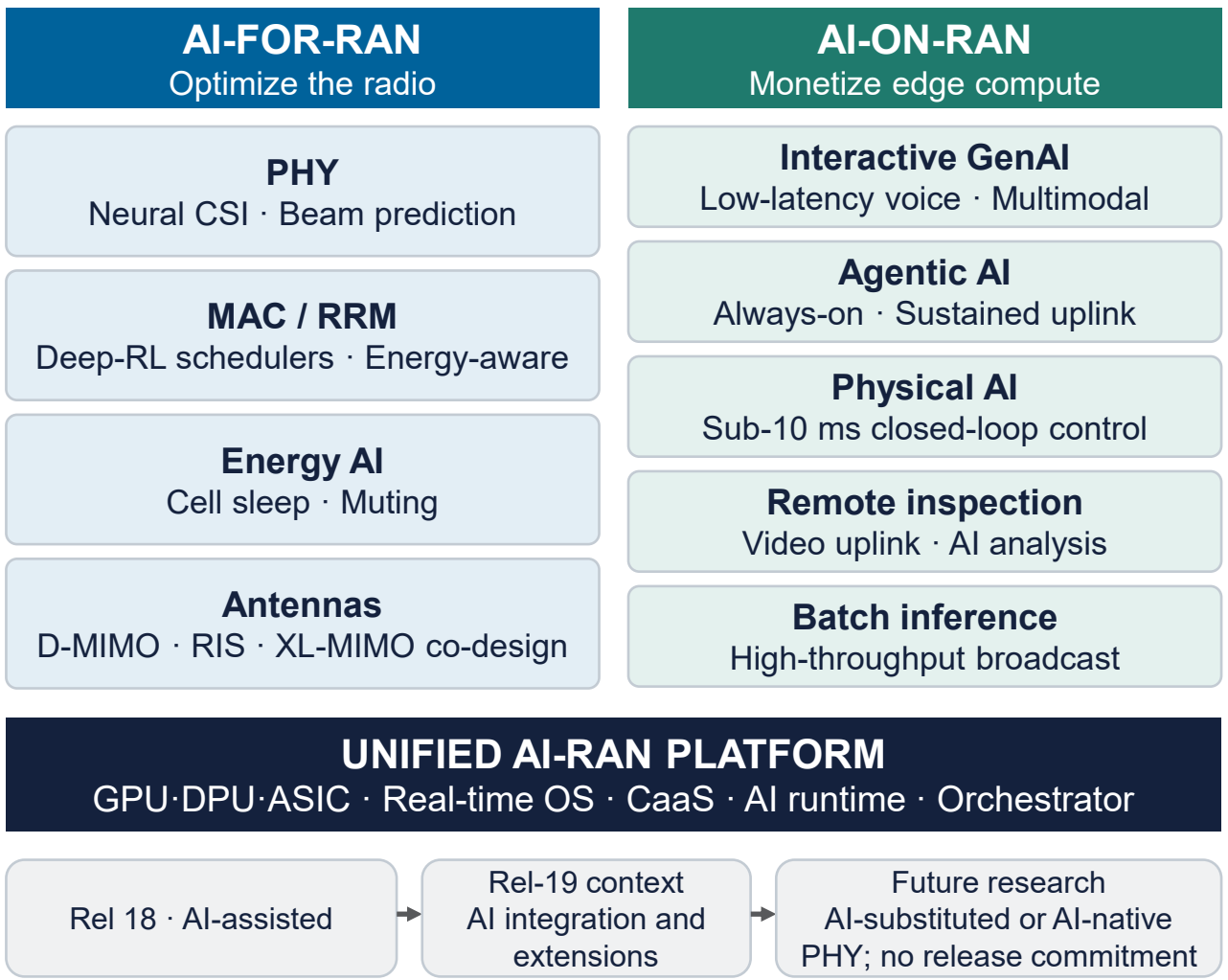

**FIGURE 14. AI-RAN: AI-for-RAN and AI-on-RAN on a Shared Platform. Dual value streams on one accelerated cloud-native platform (GPU/DPU/ASIC, real-time host OS, CaaS, AI runtime) and AI-RAN Orchestrator; release references are maturity context only, not commitments for specific AI-substituted PHY functions.**

The economic case rests on three compounding advantages over central-cloud serving: physical proximity (central round-trips add an estimated 30–80 ms against the 10 ms budget of the AI-RAN WG3 remote-inspection profile [13]); a shared cost basis (AI-for-RAN energy and capacity gains subsidize the accelerated hardware AI-on-RAN monetizes); and data sovereignty (medical images, personnel video, and transaction records frequently cannot leave the facility, making edge inference compulsory rather than preferable). Under the Control Compact, realizing these potential advantages requires the operator to own the AI substrate; purchasing inference from a vendor-managed service converts the sovereign edge advantage into a managed service at vendor-set prices. The standards context frames a possible path, not a commitment: Release 18/19 include AI-assisted CSI/beam work within 5G-Advanced, Release 20 studies 6G AI topics, and any Release 21 normative treatment of AI-substituted blocks or AI-native network management would depend on future 3GPP agreement and demonstrated gain.

### C. OFDM Under Doubly Selective Channels and 6G Radio Waveform Baseline

*Why delay–Doppler selectivity constrains OFDM.* The efficiency of CP-OFDM rests on an assumption: that each subcarrier experiences approximately flat, time-invariant fading over one symbol. That holds while the channel coherence time is long relative to the symbol duration and its delay spread is shorter than the cyclic prefix. In a doubly selective channel – significant delay spread and significant Doppler spread together – the first condition fails: relative motion makes the channel vary within the symbol, subcarriers lose orthogonality, and inter-carrier interference appears whose power grows with the Doppler spread normalized to the subcarrier spacing. Widening the subcarrier spacing suppresses that interference but shortens the symbol, raising cyclic-prefix overhead and coarsening the time–frequency granularity a scheduler can exploit. Several 6G services fall in exactly this regime – high-speed vehicular links (vehicle-to-everything, V2X, a Mission-Critical Determinism use case in Fig. 4) and rail, uncrewed-aerial corridors, and non-terrestrial links (Fig. 12) – all with large and rapidly changing Doppler.

Delay–Doppler and chirp-based waveforms respond by choosing a domain in which the same channel is sparse and quasi-static over the frame – the delay–Doppler grid for orthogonal time frequency space (OTFS) modulation and an affine (chirp) domain for affine frequency division multiplexing (AFDM) – so that each transmitted symbol experiences the full channel diversity rather than one fading realization. The cost is paid in receiver complexity, in pilot and guard design, and, for OTFS, in two-dimensional equalization; AFDM obtains comparable diversity through a one-dimensional transform whose chirp parameters must satisfy stated conditions. A unified treatment of these waveforms within one framework is given in [100], and [101] establishes the full-diversity condition and a low-complexity detector for the index-modulated affine case. These are research waveform candidates, not current 3GPP baseline technologies. Appendix G compares CP-OFDM, DFT-s-OFDM, OTFS, AFDM, index-modulated variants and AI-assisted receivers across robustness in doubly selective channels, peak-to-average power ratio, receiver complexity, sensing and positioning suitability, non-terrestrial suitability, and standardization status. From an operator's standpoint the decision is not which waveform is best in isolation, but which services justify which receiver complexity, and whether the resulting device and network costs fit the service-level objectives.

The standards implication is not to reopen the 6G radio communication-waveform family. RAN#112 cleared the waveform checkpoint and confirmed that the conclusion would not change [91], while the RAN1#125 Chair's notes record CP-OFDM for the downlink and CP-OFDM/DFT-s-OFDM for the uplink as the explicit working-group agreement [92]. The remaining engineering space is to improve operation within that baseline through numerology, reference-signal design, channel tracking and equalization, uplink low-PAPR enhancement, and feature-specific receiver processing. The RAN1#126 record on FDSS-based DFT-s-OFDM is a concrete example: change the part that limits uplink coverage or transmitter efficiency while preserving the common waveform framework [94]. A useful stress variable for this space is normalized Doppler, $f_D/\Delta f$: as Doppler spread becomes a larger fraction of subcarrier spacing, single-tap per-subcarrier equalization becomes progressively less accurate. Increasing subcarrier spacing improves Doppler tolerance but shortens the symbol and changes coverage, phase-noise, cyclic-prefix overhead and scheduling trade-offs.

DFT-s-OFDM addresses the first-order uplink problem of power-amplifier efficiency and coverage through lower PAPR. It does not by itself eliminate Doppler-induced inter-carrier interference, so high-mobility operation still depends on channel tracking, receiver design, numerology and reference-signal choices. AI-assisted receiver processing can be introduced without replacing the agreed over-the-air waveform, although individual AI use cases may affect reference-signal design, resource mapping or other physical-layer parameters. Learned channel estimation, equalization, interference cancellation, channel-state-information reconstruction or detector selection can improve operation of a standardized waveform provided performance is compared against a corresponding non-AI baseline and capability signaling, lifecycle control, fallback and multi-vendor testability are preserved. From a standardization perspective, observable behavior, capability exchange, inputs and outputs, performance requirements and fallback are stronger candidates for specification than a frozen neural architecture. Table 7 states, for each affected topic, what is agreed, what remains under study, and the wording discipline this article applies; the research waveform families of Table 18 are assessed against that baseline rather than presented as co-equal 3GPP candidates.

### *D. The AI-Native Air Interface: Three Stages*

As illustrated in Fig. 15, the path to an AI-native physical layer is presented as a three-stage author research maturity model informed by 3GPP 6G study work [37], not as an approved release commitment.

*Stage 1 – AI-Assisted Enhancement* (3GPP Release 18/19 | 2024–2026): AI augments blocks within a conventional specification – neural channel estimation, learned CSI compression, beam prediction – deployable via software on existing hardware, funding the later stages.

*Stage 2 – Candidate AI-Substituted Functions* (3GPP Release 20/21 | 2025–29): learned equalizers, neural decoders, and deep-reinforcement-learning schedulers are candidates to replace MMSE, low-density parity-check (LDPC)/turbo, and rule-based medium-access-control (MAC) scheduling, each gated on measurable gain, fallback, and lifecycle value; standardize inputs/outputs, not algorithms; requires accelerated compute. Release 20 is the 6G study phase; any

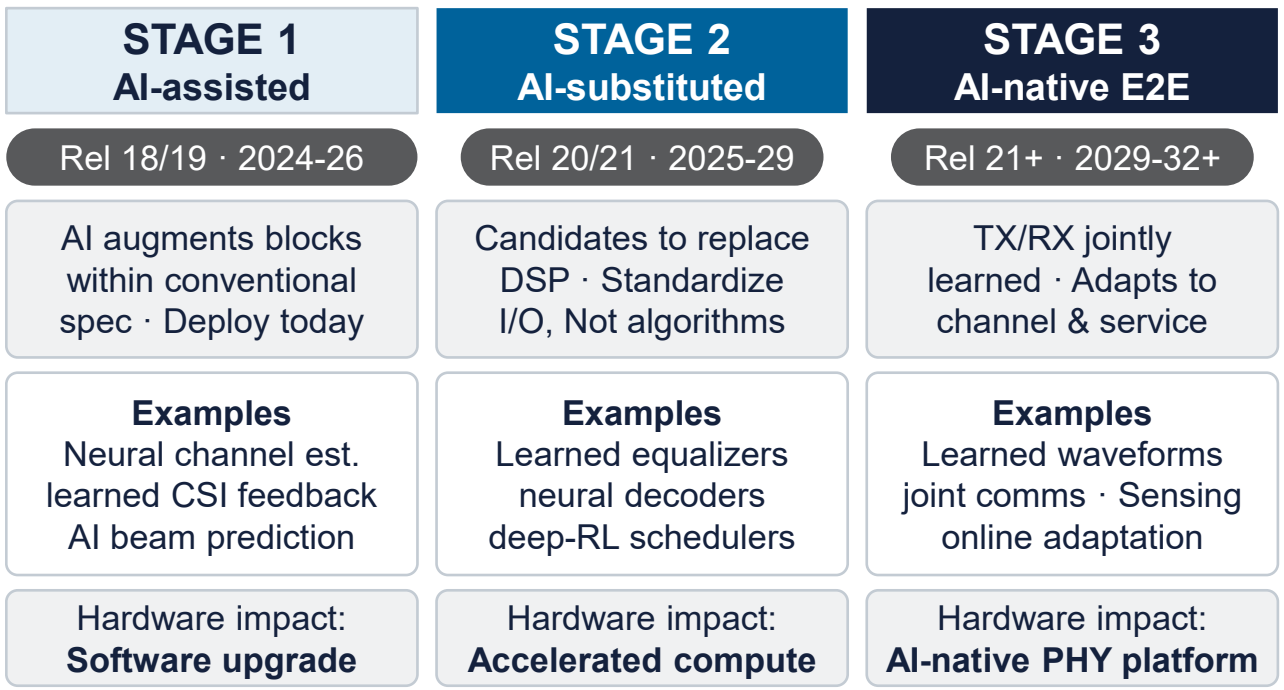

**FIGURE 15. AI-Native Air Interface – A Three-Stage Evolution. Stage 1 (Rel 18/19 AI-assisted enhancements) → Stage 2 (6G study topic; AI-substituted functions requiring accelerated compute; no approved release commitment) → Stage 3 (longer-term research toward end-to-end learned transmit-receive chains; no approved release commitment).**

normative treatment would depend on future 3GPP agreement.

*Stage 3 – AI-Native End-to-End* (3GPP Rel 21+ | 2029–32+): transmitter and receiver jointly optimized as a learned communication system, continuously adapting to channel, interference, and service – the enabling technology for cyber-physical fusion [14]. Zheng et al. [16] provide the concrete Stage 3 realization: cross-module optimization of uplink learned source-channel coding with modulation, downlink modulation with precoding and CSI feedback, and control agents performing model switching by channel state – a natural dApp target, since model switching at channel-state granularity needs the sub-10 ms E3 loop. Stage 3's critical challenge is multi-vendor interoperability of jointly learned systems across UE chipset and infrastructure vendors – an open research problem (Section XII). Device timing constrains commercialization. Applying the 5G SA precedent

TABLE 7. 6G RADIO PHYSICAL-LAYER DECISION STATUS AND REQUIRED WORDING DISCIPLINE

| Topic | Current RAN / RAN1 agreement or baseline | Remaining study or down-selection | Wording discipline applied here |
|---|---|---|---|
| Communication waveform | RAN#112 cleared the waveform checkpoint; RAN1#125 explicitly agreed CP-OFDM for the downlink and CP-OFDM/DFT-s-OFDM for the uplink [91], [92]. | Enhancements within the selected baseline, especially DFT-s-OFDM with FDSS for the uplink [94]. | Treated as a settled study baseline. Non-selected waveform families are not presented as co-equal 3GPP candidates. |
| Uplink low-PAPR | DFT-s-OFDM is retained for the uplink and FDSS-based enhancement is the active low-PAPR path in the RAN1#126 study record [94]. | Transparent and non-transparent filtering, spectral extension and truncation, modulation applicability, and net-gain evaluation. | Described as active RAN1 down-selection, not a completed normative feature. |
| Modulation | Uniform-QAM baseline retained. At RAN1#126, uplink 1024QAM was recommended for fixed wireless access; downlink 4096QAM reached no recommendation consensus; constellation shaping reached no support consensus [94], [96]. | Normative applicability and signaling only if taken into later specification work. | Recommendation and no-consensus language used exactly; none of these is called a finalized 6G feature. |
| Channel coding | LDPC for data and polar for L1 control remain the baseline; BG3 is the targeted LDPC extension beyond the NR operating range. RAN1#126 agreed the uplink BG3 capability rule [94], [95]. | Detailed BG3 realization, any additional plenary exception, and small-UCI code down-selection. | Working-group agreement kept separate from later normative specification. |
| Synchronization / SSB | RAN1#126 agreed 20 resource blocks per 6G synchronization signal block and 12 for the primary synchronization signal [94]. | PSS and SSS sequence type and length, number of symbols, periodicity and related details. | Cited to 3GPP TR 38.760-1, the current RAN1 study baseline for these agreements; TR status, not a completed normative specification. |
| Multi-carrier and MRSS | MRSS and 6G radio spectrum aggregation are in the RAN-approved 6G study scope [93], with detailed multi-carrier mechanisms under RAN1 study [94]. | Fast carrier enable and disable, cross-carrier and multi-carrier scheduling, and deployment assumptions including non-collocated cases. | MRSS, native 6G aggregation, inter-RAT aggregation and mobility are kept distinct; no preferred anchor is described as a 3GPP agreement. |

– several years between standardization and meaningful handset penetration – and typical silicon cycles, consumer AI-native services at scale are a Phase 3 objective (around 2030 and beyond). Enterprise and IoT devices, for which custom silicon is procurable, accelerate ahead – which is why Phase 2 Guarantee Economy revenue depends predominantly on enterprise, industrial IoT, and fixed wireless (Section XI).

### *E. dApps and the E3 Interface*

The Near-RT RIC 10 ms–1 s loop is insufficient for real-time spectrum sharing, ISAC feedback, and deterministic control. The dApps framework (nGRG RR-2025–05 [19]) introduces a control tier executing directly on DU/CU infrastructure, as depicted in Fig. 16, with three E3 procedures: *Registration* (capability and resource declaration to the host CU/DU, enabling lifecycle management without bespoke integration), *Subscription* (user-plane data and telemetry streams at sub-millisecond cadence over Unix-domain-socket exchange), and *Control Message* (spectrum allocation, beam steering, and slice admission inside the sub-10 ms cycle). The nGRG prototype on OpenAirInterface validated real-time spectrum sensing and sub-meter positioning over the air [19]. The mapping to Section IV is precise but separates two latency budgets: *Mission-Critical Determinism* retains the radio/user-plane latency target in Table 2, while dApps are proposed to provide a sub-10 ms E3 control-loop budget for control actions; *Sensing* requires direct user-plane access; *AI-Native Communication* benefits from in-node inference. *Deployment readiness* requires an E3-exposing CU/DU implementation, a dApp certification framework comparable to xApp onboarding, and a conflict-resolution mechanism arbitrating dApp actions against xApp policies – the Salmi-style engine [17] lifted into the dApp domain. *Highest-tier Determinism* SLAs cannot be priced credibly without a control plane that intervenes inside the radio-frame cadence.

### *F. Spectrum Strategy: Coverage, Capacity, and Sensing*

The spectrum strategy, illustrated in Fig. 17, is layered against the outcome families. 3GPP frequency range 1 (FR1, up to 7.125 GHz) is the coverage and ISAC foundation: proven 3.5/4.9 GHz mid-bands plus legacy sub-3 GHz deliver indoor penetration, reach, and the $10^6$ devices/km² density target [4]; the TPR positioning targets (0.75 m indoor factory, 6 m urban macro, 95% detection) [4] are evaluation targets whose achievability depends on bandwidth and geometry. The *upper mid-band* (7–24 GHz – an informal industry grouping, not an agreed 3GPP range) is the 6G primary capacity layer; WRC-27 Agenda Item 1.7 studies specific candidate bands – 4.4–4.8 GHz, 7.125–8.4 GHz, and 14.8–15.35 GHz – not the entire 7–24 GHz range. *mmWave* (24–100 GHz, of which 3GPP has standardized 24.25–71 GHz as *FR2*) serves dense urban, indoor, and fixed-wireless-access (FWA) hotspots, where the draft TPR peak rates (36 Gbit/s downlink, 18 Gbit/s uplink) and 1 ms HRLLC latency are approached only under favorable hotspot conditions; they are not intrinsic mmWave guarantees. *Sub-THz* (>100 GHz) remains extreme-capacity research – carrier bandwidths of up to 10 GHz for holographic and extreme XR. NTN constellations integrate with terrestrial assets – coordinated through the RIC framework – on a deployment-dependent basis for ubiquity and resilience, not as blanket coverage across all bands.

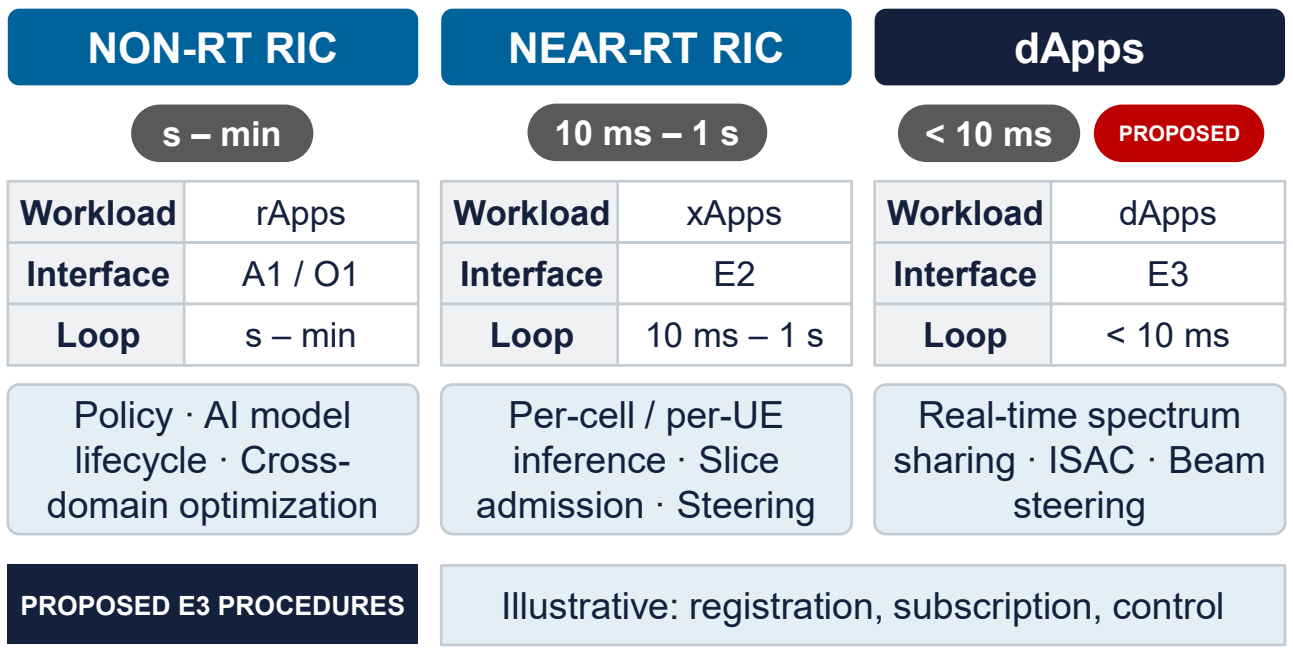

**FIGURE 16. dApps and the E3 Interface – proposed sub-10 ms edge control. Three intelligence tiers – non-RT RIC/rApps (A1/O1, s–min), Near-RT RIC/xApps (E2, 10 ms–1 s), dApps (E3, target <10 ms) – plus the E3 procedures: Registration, Subscription, and Control Message. The E3 interface and its sub-10 ms target are O-RAN nGRG industry research and a proposed control target, not a normative deployment guarantee.**

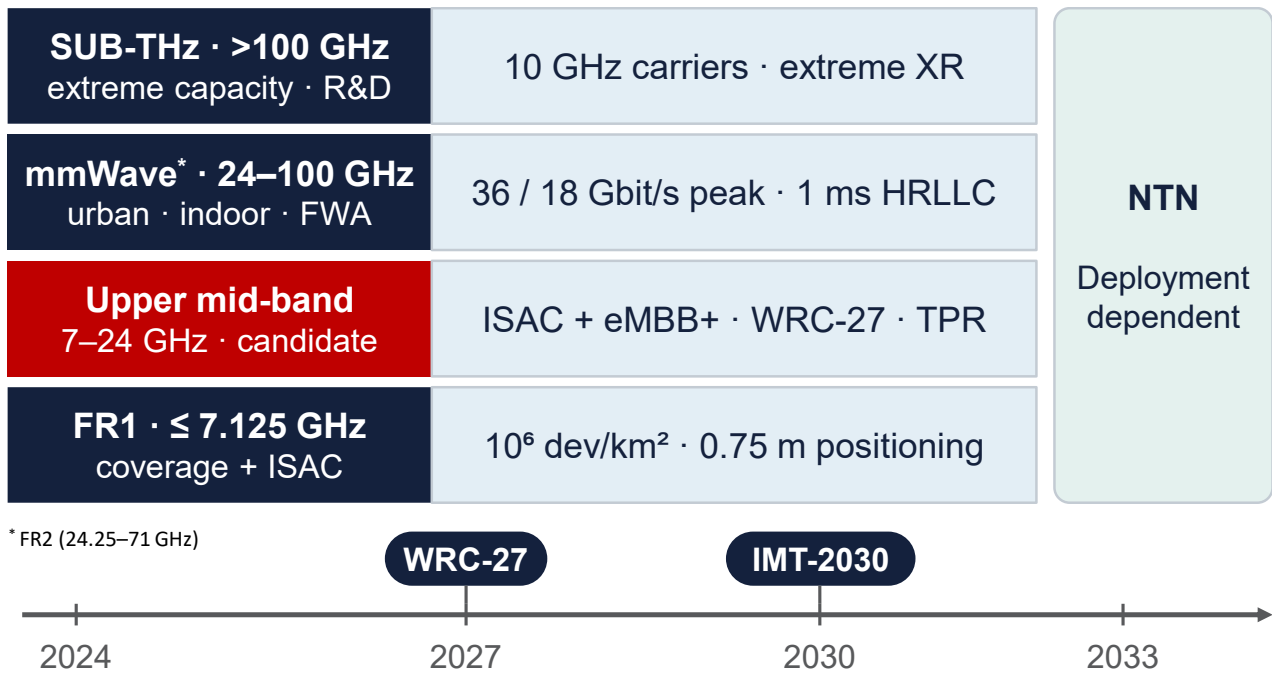

**FIGURE 17. Spectrum Strategy – Coverage, Capacity, and Sensing. Four spectrum layers set against the outcome families: FR1 (up to 7.125 GHz) as the coverage and ISAC foundation; the upper mid-band (7–24 GHz, an informal industry grouping in which WRC-27 studies only specified candidate bands) as the primary 6G capacity layer; mmWave (24–100 GHz, of which 3GPP FR2 covers 24.25–71 GHz) for dense urban, indoor, and FWA hotspots; and sub-THz (>100 GHz) for extreme-capacity research. NTN integration is deployment-dependent, not blanket coverage across all bands.**

WRC-27 contingency planning is essential: upper mid-band identification faces structurally organized regional opposition, so operators should model delayed or partial outcomes as the base case, sizing Phase 2 on existing allocations plus AI-for-RAN spectral-efficiency gains, with WRC-27 success treated as material upside rather than prerequisite. Operators and administrations must engage the European, Asia-Pacific, and African regional preparatory processes – the European Conference of Postal and Telecommunications Administrations (CEPT), the Asia-Pacific Telecommunity (APT), and the African Telecommunications Union (ATU) – now – while planning as though outcomes will be partial or late.

*Cross-layer contention between sensing and communication.* Where sensing and communication share a carrier, three contentions arise and none is resolved by spectrum policy alone. First, occupancy against capacity: sensing waveforms and sensing-dedicated resources consume time–frequency resources that would otherwise carry traffic, so a sensing objective expressed as coverage and update rate translates directly into a capacity reduction that has to be priced. Second, duty cycle against accuracy: both positioning accuracy and update interval improve with sensing duty cycle, so the sensing targets in Table 2 and the Sensing-as-a-Service objectives in Table 3 jointly imply a minimum duty cycle whose capacity cost is deployment-specific. Third, objective conflict: a sensing objective and a latency objective on the same carrier can require the scheduler to make opposite decisions in the same interval. The third case is not a radio problem but a governance one, and it is the conflict-resolution mechanism of Section VII that must arbitrate it. That in turn requires sensing objectives to be expressed in the same objective language as communication service-level objectives, with an explicit precedence rule and an auditable record of which objective was preferred and why. No such precedence rule exists in current specifications; this article proposes it as a standardization item rather than asserting a resolution.

*Service-layer arbitration for that contention*. The service layer should expose both communication service-level objectives and sensing objectives to a common arbitration function. Hard safety and mission-critical communication constraints are enforced first; within the remaining feasible region the scheduler optimizes sensing update rate, accuracy or confidence against communication throughput, latency and reliability. A sensing request that would violate a contracted latency or reliability bound is deferred, reduced in duty cycle, moved in frequency or beam, or rejected with an auditable reason; conversely a time-critical sensing objective can reserve bounded resources while the affected communication service stays within its contractual envelope. The same conflict record feeds the Section VII policy and audit plane, so that cross-domain agents cannot independently overbook radio resources. Baseline 3GPP sensing evaluation assumptions allow CP-OFDM as a starting point or benchmark while permitting other waveforms to be reported, and study 100 MHz at 4 GHz and 200 or 400 MHz around 7 GHz, which makes explicit cross-layer resource accounting concrete rather than theoretical.

### G. The Near-RT RIC Toward 6G

The Near-RT RIC evolves for 6G [18] with enhanced AI/ML support (federated learning, distributed edge training), expanded service exposure, unified data management, multi-RIC interworking, and Communication and Computing Integrated Network (CCIN) support enabling joint radio-and-compute reservation – the architectural precondition for end-to-end Guarantee-Economy SLA enforcement. Cross-RIC federation lets an enterprise-campus xApp coordinate with the surrounding macro RIC so mobility preserves the guaranteed SLO without unfiltered access to public-network telemetry – cross-domain AI coordination without cross-domain raw-data sharing, a privacy-preserving pattern that can support GDPR [57] and EHDS [58] compliance-by-design when lawful basis, purpose limitation, retention, transfer, and sector-specific obligations are satisfied. Three extensions beyond current specification are needed: federated learning at the RIC tier itself; formal certification of xApp interaction graphs to make conflict resolution auditable at third-party scale; and first-class observability of the RIC itself via eBPF instrumentation [9], [10] – the engineering path to low-latency drift detection, supporting the trust-management and tamper-evidence objectives of ETSI ZSM 017 [31].

## X. TRUST, SOVEREIGNTY, AND SUSTAINABILITY

### A. Zero-Trust for Autonomous 6G

In 4G/5G, security was a perimeter concept; agentic AI demolishes it. At Level 4 autonomy, compromising the telemetry feeding Observe, poisoning the model behind Decide, or injecting commands into Act requires only access to the closed-loop automation infrastructure – not a perimeter breach. ETSI ZSM 017 [31] maps risks across each loop stage and proposes three interlocking mechanisms, illustrated in Fig. 18: Closed Loop Trust Management (per-domain cryptographic trust certificates, immutable audit trails, inter-loop identity verification), Closed Loop Access Control (attribute-based least-privilege policies over telemetry reads and configuration writes), and Closed Loop Security Exposure (a real-time security-posture interface enabling automated anomaly detection and human audit under EU AI Act Article 14 [56]). The convergence is useful but bounded: ZSM 017 is an ETSI Group Report, not an AI Act conformity standard, so ZSM 017-aligned controls may contribute technical evidence relevant to an AI Act conformity assessment but do not by themselves establish legal conformity – a partial, not a dual, return.

ZSM 017 does not cover LLM-specific threats: prompt injection, tool poisoning, and goal misgeneralization [35].

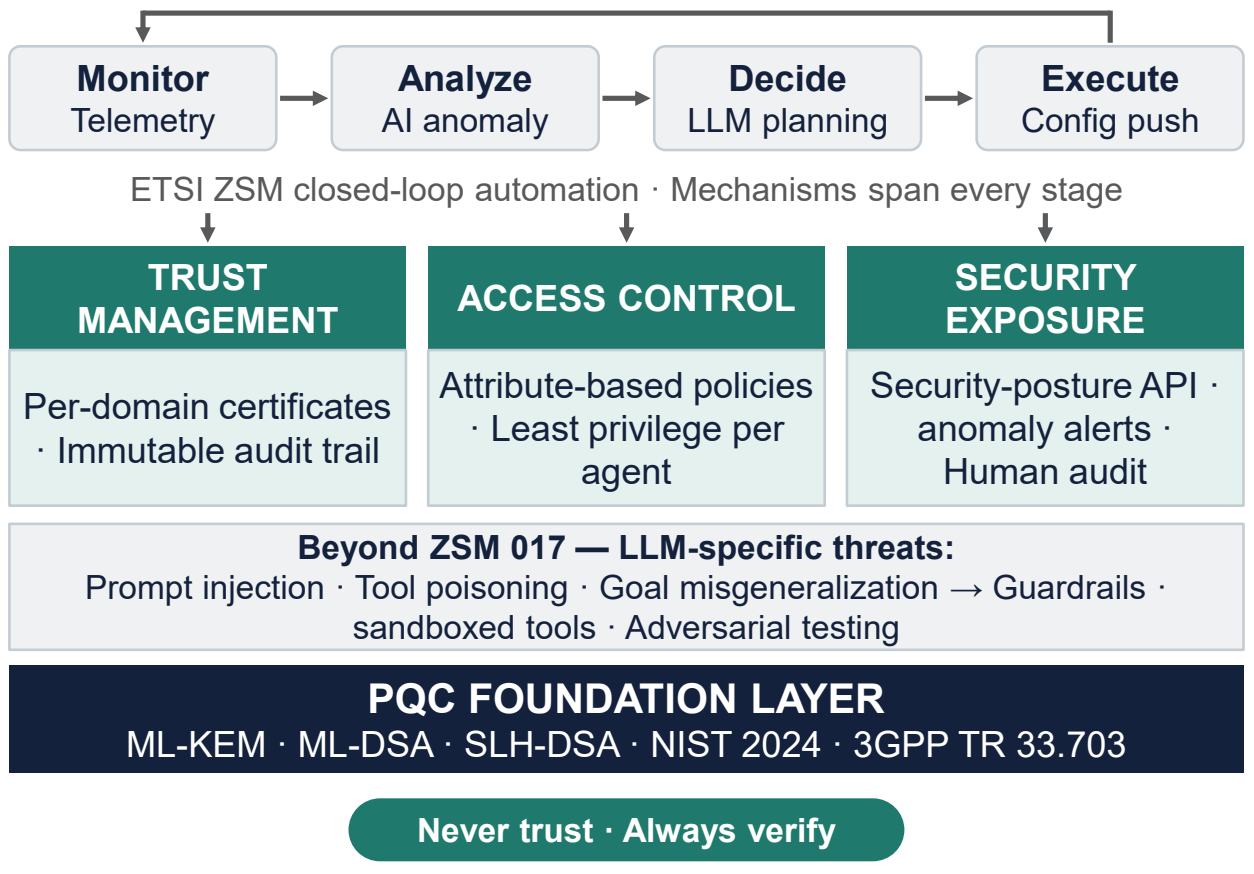

**FIGURE 18. Zero-Trust Architecture for Autonomous 6G. Three ETSI ZSM 017 mechanisms – Closed Loop Trust Management, Access Control, and Security Exposure – spanning every *Monitor* → *Analyze* → *Decide* → *Execute* stage; the PQC foundation layer (NIST 2024; 3GPP TR 33.703) underpins the trust hierarchy.**

These require adversarial testing pipelines, output guardrails validating configuration commands against policy whitelists before execution, sandboxed tool execution, and immutable logs for forensic reconstruction – deployment prerequisites, not post-deployment items, though their efficacy against production telecom attack patterns remains to be empirically validated (Section XII).

Post-quantum cryptography – NIST-finalized ML-KEM [33], ML-DSA [34], and stateless hash-based digital signature algorithm (SLH-DSA) [102] – underpins the trust hierarchy, and 3GPP is studying the transition to PQC in Release 20 (TR 33.703); with cryptanalytically relevant quantum computers expected within 6G's operational lifetime, migration is an operational priority today. Choudhary et al. [32] show the migration problem is operational rather than algorithmic: hybrid post-quantum AKA variants have measurable overheads compatible with commercial platforms, but cut-over sequencing across RAN, subscription permanent/concealed identifier (SUPI/SUCI) concealment, roaming, and lawful-intercept conformance must be coordinated. Until that study yields normative specifications, operators will sequence these steps divergently. And cryptographic resilience does not imply AI resilience [35]: an input-manipulable scheduling agent compromises service delivery even under post-quantum-strong cryptography.

### B. Data Sovereignty and Regulatory Compliance

Data sovereignty is now a primary procurement criterion for the highest-value Guarantee Economy segments. In the EU, GDPR [57], EHDS [58], and the EU AI Act [56] converge: critical-infrastructure AI may be high-risk depending on intended purpose and Article 6/Annex III classification, triggering human oversight (Article 14), robustness (Article 15), and pre-deployment conformity assessment where applicable. National-security policy is a structural accelerant – the EU 5G Toolbox [103], the US Secure and Trusted Communications Networks Act [61] and CHIPS Act [62], and analogous frameworks in the UK, Australia, Japan, Korea, and India [55] all increase scrutiny of supplier provenance and may impose high-risk-vendor restrictions depending on jurisdiction and asset class; operator-owned software materially reduces exposure to high-risk-supplier exclusion – though its open-source supply chain remains within the scope of the Cyber Resilience Act [60] and provenance regimes – broadly aligning regulation with the Control Compact.

Operational treatment is required for lawful-intercept evidence when agents initiate actions, cross-border training-data governance, and changing AI-regulatory obligations. ISAC adds bystander sensing, requiring a lawful basis, purpose limitation, and data-protection impact assessment where applicable. Regionally parameterized policy templates can support compliance operations, but static architectural rules cannot substitute for a jurisdiction- and use-case-specific legal assessment [56].

*Conformity assessment for a Level 4 loop.* Article 14 of Regulation (EU) 2024/1689 [56] requires high-risk AI systems to be designed so that natural persons can effectively oversee them, including the ability to interpret output, to decide not to use the system, and to intervene or halt it. For a closed loop operating at autonomy Level 4 this is not satisfied by an emergency stop. As Section VII.E states, disabling agents does not restore distributed network state, so an oversight capability that can only halt is not a capability to intervene. A conformity assessment of such a loop would therefore have to examine four things this article can specify but not yet evidence: that the validation record and explanation log are sufficient for a human to reconstruct why an action was taken; that the escalation path and confidence threshold of the safety-gate contract are declared and honored; that a rollback owner and a recovery objective exist for each action class; and that the blast-radius limit is enforced rather than advisory. Whether a network management system of this kind falls within a high-risk category at all depends on the use to which it is put and is a legal determination rather than an architectural one. The governance layer described here establishes auditability, not legal sufficiency, which requires a use-case-specific assessment [56].

### C. Sustainability as Competitive Architecture

Sustainability has moved from responsibility commitment to compliance criterion: ITU-T L.1470 [104] sets Paris-aligned emission trajectories for the ICT sector, the EU Corporate Sustainability Reporting Directive (CSRD) [105] mandates auditable Scope 1/2/3 disclosure, and the ITU-R TPR [4] treats energy efficiency as a formal IMT-2030 performance dimension and draft evaluation requirement, rather than a universal deployment guarantee. As illustrated in Fig. 19 and defined in Table 8, the KPI framework spans three levels. Component level: energy per bit, idle-versus-active power, cell sleep, and antenna muting (3GPP and O-RAN reporting).

System-level metrics include TEER, network carbon intensity, energy reuse factor (ERF), and power usage effectiveness (PUE) [54]; service-level treatment requires carbon per delivered outcome and auditable Scope 1/2/3 allocation. The Section III.C case reports a bounded autonomy assessment and operator estimate, not measured improvement in these system or service metrics.

Zero-trust, sovereignty, sustainability reporting, and agentic operations can share an operator-governed telemetry substrate,

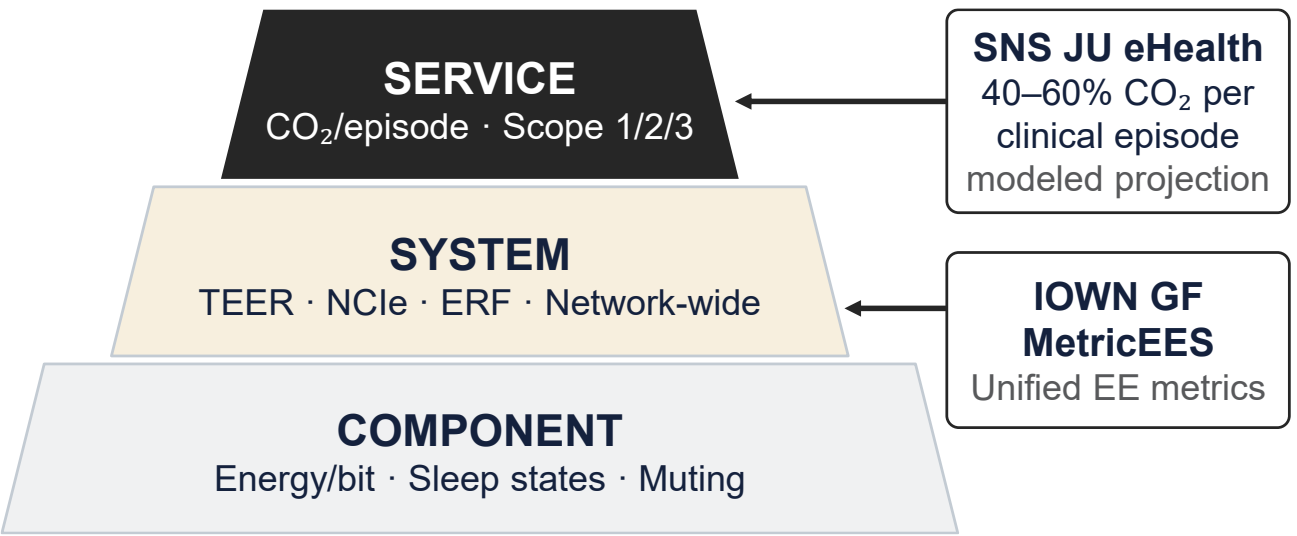

**FIGURE 19. Sustainability KPI Framework – Component, System, and Service levels.**

subject to their distinct assurance requirements. The unresolved link is auditable allocation of energy and carbon to a session, slice, API call, or business outcome; without it, sustainability cannot be priced as a verifiable service dimension.

A preliminary pathway for service-level energy attribution. Section XII identifies service-level energy and carbon attribution as unsolved. The gap is real, but a route exists and the operator holds the measurement points, so the problem can be stated as a standardization proposal rather than only as an open question. Three measurement classes are available: equipment-level energy from radio-unit and baseband counters against a declared equipment boundary; per-flow and per-workload attribution from kernel-level observation of bytes and accelerator time; and facility-level energy against a declared facility boundary. Given these, three allocation rules are candidates, each with a different failure mode. Proportional-to-traffic allocation divides measured energy by delivered bytes: simple and auditable, and it systematically under-charges low-volume latency-critical services because it ignores the capacity held idle in order to guarantee them. Marginal allocation charges a service the energy that would not have been consumed in its absence: economically the correct rule, and not additive, so allocations do not sum to measured total energy. Reserved-capacity allocation charges the energy attributable to resources held for a service whether used or not: it prices guarantees correctly and requires a declared reservation model that the network may not expose. An auditable declaration must therefore state four things – the measurement boundary, the allocation rule, the treatment of idle and reserved capacity, and the emission factor with its source and period. Without all four, a per-service figure is not comparable between operators. The dominant error sources are the unallocated shared baseline, the accuracy of accelerator energy accounting, and the temporal granularity of the emission factor.

Energy against service guarantees. The same measurements expose a trade-off the guaranteed framework must arbitrate rather than conceal. Cell and carrier sleep, and antenna muting, reduce energy at the cost of wake-up latency and reduced instantaneous capacity – which is precisely what a latency or reliability commitment forbids at the moment it is needed. A guaranteed tier is therefore also an energy commitment, and arbitration belongs in the conflict-resolution mechanism of Section VII, with the energy objective expressed in the same language as the service-level objectives against which it competes. The externally assessed energy-optimization scenario discussed in Section III.C is bounded in exactly this way: it operates under performance-triggered reversion and backoff, which is a service-protective constraint placed on an energy objective rather than unconstrained optimization.

TABLE 8. 6G SUSTAINABILITY KPI CATALOG WITH PRIMARY STANDARD OWNERS

| KPI | Primary source | Measurement boundary | Unit / status | Reference-aligned target |
|---|---|---|---|---|
| TEER | ATIS-0600015 | Equipment under a prescribed weighted load | Throughput/W; equipment benchmark, not a GHG-scope metric | Operator-defined vs. own 5G baseline |
| Network energy efficiency | ITU-R TPR §4.19; ETSI ES 203 228 | Operational network; low/no load vs. a fully-loaded reference case | Relative energy consumption, % (evaluation metric) | No universal % target – a draft-TPR evaluation method |
| NCIe | ITU-T L.1333 | Explicit network boundary + emission factor | $kgCO_2e$/GB; emissions boundary must be declared | Track vs. ITU-T L.1470 [104] Paris-aligned −45% (2020→2030) trajectory |
| ERF | Facility/ datacenter standards | Facility heat-reuse boundary | Dimensionless fraction; facility efficiency metric, not a GHG-scope metric | Applicable only where useful heat recovery exists |
| PUE | ISO/IEC 30134-2; facility standards | Datacenter/facility energy boundary | Dimensionless ratio; facility efficiency metric, not a GHG-scope metric | Minimize where facility owned/ controlled; not a radio-network KPI |
| Cell-sleep utilization | Operator / O-RAN measurement | Cell/carrier/RU time by sleep state | % time; operational utilization metric, not a GHG-scope metric | Operator-specific; not a universal standard target |
| GHG Scope 1/2/3 | GHG Protocol Corporate Standard; CSRD [105] and applicable ESRS | Corporate inventory: direct, purchased-energy, and value-chain emissions | $tCO_2e$ by scope; corporate accounting, not a network or service KPI | Report against the applicable inventory boundary and transition plan; do not infer a service-level allocation or universal operator target |
| Service carbon intensity | Research gap | Session, slice, API, or delivered outcome | $gCO_2e$ per service unit | Proposed standardization objective (no current standard target) |

## XI. FROM VISION TO DEPLOYMENT – ROADMAP AND STAKEHOLDER IMPLICATIONS

### A. Phased Deployment Roadmap

The roadmap in Fig. 20 distinguishes external standards and spectrum milestones from operator deployment targets. Its Release-21 and ITU-R dates are planning assumptions, not commitments, and the profile reflects a cloud-native operator rather than an industry average [5].

*Phase 1* (2025–2027): 5G-Advanced deployment and 6G studies. Priorities are cloud-native lifecycle automation,

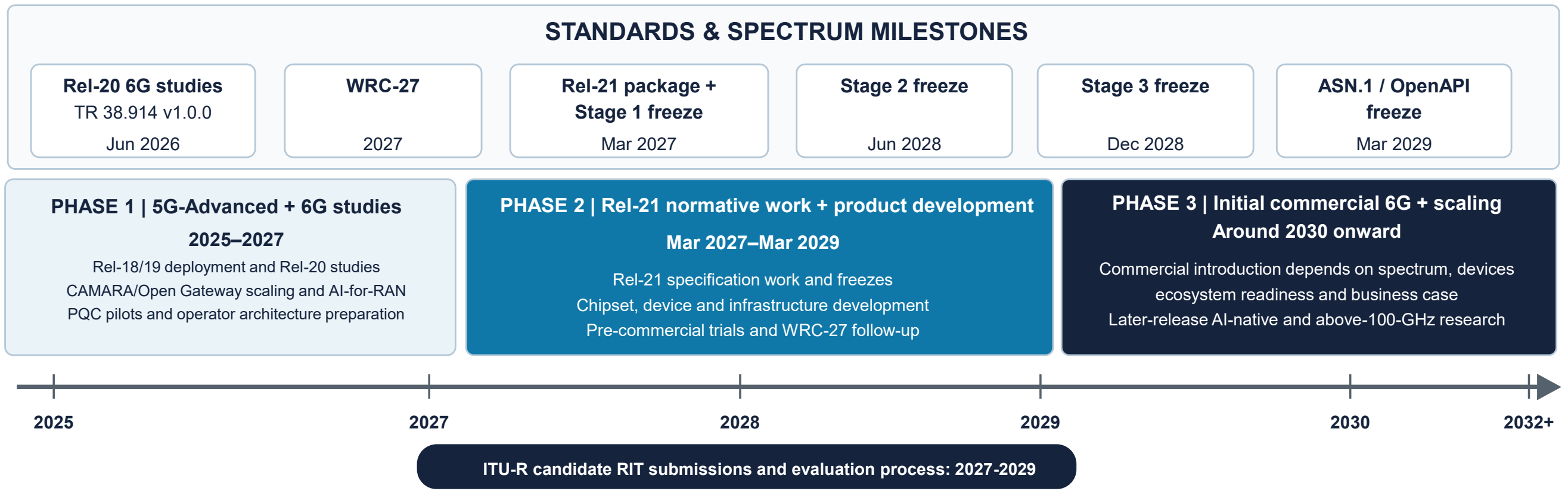

**FIGURE 20.** **6G standards, product, and commercial roadmap.**

bounded AI-for-RAN energy optimization, initial CAMARA APIs, governed agentic-operation trials, and post-quantum pilots [64], [106], [69], [71], [72], [107], [63], [108]. Procurement should use component-appropriate conformance and interoperability evidence. Investment precedes revenue and should be released against explicit deployment gates.

*Phase 2* (illustrative 2027–2029): specification and product development proceed alongside pre-commercial trials, WRC-27 follow-up, and ITU-R evaluation. Guarantee Economy pilots can use mature 5G-Advanced independently of 6G timing; delayed standards defer 6G-dependent offers.

*Phase 3* (around 2030 onward): commercial 6G scales only where spectrum, devices, interoperability, and demand are ready. AI-RAN, governed autonomy, dApps, expanded APIs, and higher-frequency operation remain gated by evidence and later-release maturity.

Progression between deployment phases is gated, not assumed: it requires spectrum and regulatory availability; interoperable chipset and UE evidence within stated power and thermal budgets; measured network-energy and model-assurance performance; and signed enterprise demand with an independently verifiable SLO. Failure of any gate triggers a 5G-Advanced-only commercialization path and defers the 6G-dependent offer.

Standards dependency against operator control. New spectrum, 6G radio/core specifications, agent identity and regulatory enablers, E3/dApp conformance, and service-level energy attribution depend on external decisions. Data and AI ownership, procurement rights, MCP-mediated tool governance, bounded automation, and measurable commercial offers remain under operator control. Table 9 separates these tracks by phase.

### *B. Implications for Stakeholders*

The operator-controlled 6G model has differentiated implications for nine stakeholder groups. The discussion below frames them as analytical implications of the architecture, economics, and standardization trajectory developed in Sections III–X.

Stakeholder implications follow directly from the control and evidence boundaries. Operators should stage ownership and automation by unit economics; vendors should compete on interoperable and auditable interfaces; hyperscalers remain partners where data and control terms are explicit; regulators should assess scenario-specific execution authority and oversight; and SDOs should advance interoperable management and exposure alongside radio specifications. Buyers need independent measurement and enforceable remedies, device suppliers need feasible power and certification budgets, academia should prioritize field-grounded safety and interoperability [109], [110], and investors should release capital against measured operational and customer-demand gates.

TABLE 9. ROADMAP DEPENDENCIES: EXTERNALLY GATED AGAINST OPERATOR-CONTROLLED ITEMS

| Deployment phase | Externally gated: SDO, spectrum, regulation | Operator-controlled: can proceed independently |
|---|---|---|
| Phase 1 (2025–2027) | Release-20 6G study conclusions; ITU-R IMT-2030 evaluation work; WRC-27 preparatory studies; national and regional coexistence studies; O-RAN, ETSI and CAMARA specification evolution. | Deploy AI-for-RAN and energy automation; build the governed RAN–core–cloud data layer; deploy eBPF observability; prototype MCP adapters over existing SBI, O-RAN and OAM interfaces; scale Open Gateway APIs; run post-quantum and assurance pilots. |
| Phase 2 (2027–2029; specification and product-development window) | Release-21 normative feature scope and freeze dates as finally agreed by 3GPP; WRC-27 outcomes; the corresponding band, RF and conformance work; chipset and device profiles tied to stable specification baselines; the IMT-2030 candidate-RIT process. | Commercialize outcome and SLA products on mature 5G-Advanced; run enterprise, fixed-wireless and industrial pilots; harden MCP governance, identity and rollback; validate digital-twin safety, service assurance and cross-domain policy arbitration. |
| Phase 3 (around 2030 onward) | Commercial 6G depends on mature normative specifications, assigned spectrum, interoperable devices and chipsets, and national regulation. Later waveform and AI-native physical-layer features remain feature-specific gates. | Scale only the 6G features that pass measured coverage, capacity, power, total-cost and demand gates. If an external dependency slips, preserve the 5G-Advanced commercial path and defer the affected 6G-specific feature rather than the whole operator transformation. |

TABLE 10. TRANSFERABILITY BY OPERATOR ARCHETYPE

| Operator archetype | Applicable Control Compact stance | Principal structural barrier | Dominant transformation cost driver | Which claims transfer |
|---|---|---|---|---|
| Greenfield cloud-native | Own the control plane, AI substrate and data layer from the outset. | Coverage and scale economics before the subscriber base matures. | Platform and talent build-out. | The evidenced claims; the platform, tiers and willingness to pay remain unvalidated here too. |
| Brownfield incumbent | Own the control plane for new deployments; federate through migration; consume where the legacy estate cannot be opened. | Legacy network functions and vendor contracts that do not expose the interfaces or data the Compact requires, and an organization structured around appliance operations. | Parallel running and migration sequencing rather than new build; write-down of installed assets. | Conditionally. The ownership principles and the governance obligations transfer. The transformation economics and the achievable autonomy timeline do not, and are the subject of the investigation noted in Section XII. |
| Low-ARPU market | Own the data layer and the automation that reduces operating cost; consume more of the platform. | Absolute capital availability, and a small addressable base for premium-priced guarantees. | Capital intensity relative to revenue per subscriber; the assurance cost in (8) does not scale down with ARPU. | The operations and automation argument transfers and may matter more. The premium pricing of the Guarantee Economy largely does not, since (10) requires a premium the market may not support. |
| MVNO and wholesale | Own the customer relationship, identity and commercial layer; consume radio and core. | No control of the enforcement points on which the guaranteed outcomes depend, and measurement boundaries that sit inside the host network. | Contractual and integration cost of obtaining measurement and remedy rights from the host. | The outcome-definition discipline transfers. Ownership of the control plane does not, and guarantees can be resold only to the extent the host contractually guarantees them. |

### C. Scope, Limitations, and Transferability

The evidence comes predominantly from a greenfield, cloud-native operator with an integrated digital ecosystem. That context improves the feasibility of disaggregation, in-house platform development, and ecosystem monetization, so transformation economics and deployment pace cannot be generalized directly. More transferable principles are ownership allocation by layer, measurable service boundaries, and the identity, validation, audit, and rollback obligations of governed agents. The complete platform, its performance and security, and customer willingness to pay remain unvalidated. Table 10 summarizes transferability by operator archetype.

Multi-provider guarantee chains require an agreed measurement boundary, custody of the obligation, breach-attribution evidence, and remedy alignment. Roaming, MVNO, and cloud-integrated services place the technical and contractual boundaries in different domains. Because current agreements do not consistently exchange auditable per-session evidence, the six-tier catalog is scoped initially to a single operator.

## XII. OPEN RESEARCH QUESTIONS

Fourteen open questions will determine the pace of the transition.

(1) *Brownfield transformation economics*: whether cloud-native infrastructure produces full-lifecycle savings during legacy migration. A New Energy and Industrial Technology Development Organization (NEDO) project awarded to KDDI and Rakuten Mobile targets joint optimization of virtualized base stations and computing infrastructure, with an objective of approximately 40% reduction in datacenter and RAN power consumption by fiscal 2030 [111]. The target is not a measured result; transferability depends on disclosed baselines and costs.

(2) *Formal verification for LLM agents in safety-critical loops*: no accepted method links agent reasoning quality, calibrated action-conditional prediction, and SLA impact at operator scale. Network predictors and world models quantify selected errors [112], [113], [81], while safety shields, configuration verification, and predictive certification provide partial precedents [114], [115], [116], [117], [82], [118]. Conformal methods add coverage tools but face selection and distribution-shift limits in closed loops [117], [119], [120]. The Section VII gate therefore requires testing of escalation rate, conditional coverage, OOD recall, attack blocking, drift latency, conflict handling, and recovery—not an assumed safety guarantee.

(3) *Guarantee Economy scale conditions*: whether jurisdiction-level legal infrastructure for SLA enforcement, standardized SLA term frameworks, and independent performance verification can be established within the Phase 2 window.

(4) *Multi-vendor training compatibility for Stage 3 air interfaces*: jointly learned transmitter-receiver systems require shared training or standardized learned-component representations across competing chipset vendors – no 3GPP precedent exists, and the answer determines whether Stage 3 is a platform technology or a vertically integrated proprietary capability; the UE side additionally lacks committed neural-accelerator, model-update, and certification roadmaps.

(5) *Multi-agent governance at network scale*: emergent behavior grows non-linearly with coordinating agents (federated RICs, cross-vendor dApps, enterprise-operator agent interaction); formal methods for specifying and verifying coordination protocols under the ZSM 017 [31] threat model and the adversarial taxonomy [35] remain insufficient.

(6) *Post-quantum migration sequencing*: hybrid AKA variants are quantified [32], but no consensus cut-over criteria exist – joint 3GPP SA3/NIST/operator work is required.

(7) *Adversarial robustness benchmarks*: the field lacks a MITRE ATT&CK-equivalent for AI-native RAN – a versioned catalog of production-observed attack techniques with mapped defenses – without which procurement cannot demand evidence-based robustness; relatedly, federated explainability that preserves audit-grade rationale [36] across operator and national boundaries without leaking training data remains unsolved.

(8) *Service-level energy attribution*: MetricEES [54], ITU-T L.1333, ATIS TEER, ISO/IEC 30134-2, and ETSI ES 203 228 cover equipment, network, and facility levels, but no standard ties energy to delivered business outcomes. Without such a standard, sustainability cannot be priced as a first-class commercial dimension, and the Smart Networks and Services Joint Undertaking (SNS JU) eHealth result [68] cannot be replicated systematically.

(9) *Waveform selection under mixed mobility*: define the Doppler, delay, carrier, bandwidth, mobility, MIMO, and phase-noise envelope in which the CP-OFDM/DFT-s-OFDM baseline needs enhanced tracking, numerology, or receiver processing [100], [101].

(10) *Pilot and guard design and resulting overhead*: pilot and guard design for delay–Doppler and affine waveforms at the frame sizes and latency budgets the Table 2 outcomes require, and the resulting overhead relative to CP-OFDM at equal spectral efficiency – the quantity that decides whether robustness in doubly selective channels is affordable in a commercial deployment rather than a link study.

(11) *Low-PAPR uplink for coverage-limited 6G bands*: quantify cell-edge gain after power-amplifier backoff, error vector magnitude (EVM), spectral-mask, device-power, receiver-loss, and implementation-complexity effects, rather than judging FDSS from one link-level metric.

(12) *Joint waveform and sensing design*: whether a waveform selected for range–Doppler sensing performance remains compatible with the communication service-level objectives contracted on the same carrier, and how the two objectives are arbitrated – the standardization gap identified in Sections IX.F and VII.

(13) *A common evaluation methodology for 6G radio physical-layer enhancements*: which reference-signal, channel-estimation, equalizer, Doppler, phase-noise, power-amplifier, mobility, MIMO and ISAC assumptions must be fixed so that competing proposals are compared against the same CP-OFDM/DFT-s-OFDM baseline rather than against each other's simulation set-ups.

(14) *Interoperability and fallback for AI-assisted receivers*: which items require standardization for multi-vendor operation – capability exchange, model input and output semantics where exposed, performance monitoring, lifecycle control, fallback triggers and test vectors – while leaving the internal receiver algorithm and training method implementation-specific.

## XIII. CONCLUSION

This article proposes an operator-governed approach to 6G architecture that connects software and data control with verifiable service outcomes. Its commercial implications remain hypotheses to be tested through operational measurements and customer-facing pilots.

The Control Compact gives architectural and procurement decisions a strategic rationale; the six-tier Guarantee Economy translates control into contractible service products; and the Network MCP Platform remains a key conceptual contribution connecting agent-facing tools, operator controls, and telecom execution interfaces. The contribution is this integration and governance framework, not invention of MCP or the other building blocks.

The operational substrate has complementary evidence: the scoped external autonomy assessment discussed in Section III.C, a nationwide RIC deployment [5], and LLM-assisted infrastructure optimization [8]. Both real-network examples from Rakuten Mobile illustrate agentic workflows using MCP tools. RAN second-line (L2) support combines diagnosis, selected recovery, and outcome verification; IMS availability investigation combines evidence collection, adaptive investigation, and reporting. These observations illustrate MCP-enabled agentic operations. They do not validate the complete Network MCP Platform, its security or performance guarantees, or the commercial viability of the six-tier catalog.

A practical next step is to evaluate a bounded operational domain against the Control Compact, measure an MCP-enabled agentic workflow against a matched direct-tool baseline, and test a service offer with explicit SLOs and remedies. Such evaluations can proceed with existing infrastructure. They would establish where the proposed integration yields operational or commercial value and identify the conditions required for broader deployment.

## AI USE DECLARATION

The authors used Claude (Anthropic) and Codex (OpenAI) for editing and proofreading, literature-search assistance, and document review. The tools also produced first drafts of the figure and table captions and of the workflow descriptions in Section VII.D, which the authors revised and verified against the underlying material. Every reference was checked by the authors against its original source. No section of this article contains AI-generated scientific content, and all content, claims, analysis, and conclusions are the sole responsibility of the named authors.

## APPENDIX A. CAMARA/OPEN GATEWAY 6G API CANDIDATES

Table 11 summarizes the CAMARA and Open Gateway API candidates for 6G.

## APPENDIX B. CLAIM PROVENANCE

Table 12 classifies quantitative claims into one of four evidence classes – standard requirement, literature result, production measurement, and author design target – and states, for each, the source, the measurement boundary, and the validation status. Unstated boundaries are not inferred, and production measurements are not described as independently verified unless the source establishes that status. Withdrawn claims are retained in the table so that the withdrawal is explicit rather than silent. Two kinds of claim map onto these four classes only approximately, and the mapping is disclosed here rather than left to inference: analytical results derived in Appendix C under stated assumptions are carried as author design targets, since a derivation is not a measured result; and an external organization's published objective or trajectory is carried as a standard requirement, since it is normative for the body that set it rather than a requirement on this operator. In both cases the measurement boundary states the distinction. The table classifies the values for which these four classes are material; counts internal to the article's own structure, such as the number of contributions or service tiers, are deliberately excluded.

TABLE 11. CAMARA / OPEN GATEWAY 6G API CANDIDATES

| API family | Primary use case | Exposed objective / SLO | TR 22.870 family | Maturity / status |
|---|---|---|---|---|
| Quality on Demand (QoD) | Guaranteed bandwidth/ latency for enterprise apps | Guaranteed bitrate; latency profile; priority class | URLLC+; eMBB+ | Mature (CAMARA GA) |
| Location Verification / Location Retrieval / Geofencing Subscriptions | Asset tracking, geofencing, logistics | Verify/retrieve device location; area subscription | Positioning; location | Mixed: Location Verification stable; Location Retrieval initial; check each API at submission |
| Number Verification / SIM Swap | Authentication; fraud prevention | Real-time verification; response SLA to be contractually defined | Sovereign Trust | Mature (CAMARA GA) |
| Device Reachability Status / Device Roaming Status | IoT reachability and roaming state | Status update latency; reachability accuracy | Massive IoT | Mature (CAMARA GA) |
| Simple Edge Discovery | Discover nearest edge for app deployment | Nearest MEC endpoint resolution | Edge / MEC | Mature (CAMARA GA) |
| Model As A Service / Optimal Edge Discovery | Edge AI-service management and edge selection | Deployment-specific latency, throughput, and availability objectives | AI/ML as a service; AI NaaS | Initial (CAMARA; sandbox; status checked Jul. 2026) |
| Network Slice Booking / Network Slice Assignment | Enterprise slice reservation and device assignment | Reservation and assignment requests; performance targets are deployment-specific | NaaS; determinism | Initial (CAMARA; sandbox; status checked Jul. 2026) |
| ISAC Sensing exposure | Environmental sensing, occupancy, velocity | Sensing objective, authorization, and assurance (not internal radio timing) | ISAC | Author-proposed; not a current CAMARA API |
| AI-RAN Exposure | Third-party real-time RAN control | Objective, authorization, policy, status, assurance (not internal E3/xApp/dApp timing) | Programmable RAN | Author-proposed; not a current CAMARA API |
| Energy Footprint Notification | Report E2E energy consumption and carbon footprint for an application service | Declared-boundary energy and carbon information; no universal per-session attribution standard | Sustainability | Initial (CAMARA; sandbox; status checked Jul. 2026) |
| Sovereign Attestation | Verifiable data residency and jurisdiction | Data-residency assertion and jurisdictional evidence; cryptographic proof | Sovereign Trust | Author-proposed; not a current CAMARA API |

## APPENDIX C. ANALYTICAL MODELS

This appendix derives four results used in the body: the integration and trust-relationship savings of router mediation (A), the enforcement-loop latency budget (B), router-tier availability under $m + k$ redundancy (C), and the break-even economics of a guarantee tier (D). Each derivation states its assumptions and limits; none is a measurement.

### *A. Integration and Trust-Relationship Complexity*

Let $N$ be the number of agent-side consumers – agents or workflows requiring access to network operations – and $M$ the number of domain-side tools and interfaces they must reach. Under direct bilateral integration each consumer requires its own adapter to each tool, so the number of integrations $I$ is:

$$I_{\text{direct}} = NM \tag{1}$$

Under router mediation each consumer integrates once with the router and each tool is adapted once behind it, so

$$I_{\text{router}} = N + M + C_r \tag{2}$$

where $C_r$ is the router's own integration and operational cost expressed in adapter-equivalents. Mediation is advantageous while $NM > N + M + C_r$. For fixed $C_r$ and $N = M$, the ratio of the integration counts (1) and (2) approaches $N/2$, so the benefit grows linearly in the number of consumers rather than being a fixed saving. The defensible result is marginal: with $M$ and $C_r$ held fixed, onboarding one further agent class costs $M$ adapters under direct integration but one under mediation, so the router reduces the marginal onboarding cost from $O(M)$ to $O(1)$. It centralizes adapter work rather than eliminating it, and the unconditional total-effort reduction remains unestablished.

*Trust relationships scale separately*. For $K$ principals that must mutually authenticate without a common trust domain, the number of pairwise relationships is:

$$T_{\text{pair}} = K(K-1)/2, \qquad T_{\text{hier}} = K \tag{3}$$

where $T_{\text{hier}}$ is the count when a hierarchical trust domain issues attested workload identities and each principal establishes trust only with the domain authority. This is the precise sense in which trust relationships fall from $O(K^2)$ to $O(K)$.

*Assumptions and limits*. (a) Every consumer must reach every tool; where the reachability matrix has density $\rho < 1$, (1) becomes $\rho NM$ and the advantage scales with $\rho$. (b) The router is a single logical trust and policy point; federating several routers reintroduces pairwise relationships between them. (c) Adapter effort per tool is independent of the number of consumers. (d) These expressions count integration and trust relationships. They are not a performance result; the extra hop that mediation adds is accounted for in (4).

### B. Enforcement-Loop Latency Budget

Decompose the closed loop into named components: detection $T_{det}$, classification $T_{cls}$, policy decision $T_{dec}$, model reasoning $T_{LLM}$ where an agent is engaged, distributed execution $T_{exe}$, and verification $T_{ver}$. Then

$$T_{enf} = T_{det} + T_{cls} + (T_{dec} + T_{LLM}) + T_{exe} + T_{ver} \quad (4)$$

Execution across $D$ domains behaves differently according to whether commits proceed in parallel or must be ordered:

$$T_{exe} = \max_d (T_{prop,d} + T_{commit,d}) \text{ or } \Sigma_{d=1}^{D} (T_{prop,d} + T_{commit,d}) \quad (5)$$

where $T_{prop,d}$ and $T_{commit,d}$ are the propagation and commit times in domain $d$.

The consequence is the useful one. Under ordering constraints, the execution term grows linearly in the number of domains, so for multi-domain enforcement dependency ordering, not per-call latency, can come to dominate the budget. A sub-second containment claim therefore requires a bound on each term of (4) and an explicit statement of which branch of (5) applies; a single end-to-end figure does not establish it, and neither does a fast detection term. Table 13 lists what bounds each component and where it must be measured. No values are proposed here, because none has been measured.

### C. Router Availability Under m + k Redundancy

Let a single router instance have availability $p$ = MTBF/(MTBF+MTTR), where MTBF is the mean time between failures and MTTR the mean time to repair, and let $n = m + k$ instances be deployed, $k$ of them spare, and let the service require at least $m$ of them. If failures are independent, the availability of the router tier is:

$$A_{tier} = \Sigma_{j=m}^{n} \, C(n, j)\, p^j\, (1 - p)^{n-j} \quad (6)$$

where $C(n, j) = n!/[j!(n - j)!]$ is the binomial coefficient, so each term is the probability that exactly $j$ of the $n$ instances are available. The tier does not stand alone. With a shared policy service of availability $q$ and a shared control-plane network of availability $s$ in series,

$$A_{eff} = qsA_{tier} \quad \text{and} \quad A_{path} = 1 - (1 - A_{eff})(1 - b) \quad (7)$$

where $b$ is the availability of the SBI-only bypass path retained as a degraded mode, and $A_{path}$ is the probability that some authorized execution path exists. Both product forms in (7) assume that the policy service, the control-plane network, the router tier and the bypass path fail independently; a shared dependency requires the joint probability instead.

Two results follow, and both matter more than any number. First, $A_{eff} \leq \min(q, s)$ irrespective of $n$: adding router replicas cannot raise availability above that of the least available shared dependency, so redundancy in the mediation tier is not a substitute for redundancy in the policy service. Second, the independence assumption behind (6) is the one most likely to fail in practice – correlated failure through a shared control plane, a shared configuration push, or a common software defect – so a correlated-failure factor has to be measured rather than assumed, and (6) should be read as an optimistic baseline rather than as a bound.

### D. Guarantee-Tier Expected Value

Per served unit over one observation window, let $r$ be the premium over the best-efforts baseline, $c$ the incremental cost of delivering the guarantee, $\beta$ the probability of breach within the window, $\lambda$ the remedy expressed as a multiple of $r$, $a$ the attach rate, and $B$ the addressable base.

Let $\pi$ denote the contribution of one served unit over the window (premium net of the remedy paid on breach and of the incremental cost), and $\Pi$ the expected contribution across the portfolio. Then:

$$E[\pi] = r(1 - \beta\lambda) - c \quad (8)$$

$$\Pi = aB\,[\, r(1 - \beta\lambda) - c \,] \quad (9)$$

Setting (8) to zero gives the break-even breach probability and the break-even premium:

$$\beta^* = (1 - c/r)/\lambda\,, \qquad r^* = c/(1 - \beta\lambda) \quad (10)$$

with sensitivities $\partial E[\pi]/\partial\beta = -r\lambda$ and $\partial E[\pi]/\partial\lambda = -r\beta$, valid for $r > 0$, $\lambda > 0$ and $\beta\lambda < 1$. The tier is viable only while $\beta < \beta^*$. Because $\beta^*$ falls as $c/r$ rises, a tier whose assurance cost approaches its premium has almost no tolerance for breach; and a remedy denominated as a multiple of the premium, $\lambda > 1$, reduces that tolerance proportionally. This is the structural reason the highest-determinism tiers are the hardest to price, not merely the hardest to deliver.

*Worked example.* Take two tiers over a one-month observation window, with every parameter illustrative and none measured. An assured-throughput tier prices a premium $r$ = 100 currency units per served unit, incurs $c$ = 60 in incremental assurance cost, and denominates its remedy at the full premium, $\lambda = 1$; a deterministic-latency tier prices the same premium at $c = 90$ with a three-times-premium remedy ladder, $\lambda = 3$. By (10) their break-even breach probabilities differ by more than an order of magnitude – $\beta^* = 40\%$ against $\beta^* = 3.33\%$ – so at an assumed $\beta = 2\%$ the first tier returns $E[\pi] = 38$ per unit and the second only 4, and by (9) a 5% attach rate over a base of $B$ = 10,000 units yields 19,000 against 2,000 per window. The sensitivity $\partial E[\pi]/\partial\beta = -r\lambda$ makes the asymmetry concrete: one additional percentage point of breach probability costs the first tier 1 unit of margin and the second 3, removing three-quarters of what it had. The example fixes no tariff and asserts no demand; it shows that, at equal premiums, breach tolerance is set jointly by the cost-to-premium ratio $c/r$ and the remedy multiple $\lambda$ – here a factor of four and a factor of three in the twelvefold gap, and that $\beta$ and $a$ are the two quantities a pilot must establish before either tier is offered.

*Limits.* The model prices the contractual remedy only. It does not price the reputational or contractual consequence of repeated breach, nor the option value of the guarantee to the buyer. $\beta$ must be estimated per service-level objective from measured distributions, which this article does not supply; and $a$ has no evidential basis here at all. The model therefore bounds the commercial question and identifies which single quantity – the attach rate at a disclosed premium and remedy – a pilot would have to establish first.

## APPENDIX D. COMPARISON WITH ADJACENT ARCHITECTURES AND LITERATURES

Table 14 locates the contribution relative to adjacent frameworks and literature; it does not claim measured superiority.

## APPENDIX E. COST OF OWNERSHIP UNDER THE CONTROL COMPACT

Table 15 makes the Control Compact a per-component, lifecycle decision by pairing each stance with its costs and exit conditions.

## APPENDIX F. THREAT ANALYSIS AND THE SAFETY-GATE PARAMETER CONTRACT

Table 16 maps agentic-operation threats to proposed controls, signals, recovery, residual risk, and validation status. All remain unevaluated; execution still depends on the independent validation layer rather than model correctness.

Table 17 defines which gate parameters must be declared and measured before autonomous execution. Unknown or exceeded values close the gate; no unmeasured telecom-loop parameter is presented as validated.

## APPENDIX G. CANDIDATE 6G WAVEFORMS AND SIGNAL-PROCESSING ARCHITECTURES

Table 18 compares candidate waveforms and receivers from an operator decision perspective. It reports no performance ranking because results depend on channel, frame, and receiver assumptions [100], [101].

TABLE 12. PROVENANCE OF QUANTITATIVE CLAIMS

| Claim | Value as stated | Class | Source | Measurement boundary | Validation status |
|---|---|---|---|---|---|
| Unique mobile subscribers | > 5 billion | Literature result | Industry report [1] | Global, as published | Published |
| Immersive user-plane latency | ≤ 4 ms | Standard requirement | ITU-R TPR [4] | Radio interface, TPR evaluation configuration | Draft normative requirement |
| Immersive user-experienced data rate | 5th-percentile DL 300 / UL 50 Mbit/s (Dense Urban) | Standard requirement | ITU-R TPR [4] | Dense Urban test environment, 5th percentile | Draft normative requirement |
| Immersive peak data rate | DL 36 / UL 18 Gbit/s | Standard requirement | ITU-R TPR [4] | Peak spectral efficiency at maximum bandwidth | Draft normative requirement |
| HRLLC user-plane latency | ≤ 1 ms | Standard requirement | ITU-R TPR [4] | Radio interface | Draft normative requirement |
| HRLLC reliability | $1 - 10^{-5}$ | Standard requirement | ITU-R TPR [4] | 32-byte L2 PDU at coverage edge | Draft normative requirement |
| Jitter for critical and enterprise tiers | < 1 µs | Author design target | Author target, vertical-sourced; no TPR figure exists | Packet-delay variation; boundary not defined in any standard | Unvalidated author target |
| Connection density | $10^6$ devices/km² | Standard requirement | ITU-R TPR [4] | Per km², TPR evaluation configuration | Draft normative requirement |
| Positioning accuracy | ≤ 0.75 m indoor / ≤ 6 m urban macro | Standard requirement | ITU-R TPR [4] | TPR positioning test environments | Draft normative requirement |
| Sensing detection probability | 95% at 5% false alarm | Standard requirement | ITU-R TPR [4] | TPR sensing evaluation configuration | Draft normative requirement |
| Sensing update interval | < 1 s | Author design target | Author and vertical target | Per sensing objective; not a TPR figure | Unvalidated author target |
| Edge inference latency | < 10 ms | Author design target | Author target informed by AI-RAN WG3 deployment profiles [13] | Device to RAN-local (dApp-tier) inference; edge-cloud paths excluded | Unvalidated author target |
| AI-native uplink traffic share | ~ 26% | Literature result | Industry forecast [53] | Forecast basis as published | Published forecast, not a measurement |
| Central-cloud round-trip penalty | 30–80 ms (estimated) | Author design target | Author's estimate against the AI-RAN WG3 10 ms profile [13] | Central round trip versus edge serving; path not specified | Unvalidated estimate |
| dApp E3 control-loop budget | < 10 ms (target) | Literature result | O-RAN nGRG research report [19] – contributed research, not normative | E3 control loop on DU/CU infrastructure | Prototype-demonstrated in [19]; not a normative requirement |
| Near-RT RIC control loop | 10 ms – 1 s | Standard requirement | O-RAN architecture description [98] | E2 control loop; specified as an order-of-magnitude range and use-case dependent, not a hard bound | Normative O-RAN architecture description (ETSI PAS) |
| Slice provisioning time | < 5 min | Author design target | Operator product target | End-to-end provisioning: boundary not defined | Unvalidated commercial target |
| Slice integrity | ≥ 99.9% | Author design target | Author commercial target | Observation window not defined | Unvalidated commercial target |
| Premium consumer downlink rate | ≥ 500 Mbit/s | Author design target | Author commercial target | Per subscriber, radio conditions not specified | Unvalidated commercial target |

| Claim | Value as stated | Class | Source | Measurement boundary | Validation status |
|---|---|---|---|---|---|
| Premium consumer end-to-end latency and availability | ≤ 20 ms; 99.99% | Author design target | Author commercial target against an application budget [97] | Application end to end; observation window not defined | Unvalidated commercial target |
| Enterprise determinism reliability | to 99.9999% (≈ $1 - 10^{-6}$) | Author design target | Author commercial target, above the TPR $1 - 10^{-5}$ baseline [4] | Device to edge | Unvalidated; exceeds the TPR baseline |
| AI inference model availability | 99.99% | Author design target | Author commercial target | Model-serving endpoint; window not defined | Unvalidated commercial target |
| FY2025 commercial results | Revenue JPY 374.7 bn (+32.0% YoY); net ARPU JPY 2,467; > 10 M subscribers | Production measurement | Operator corporate disclosure [5] | Consolidated segment reporting, FY2025 | Audited corporate disclosure; commercial context only, not architectural proof |
| Rakuten Ichiba GMS uplift for mobile subscribers | ~ 48.8% higher annual average GMS | Production measurement | Operator corporate disclosure [5] | Annual average Rakuten Ichiba GMS, Rakuten Mobile subscribers versus non-subscribers, FY2025; an association, not a measured causal effect of connectivity | Audited corporate disclosure; ecosystem commercial context only, not evidence of demand for the proposed service tiers |
| Stated willingness to pay for guaranteed reliability | ~ half of surveyed 5G users | Literature result | Alliance member research [13] | Survey-stated willingness as published; not an independent demand study | Published industry research; not evidence that a buyer accepts a disclosed premium or an enforceable SLA (Section V.C) |
| Kubernetes resource-optimization saving | 61% average across 68 microservices in 5 namespaces (21–88% per namespace) | Production measurement | Peer-reviewed study [8] | Named namespaces; potential cost reduction under human approval | Peer-reviewed; reported as potential and human-in-the-loop |
| RIC network power reduction | 15–20% | Production measurement | Operator-reported estimate; no published locator | Not stated in the source; baseline, period and network boundary undeclared | Not independently verified; not additive with the row below |
| RAN energy conservation | ~ 20% (estimate) | Production measurement | Operator announcement [49] | Not stated – baseline kWh, observation period, site and cell population, traffic normalization and service-quality constraints all outstanding | Operator-reported estimate, not an independently measured benefit |
| Autonomous-network autonomy assessment | Level 4, score 4.0 | Production measurement | TM Forum ANLAV report [47]; public register [50] | GB1059H; intra-RAT intra-frequency and inter-RAT multi-site in scope, intra-RAT inter-frequency out of scope | External assessment of method and supporting evidence, not of network performance |
| Open RAN automation footprint | ~ 150,000 cells | Production measurement | Operator operational record | Not stated; cell definition, period and inclusion criteria undeclared | Appears in an author biography only; not yet carried as an evidenced claim in the body |
| Post-5G energy-reduction objective | ~ 40% by FY2030 | Standard requirement | Project objective [111] | Project objective, not a result; baseline not disclosed | Forward-looking target of a research program |
| MCP-mediated router overhead | single-digit milliseconds | Author design target | Author design target | MCP-mediated versus direct tool invocation on a matched network path | Explicitly unverified (Section VII.E) |
| Containment time | sub-second | Author design target | Author design target | Detection, classification, policy decision and distributed enforcement combined | Not measured; decomposed analytically in Appendix C, (4)–(5) |
| Kernel-plane detection latency | sub-millisecond | Author design target | Author design target | eBPF syscall-trace observation only; excludes classification and enforcement | Not measured (Section VII.F) |
| Trust-relationship scaling | $K(K - 1)/2 \rightarrow K$, under one hierarchical trust domain | Author design target | Derived in Appendix C, (3), under stated assumptions | Counts trust relationships between principals, not runtime message complexity | Derivation under stated assumptions; not measured |
| Integration count under router mediation | $NM \rightarrow N + M + C_r$, advantageous while $NM > N + M + C_r$ | Author design target | Derived in Appendix C, (1)–(2), under stated assumptions | Counts integrations, not runtime message complexity or total effort | Derivation under stated assumptions; not measured. The unconditional $O(N^2) \rightarrow O(N)$ total-effort form is not established (Section VII.A) |
| LLM engagement in production closed loops | withdrawn | Author design target | Previously an author assumption with no stated analytical basis | Would require model-call rate per completed task | Withdrawn pending measurement (Section VII.E) |
| eBPF telemetry overhead reduction | withdrawn | Author design target | Previously cited to a source that does not benchmark the configuration claimed | Would require a same-workload comparison against a user-space sidecar baseline | Withdrawn by the authors pending a same-workload comparison |
| Sector decarbonization trajectory | − 45% (2020 → 2030) | Standard requirement | ITU-T L.1470 [104] | Sector-level trajectory, not an operator target | Normative sector trajectory |
| Candidate spectrum under study (Japan) | 7.125–8.4 GHz; 14.8–15.35 GHz | Standard requirement | National administration study material [40] | Study status only; no allocation decision | Under study; no allocation decision |
| Latent predictive models on network measurements | qualitative | Literature result | [81], [82] | Configuration-effect forecasting on one operator's data; channel dynamics from CSI | Neither work gates actions |
| Latent screening model used as a safety gate | no value asserted | Author design target | [80], [83] conceptual only | Escalation fraction, coverage on the admitted subset and OOD recall are all unmeasured | Not evaluated; gating exists in four other forms ([114]–[116], [118]), none on calibrated predicted performance |
| Joint-embedding architectures for control over capacity-limited networks | qualitative | Literature result | [113] | Models the controlled plant, the network being the capacity-limited channel | Cited only for joint-embedding prediction in networked control |

| Claim | Value as stated | Class | Source | Measurement boundary | Validation status |
|---|---|---|---|---|---|
| Gate parameter definitions and admission conditions | see Table 17 | Author design target | Values set at commissioning against the Section VII.E instrumentation | Each row of Table 17 gives the condition, its basis and its escalation trigger | Three quantities carry values with a stated basis; $q$, $t_{eff}$ and $H_{cal}$ are named as unmeasured |
| Learned network model accuracy | 6.24% mean relative delay error | Literature result | [112] | Mean relative error on delay only | Published result; no tail-quantile or generalization claim is made from it |
| Digital twin synchronization requirements | no numeric bound | Standard requirement | ITU-T Y.3090 [79] | Requirements and architecture only | Sets no numeric divergence bound |
| Threat controls and residual risks (Table 16) | no value asserted | Author design target | This article | Injection-corpus block rate, induced-drift detection latency and concurrent-intent conflict rates | Not evaluated; no adversarial evaluation performed |
| Service-level carbon per clinical episode | 40–60% $CO_2$ per episode | Literature result | SNS JU eHealth study [68] | Modelled projection per clinical episode; not a measured operational result and not a network or operator KPI | Published modelled projection; not independently verified |
| 6G synchronization signal block size | 20 resource blocks per SSB; 12 for the PSS | Standard requirement | 3GPP RAN1 study agreement [94] | RAN1 working agreement recorded in a TR under study; not a complete normative specification | Under study; sequence type and length, symbol count and periodicity still open |
| Throughput/scalability | no value asserted | Author design target | Section VII.E evaluation design | MCP-mediated execution versus direct invocation of the same tool on a matched network path, payload, authentication and logging | Not evaluated; sample sizes and predeclared acceptance criteria not yet set |
| Failure-recovery success and duration | no value asserted | Author design target | Section VII.E failure and recovery scenarios | Detection, reconciliation of in-flight operations and steady-state restoration, per scenario | Not evaluated; awaiting fault-injection tests. One RAN recovery and one IMS access failure are illustrative records only |
| Agent decision quality | no value asserted | Author design target | Section VII.E evaluation design | Decision outcomes against expert-labelled outcomes; no labelled decision set exists yet | Not evaluated; requires expert labelling and human-in-the-loop adjudication |

TABLE 13. ENFORCEMENT-LOOP COMPONENTS AND MEASUREMENT POINTS

| Component in (4) | What bounds it | Where it is measured | Status |
|---|---|---|---|
| Detection, $T_{det}$ | Telemetry path and sampling interval; kernel-plane observation where available | From event occurrence in the network to events available to the classifier | Not measured |
| Classification, $T_{cls}$ | Analytic or model inference cost; feature availability | Classifier input to labelled event | Not measured |
| Policy decision, $T_{dec}$ | Policy evaluation and authorization round trip | Labelled event to authorized action | Not measured |
| Model reasoning, $T_{LLM}$ | Model tier, context size, retries; zero when the loop is deterministic | Agent request to agent decision | Not measured; engagement rate withdrawn |
| Distributed execution, $T_{exe}$ | Per-domain propagation and commit; dependency ordering across domains, per (5) | Authorized action to committed state in each domain | Not measured |
| Verification, $T_{ver}$ | Observation interval needed to confirm the intended outcome | Committed state to verified outcome | Not measured |

TABLE 14. POSITIONING AGAINST ADJACENT ARCHITECTURES AND LITERATURES

| Approach or literature | What it provides | Evidence status | What this article adds | Ref. |
|---|---|---|---|---|
| ETSI ZSM closed-loop framework | Closed-loop automation framework, management-domain decomposition, end-to-end service management | Normative group specifications; a framework, not deployment evidence | Allocates ownership of each closed-loop element through the Control Compact, and binds loops to contractible outcomes with defined remedies | [22]–[24] |
| TM Forum Autonomous Networks Levels | Level taxonomy and an assessment method for autonomy | Assessment methodology, with per-scenario external validation | Uses one external Level 4 assessment as a bounded evidence class and states explicitly what it does not validate | [6], [47] |
| NGMN agentic AI framework | Level 3–5 agentic operating-model taxonomy for operators | Industry framework; no measured deployment | Adds the tool-mediation, workload-identity, validation and audit subsystems such a loop requires, and names the charging and lawful-intercept gaps that remain open | [25] |
| NTT DOCOMO 6G vision | Cyber-physical fusion vision and service direction | Operator vision document | Converts an outcome vision into priced tiers with service-level objectives, breach ladders and stated enforcement prerequisites | [14] |
| O-RAN nGRG research baselines | Near-RT RIC evolution, dApps and the E3 interface, scalable RAN | Contributed research reports; an over-the-air prototype for dApps | Positions the RIC and dApp tiers as the execution path beneath an agent-facing layer that replaces no standardized interface | [18]–[20] |
| AI-RAN Alliance architecture and roadmap | AI-for-RAN and AI-on-RAN platform classes and deployment profiles | Alliance architecture and working-group reports | Embeds the platform in an ownership, commercial and governance framework grounded in operator deployment | [11]–[13] |
| Intent-based networking | Intent as declarative goals and outcomes; intent lifecycle, abstraction and assurance | Primary concept definition and a systematic survey | Treats intent handling as one subsystem under operator-owned identity, validation and audit, and binds intents to commercial remedies rather than to configuration alone | [42], [43] |
| LLM agents for network management | LLM-generated configuration; bounded policy-governed agents above the deterministic control plane; device–edge–core feasibility; MCP token- and | Peer-reviewed experiments and feasibility studies, plus one preprint cited for prior-art positioning rather than as a measurement base. None | Specifies the operator-side controls such agents require in production: tool mediation, workload identity, risk-tiered validation, and named rollback ownership | [28], [29], [121], [122], [123] |

| Approach or literature | What it provides | Evidence status | What this article adds | Ref. |
|---|---|---|---|---|
| | latency-cost profiling; network-aware MCP tool routing | measures MCP overhead inside a 3GPP or O-RAN enforcement loop | | |
| Network slicing and SLA economics | Slice admission control, SLA-constrained resource reservation, and slice pricing | Peer-reviewed optimization literature: one analytical pricing-and-admission framework, one measurement-driven overbooking study | Both optimize expected operator profit under SLA-violation risk, [124] through demand forecasting and slice pricing and [125] through admission control and overbooking. The Guarantee Economy proposes in addition a contracted outcome per tier with enforcement and attestation, intended to lower the breach probability $\beta$ that (8) prices; neither that effect nor any advantage over these studies is evaluated here; Appendix C, (8)–(10), states the break-even condition. A broader review of this literature is outstanding and is declared as such rather than implied to be complete | [124], [125] |
| Digital twins and latent world models for control validation | Predictive validation of proposed actions before execution | Foundational results outside telecommunications; no demonstrated safe telecom control-loop validation | Specifies the safety-gate parameter contract such a layer must satisfy (Appendix F, Table 17) and reclassifies the unsupported parts as research directions | Section VII.E |

TABLE 15. OWN, FEDERATE, CONSUME – OWNERSHIP COST AND ITS LIMITS

| Stance | Software maintenance burden | Security responsibility | Integration complexity | Talent profile and scarcity | Lifecycle cost drivers | When the stance stops being right |
|---|---|---|---|---|---|---|
| Own | Continuous ownership of upgrade, patch, regression and end-of-life for the component, at the upstream release cadence. | Operator carries vulnerability management, secure development lifecycle, incident response and, where applicable, conformity assessment. | Concentrated in the operator's platform team and re-incurred at every vendor and release change; highest initial cost, lowest marginal cost per subsequent change. | Platform, site-reliability, data and ML engineering at a level that competes directly with hyperscalers for the same scarce staff; retention is the binding constraint. | Headcount, test infrastructure, release engineering, and the servicing of accumulated technical debt. | When the component confers no differentiation; when the upstream release cadence exceeds the operator's absorption capacity; or when the required staff cannot be retained at any defensible cost. |
| Federate | Shared across participants, with governance overhead substituting for engineering overhead. | Shared, with the least mature participant setting the practical floor; assurance depends on what participants will disclose to each other. | Interoperability, versioning and conformance testing dominate; bilateral exceptions erode the benefit quickly. | Standards, architecture and commercial-governance capability rather than deep implementation depth. | Participation cost, conformance testing, and the coordination latency of collective decisions. | When the federation cannot converge within the operator's planning horizon, or when a single participant's failure would be attributed to the operator by its own customers. |
| Consume | Carried by the supplier; the operator carries integration, monitoring and exit only. | The supplier's posture becomes the operator's risk; contractual transfer is limited to what the operator can actually audit. | Lowest build cost and highest switching cost; complexity reappears as data-egress and exit-path work. | Procurement, assurance, and exit-planning skills; commodity integration engineering. | License escalation, egress and exit cost, and the gradual loss of configuration control. | When the component becomes the boundary at which the operator's own service-level objectives must be enforced, or when data-access rights sufficient to train, audit or verify cannot be secured contractually. |

TABLE 16. THREATS, PROPOSED CONTROLS, AND RESIDUAL RISK

| Threat | Attack surface | Proposed control | Detection signal | Recovery | Residual risk | Validation status |
|---|---|---|---|---|---|---|
| Prompt injection and tool poisoning | Agent context: tickets, logs, alarms, and third-party tool descriptions and outputs | Untrusted-content isolation; per-intent tool allow-lists; parameter schema validation; no privilege inherited from content | Tool-call sequences inconsistent with the intent class; schema rejections; parameters outside policy ranges | Suspend the workflow, revoke the session identity, and replay the audit log to establish what was executed | Injection that induces an in-policy but harmful action sequence is not caught by schema validation | Not evaluated; adversarial testing required |
| Confused deputy and authority escalation | The router mediating calls on behalf of multiple principals | Per-request workload identity with short-lived credentials; authorization evaluated against the requesting principal, never against the router | Authorization decisions in which principal and effect scope diverge | Revoke credentials, re-authorize, and audit executed writes | Correct-but-overbroad grants remain possible wherever intent scopes are coarse | Not evaluated |
| Model drift and silent degradation | Model updates, input distribution shift, retrieval-corpus change | Version pinning with staged rollout; shadow evaluation against a labelled corpus; decision-quality monitoring | Falling agreement with expert labels; rising escalation or rollback rate | Pin to the last known-good version; fall back to deterministic automation | No agreed decision-quality metric exists, and the labelled corpus itself must be maintained as the network changes | Not evaluated; corpus and metric both open (Section VII.E) |
| Conflicting concurrent intents | Multiple agents, xApps and dApps acting on overlapping resources | Resource locks; priority rules; pre-execution conflict checks; human escalation on unresolved conflict | Lock contention; conflicting proposed deltas on the same managed object | Abort the lower-priority intent and reconcile | A fixed priority hierarchy prevents neither priority inversion nor starvation | Not evaluated |
| Compromised or replayed agent identity | Workload credential issuance, attestation, and transport | Attested short-lived identities; audience-bound tokens; replay protection | Credential use outside the expected attestation, audience or validity window | Revoke, rotate, and review every action taken under the credential | Revocation propagation time is a deployment property and is not bounded by this architecture | Not evaluated; revocation timing unvalidated |
| Unsafe action on stale state | Digital twin or telemetry lagging the live network | A declared state synchronization error bound with a fail-closed default, distinct from a staleness timeout; pre-commit re-read | Twin-to-network divergence exceeding the declared bound | Abort, resynchronize, and escalate | Divergence is observable only for state the network actually exposes for comparison | Not evaluated; bound to be declared per domain |

TABLE 17. SAFETY-GATE PARAMETER CONTRACT

| Parameter | Specification – value or basis | Escalation trigger |
|---|---|---|
| Tier assignment and screening latency, $T_{tier}$ | Twin-routed, not screenable, if the blast radius spans more than one administrative domain, if no automated inverse action exists, if the template lies outside the screening model's training support, or if any KPI in the radius has less margin to its objective than the tier's interval half-width. $T_{tier}$ within the Fig. 7 Decide budget, values set at commissioning. | Any condition met, or 99th-percentile $T_{tier}$ above that budget. |
| Escalation fraction, $p$ | $p = q + p'(1 - q)$: $q$ non-screenable by the tier-assignment rule, $p'$ escalated by the state, horizon, confidence or support conditions. All three are unmeasured, so $p$ is set once $q$ is measured (Section XII). | $p$ rising against its commissioning baseline. |
| State synchronization error | Per state class. Topology, inventory and blast-radius attributes must match exactly. KPI state admitted only if $\varepsilon_{sync} = \max_j \lvert\hat{x}_j - x_j\rvert / m_j^0 \leq \varepsilon_{max}$, with $m_j^0$ the margin of KPI $j$ at the nominal operating point; counters compared as rates. $\varepsilon_{max}$ set per class at commissioning; no measured value is claimed. | $\varepsilon_{sync} > \varepsilon_{max}$, or any exact-match class mismatched. |
| State age | Bounded separately from $\varepsilon_{sync}$: a twin matching now may have diverged unobserved since reconciliation. $\tau_{max}$ is an integer multiple of the Fig. 7 Observe period, the multiple set at commissioning. | Age > $\tau_{max}$. |
| Prediction horizon, $H$ | Two-sided: $t_{eff} \leq H \leq H_{cal}$, with $t_{eff}$ the settling time of the affected KPI – not the action's execution duration – and $H_{cal}$ the longest horizon at which coverage is measured. Both are unmeasured, so $H$ is not fixed; coverage-affecting classes such as tilt and handover hysteresis stay non-screenable until $H_{cal}$ is measured. | $t_{eff} > H_{cal} \rightarrow$ twin, otherwise human review. |
| Confidence threshold, $1 - \alpha$ | Admit only if the whole interval lies inside the envelope over the whole horizon. Nonconformity is the maximum scale-normalized deviation across KPIs and horizon steps; being a scalar over a finite index set, one quantile of it bounds every coordinate at once, so coverage is simultaneous and no multiplicity correction is required. Requires coverage conditional on selection, not marginal coverage [117], [119]. $1 - \alpha = 0.99$, uniform and operator-set; calibration then requires at least $\lceil 1/\alpha \rceil - 1 = 99$ exchangeable points per action class. | Interval straddles an envelope boundary. |
| Calibration monitoring | Empirical against nominal coverage over a rolling window, as a ratio to nominal miscoverage so the monitor binds at any $\alpha$; reported to subsystem (D) as a drift signal. $W = 10^4$ gated actions, ratio $\leq 2$ – 200 against 100 expected, about ten standard errors, if errors are independent. | Ratio > 2 over $W$. |
| Out-of-distribution support | Scalar novelty score on the proposed state, thresholded on held-out in-distribution data. Fixes the false-escalation rate only; false admission of a genuinely novel state has no distribution-free bound. Threshold set at commissioning, no rate claimed; false admission to be estimated against withheld templates and topology regions (Section XII). | Score beyond threshold. |
| Fail-closed default | Any gate component unavailable → action denied, the audit record distinguishing unavailability from validated rejection; human review queued under a bounded lease, denied on expiry. Exception: pre-authorized, twin-pre-validated containment templates – throttle, quarantine, credential revocation – stay executable while the validator is down, since strict fail-closed would let an adversary suppress containment by stalling the gate; rate-limited, leased, and logged as a priority-1 incident. | Either budget exceeded. |

TABLE 18. CANDIDATE WAVEFORMS AND RECEIVER ARCHITECTURES

| Waveform or technique | Behavior in doubly selective channels | PAPR | Receiver complexity | Sensing and positioning suitability | NTN suitability | Standardization status |
|---|---|---|---|---|---|---|
| CP-OFDM | Per-subcarrier flat-fading assumption holds only while Doppler spread is small relative to subcarrier spacing; inter-carrier interference grows with normalized Doppler | High | Low: FFT plus single-tap per-subcarrier equalization | Well established for wideband radar-style processing; range resolution set by bandwidth | Requires Doppler pre-compensation and wider subcarrier spacing | 5G NR baseline; agreed 6G downlink and uplink waveform (RAN1#125) |
| DFT-s-OFDM | Same time-frequency behavior as CP-OFDM; the single-carrier property does not remove inter-carrier interference | Low | Low to moderate | Usable, with less flexible resource shaping | Favored for power-limited uplink | NR uplink; agreed 6G uplink waveform (RAN1#125); FDSS enhancement under down-selection |
| OTFS | Multiplexes in the delay–Doppler domain, where the channel is sparse and approximately time-invariant over the frame; each symbol experiences full channel diversity | Moderate to high | High: two-dimensional equalization over the delay–Doppler grid | Delay–Doppler representation aligns naturally with range–Doppler estimation | Robust to large Doppler spread | Research waveform; not selected for the 6G radio |
| AFDM | Chirp-based multicarrier via the discrete affine Fourier transform; with correctly set chirp parameters the channel becomes sparse and full delay–Doppler diversity is achievable | Comparable to OFDM | Moderate: a one-dimensional transform, with diversity conditions on the pre-chirp parameter | Chirp structure suits joint communication and sensing | Strong under high mobility and large Doppler | Research waveform; not selected for the 6G radio |
| Index-modulated variants (AFDM-IM, OTFS-IM) | Activate a subset of chirps or delay–Doppler bins and carry additional bits in the activation pattern; full diversity is preserved under stated conditions | Variant-dependent | Joint activation-pattern and symbol detection; low-complexity detectors exist [101]; spectral efficiency is lost to inactive resources | Inherits the suitability of the host waveform | Inherits the host waveform; energy efficiency may favor power-limited links | Research [101] |
| AI-assisted receivers | Applied on top of any of the above; learned estimation and equalization can track within-symbol variation without changing the waveform | Unchanged by the receiver | Shifts cost from closed-form estimation to inference; it requires accelerated compute in the receive path | Can improve parameter estimation; introduces a model-validity question for safety-relevant sensing | Promising where Doppler statistics are predictable from ephemeris | Release 18/19 AI-assisted features; 6G study. The position in Section VIII is to standardize inputs and outputs, not algorithms |

## DISCLOSURE

The assumptions and views reported herein are solely those of the authors and do not necessarily represent those of Rakuten or its affiliates. The Control Compact and Guarantee Economy framing used here also appears in a single-author preprint by the corresponding author [126]. The present article is a distinct six-author contribution: it specifies the agentic enforcement architecture, derives the supporting analytical models, and classifies the provenance of the material quantitative claims.

## ACKNOWLEDGMENT

This article draws on the authors' knowledge, work experience and contributions to the ICT sector, together with the work of the ITU-R, 3GPP, NGMN Alliance, O-RAN Alliance, ETSI, AI-RAN Alliance, the Linux Foundation, and IOWN Global Forum, cited throughout. The use of AI tools in preparing this article is declared in the AI Use Declaration following Section XIII. The authors thank the Rakuten Mobile Inc. engineering and operations teams, whose work operating a large-scale, fully cloud-native Open RAN mobile network provides much of the deployment context discussed in this article. The authors also acknowledge Sharad Sriwastawa of Rakuten Mobile and Rakuten Symphony for the opportunity to work on this subject and for his invaluable support and contribution.

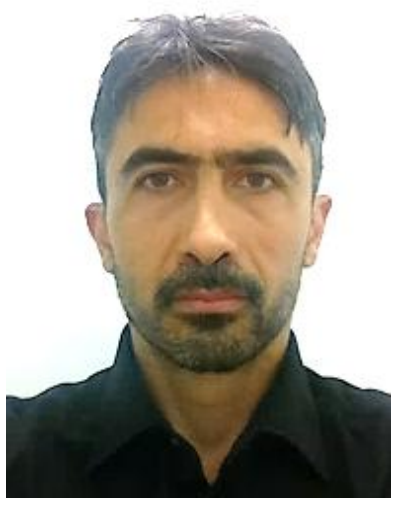

**DAVID SOLDANI** (Senior Member, IEEE) received the M.Sc. degree (magna cum laude) in engineering from the University of Florence, Italy, in 1994, and the D.Sc. degree in technology (Hons.) from the Helsinki University of Technology, Finland, in 2006. He has held professorships at the University of Surrey, the University of Technology Sydney, and the University of New South Wales, and executive roles in technology, advanced research, information security at Nokia, Huawei, and Rakuten. He is currently SVP Advanced Research and Innovation with Rakuten Mobile Inc., Tokyo, Japan.

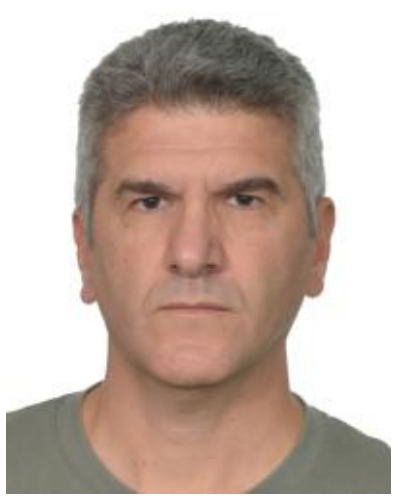

**PETRIT NAHI** is Chief AI Consultant in Rakuten Mobile's AI and Data Division and Senior Member of Technical Staff in the CTO Office. He built and led Rakuten Mobile's original Data Science team, which evolved into the AI and Data Division, establishing the company's data and AI foundations for large-scale network automation and autonomy. He has played a key role in applying AI/ML to live Open RAN network automation, supporting predictive and reactive models across approximately 150,000 cells and Rakuten Mobile's progress toward Level 4 energy autonomy. His doctoral research at Queen Mary University of London focused on agent-based control of mobile network coverage, an early form of AI-driven network control. He has held technical, product, executive, and board-level roles across Rakuten Mobile, NETSCOUT, Tektronix Communications, Newfield Wireless, and AIRCOM, and formerly served as Chairman of the Board of Kosovo Telecom.

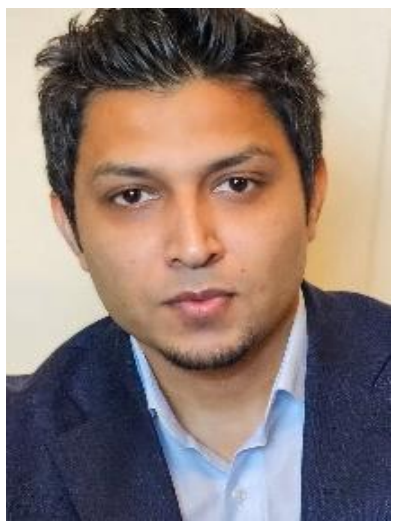

**AWN MUHAMMAD** is Head of Technical Standards at Rakuten Mobile, with over 18 years of experience in mobile networks, including RAN planning, feature development, deployment, and standardization. He represents Rakuten Mobile as Alternate Director on the O-RAN ALLIANCE Board and actively contributes to O-RAN ALLIANCE and 3GPP RAN activities, with a focus on Open RAN, energy efficiency, 5G Advanced, and 6G evolution.

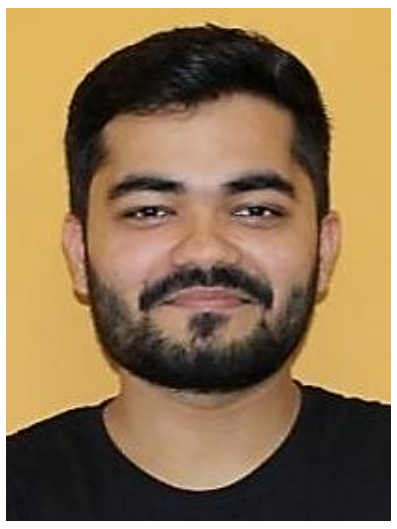

**NIKHIL DWIVEDI** received the B.Tech. degree in Electrical Engineering from Delhi Technological University (DTU), Delhi, India, and the M. Tech. degree in Artificial Intelligence from the Indian Institute of Technology Jodhpur (IIT Jodhpur), India. He is currently a Research Engineer in the Next Generation Advanced Research Department at Rakuten Mobile Inc., Japan. His work focuses on developing artificial intelligence and deep learning solutions for next-generation telecommunications and cybersecurity, with an emphasis on foundation models, trustworthy AI, and translating research into production-scale systems. His research interests include artificial intelligence, machine learning, foundation models, AI for telecommunications, cybersecurity, and open-source AI technologies.

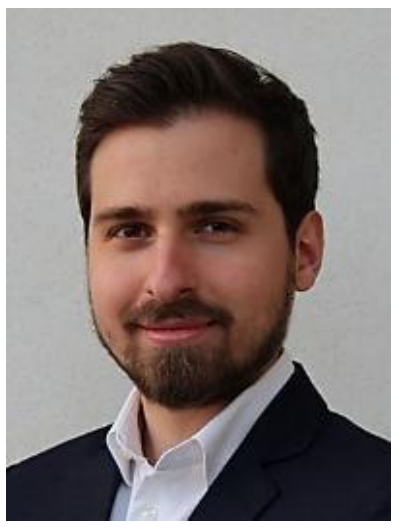

**FRANCESCO MONACO** received the Master of Science (M.Sc.) degree in Computer Engineering from Politecnico di Torino, Italy, in 2022, where he has also collaborated as a Research Assistant. He joined Rakuten Mobile Inc., Japan, in 2024, where he currently serves as an Architect/Tech Lead and Operations Manager in the Next Generation Advanced Research Department. He has been a key contributor to the conception, development, and large-scale deployment of the Sauron eBPF Platform across the Rakuten Mobile network. His interests include high-performance networking, programmable data plane, cloud-native technologies, and production-scale network observability systems.

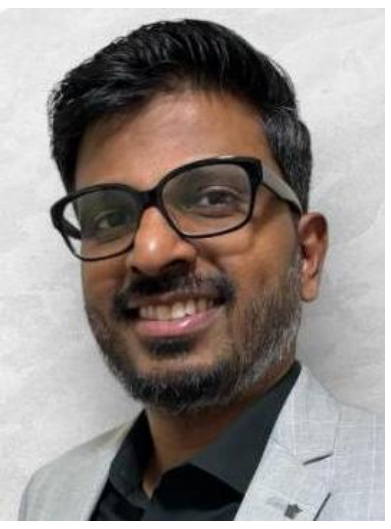

**FRANCIS JEBAMANI** is the Head of AI Products & Engineering at Rakuten Mobile Inc., with over two decades of experience in the Telco industry. His career spans Radio Access Networks (RAN), Multi-access Edge Computing (MEC), Core Networks, and AI-driven products, with a focus on delivering intelligent solutions that enhance network performance, energy efficiency, operational efficiency, and end-user experience. Before joining Rakuten Mobile, he spent nearly five years at Reliance Jio, where he contributed to the architecture, engineering, and scaling of the company's nationwide 4G Core network from the ground up. He currently leads the development of AI-powered products that accelerate intelligent automation and evolution toward autonomous networks. He also played a key role in achieving TM Forum Autonomous Networks Level 4 for Energy Efficiency. His professional interests include AI for telecommunications, network intelligence, autonomous network operations, and energy-efficient AI solutions.